\documentclass[fleqn,usenatbib]{mnras}

\usepackage{newtxtext,newtxmath}

\usepackage[T1]{fontenc}

\DeclareRobustCommand{\VAN}[3]{#2}
\let\VANthebibliography\thebibliography
\def\thebibliography{\DeclareRobustCommand{\VAN}[3]{##3}\VANthebibliography}

\usepackage{graphicx}	
\usepackage{amsmath}	
\usepackage{subcaption}

\defcitealias{Martinez-Delgado_et_al_2023}{MD23}
\defcitealias{Miro-Carretero_et_al_2023}{MC23}
\defcitealias{Miro-Carretero_et_al_2024}{MC24}

\title[STRRINGS streams]{STRRINGS: STReams in Residual Images of Nearby GalaxieS}

\author[E. Sola et al.]{
Elisabeth Sola$^{1}$\thanks{Corresponding author: es2074@cam.ac.uk},
David Chemaly$^{1}$,
Vasily Belokurov$^{1}$,
Oliver M\"{u}ller$^{2,1,3}$,
Anke Ardern-Arentsen$^{1}$,
\newauthor Elliot Y. Davies$^{1}$, 
J\'ulia Laguna-Miralles$^{1}$,
GyuChul Myeong$^{1}$,
Konstantinos Panagiotakis$^{4}$,
HanYuan Zhang$^{1}$,
\newauthor 
Denis Erkal$^{5}$,
Sergey E. Koposov$^{6,1}$,
Dustin Lang$^{7}$,
Jacob Nibauer$^{8}$
\\
$^{1}$ Institute of Astronomy, University of Cambridge, Madingley Rd, Cambridge CB3 0HA, UK\\
$^{2}$ Institute of Physics, Laboratory of Astrophysics, Ecole Polytechnique F\'{e}d\'{e}rale de Lausanne (EPFL), 1290 Sauverny, Switzerland\\
$^{3}$ Visiting Fellow, Clare Hall, University of Cambridge, Cambridge, UK\\
$^{4}$ Università Degli Studi di Padova - Dipartimento di Fisica e Astronomia G. Galilei \\
$^{5}$ Department of Physics, University of Surrey, Guildford GU2 7XH, UK\\
$^{6}$ Institute for Astronomy, University of Edinburgh, Royal
Observatory, Blackford Hill, Edinburgh EH9 3HJ, UK\\
$^{7}$ Perimeter Institute for Theoretical Physics, 31 Caroline St. North, Waterloo, ON N2L 2Y5, Canada\\
$^{8}$ Department of Astrophysical Sciences, Princeton University, 4 Ivy Lane, Princeton, NJ 08544, USA\\
}

\date{Accepted 2025 September 25. Received 2025 September 24; in original form 2025 August 1}

\pubyear{\the\year{}}

\begin{document}
\label{firstpage}
\pagerange{\pageref{firstpage}--\pageref{lastpage}}
\maketitle

\begin{abstract}

Tidal features from galaxy mergers, particularly stellar streams, offer valuable insights into galaxy assembly and dark matter halo properties. This paper aims to identify a large sample of nearby stellar streams suitable for detailed modelling and comparison with simulations to enable population-level constraints on halo properties.
We visually inspect and compile a tidal feature catalogue for  $19,387$ galaxies with redshift $z \leq 0.02$ from the Siena Galaxy Atlas 2020 using original, model, and residual images from the DESI Legacy Imaging Surveys. Residual images, produced by subtracting models of all sources, enhance the detectability of faint asymmetries such as tidal features.
We find that $11.9 \pm 0.2\%$ of galaxies host detectable tidal features, more frequently around early-type than late-type galaxies. The tidal feature fraction increases with stellar mass, from $2.4 \pm 0.4\%$ at $\sim10^8$M$_\odot$ to $36.5 \pm 1.2\%$ at $\sim 5\times10^{11}$M$_\odot$. From this, we present the first release of STRRINGS: STReams in Residual Images of Nearby GalaxieS, a subsample of 35 galaxies with long, narrow streams suitable for modelling. We segment these streams and derive their geometry, surface brightness, colours, and stellar masses.
The median $g$-band surface brightness is 26.8 mag$\,$arcsec$^{-2}$, reaching 27.5 mag$\,$arcsec$^{-2}$ for the faintest stream. Mass ratios are consistent with minor mergers, and we identify five potential dwarf galaxy progenitors. Our streams are typically longer (median 124 kpc) than the literature, with comparable widths. Stream mass correlates with length and colour, and wider streams lie at larger galactocentric radii.
STRRINGS will be expanded and used to constrain halo properties in future work.

\end{abstract}




\section{Introduction} \label{section:introduction}

In the framework of the $\Lambda$CDM paradigm, the hierarchical growth of galaxies is driven mostly by successions of mergers and accretion of gas \citep[e.g.,][]{White_and_Rees_1978,Davis_1985,White_and_Frenk_1991,Moore_et_al_1999,Cole_2000,Springel_2006}. Merger events play a fundamental role in shaping galaxies, as they can drastically modify their properties, such as morphology, gas content, star formation, kinematics, active galactic nuclei activity, black hole merging \citep[e.g.,][]{Balcells_and_Quinn_1990,Barnes_2004,DiMatteo_2009,Cooper_2010,Bois_2011,Conselice_2014,Somerville_and_Davé_2015,Knapen_2015,Rodriguez-Gomez_2017,Naab_and_Ostriker_2017,Blumenthal_and_Barnes_2018,Pearson_2019,Yoon_2022,Bilek_2023,Comerford_2024}. A key prediction of hierarchical models is the formation of tidal features, which are stars and gas stripped from their galaxy during gravitational interactions.  

Tidal features hold the imprint of the late assembly history of galaxies, as their properties are a function of the type of merger that happened. Numerical simulations help to shed light on the relations between tidal features properties and the mass ratios, morphology, impact parameters, or velocities of the merging galaxies \citep[e.g.,][]{Johnston_2008,Amorisco_2015,Pillepich_2015,Pop_2018,Mancillas_2019,Karademir_2019}. For instance, major mergers leave behind tidal tails \citep[e.g.,][]{Arp_1966,Toomre_and_Toomre_1972,Barnes_1988,Duc_and_Renaud_2013}, whose material comes from the disturbed host galaxy. Streams originate from disrupted satellite galaxies during minor mergers \citep[e.g.,][]{Lynden-Bell_1995,Ibata_2001,Bullock_and_Johnston_2005,Belokurov_2006,Martinez-Delgado_2010,Newberg_and_Carlin_2016}, while shells are formed mostly during radial mergers \citep[e.g.,][]{Quinn_1984,Prieur_1990,Ebrova_2013,Pop_2018}.

Tidal features, particularly stellar streams from minor mergers, also serve as valuable probes of both the host galaxy and its dark matter halo. The stars within these streams approximately trace the orbit of their progenitor \citep[e.g.,][but also \cite{Sanders_and_Binney_2013}]{Law_2005,Binney_2008,Eyre_and_Binney_2009}, which evolves over time within the gravitational potential of the host galaxy. This makes it possible to investigate the gravitational potential, mass and dark matter halo distribution of the galaxy \citep[e.g.,][]{Johnston_1999,Ibata_2001,Ibata_2003,Helmi_2004,Johnston_et_al_2005,Fellhauser_2006,Newberg_2010,Koposov_2010,Varghese_2011,Law_and_Majewski_2010,Lux_2013,Erkal_et_al_2016,Bovy_2016,Bonaca_2018}. 

The advent of photometric \citep[e.g., SDSS, 2MASS, DES, ][]{York_2000,Skrutskie_2006,DES_2016}, spectroscopic \citep[LAMOST, APOGEE, ][]{Cui_2012,Majewski_2017} and astrometric surveys \citep[Gaia,][]{Gaia_2016} allowed for a precise characterisation of the stream system of the Milky Way (MW) \citep[e.g., see review by][]{Bonaca_and_Prince-Whelan_2025}. When the 6D phase-space information is available, tight constraints can be put on the galaxy mass and dark matter halo shape \citep[e.g.,][]{Malhan_and_Ibata_2019,Dodd_2022,Koposov_2010,Koposov_2023}. 
Constraints can also be derived using reduced-dimensional datasets. For example, \cite{Erkal_et_al_2019} utilised 5D information on the Orphan stream (on-sky position, distance, and proper motions) and showed that the observed misalignment between stellar velocity vectors and the stream track could be attributed to the significant gravitational influence of the Large Magellanic Cloud (LMC).
They used this 5D information and simultaneously fitted the Milky Way and LMC potential to provide a mass estimate of the LMC. \cite{Fardal_et_al_2013} used radial velocities and distance measurements of the Giant Stellar Stream to infer the mass of M31's halo.

Expanding such studies beyond the Local Group represents the next step in probing the distribution of dark matter halo properties across large statistical samples of galaxies. However, at such distances, individual stars can no longer be resolved, and tidal features can only be identified through their diffuse light. Photometric surveys provide only two-dimensional spatial positions and brightness measurements along the stream, while follow-up spectroscopic observations, which are necessary for obtaining velocity information, remain very limited.
 A first attempt to develop a framework for modelling extragalactic streams was carried out by \cite{Johnston_et_al_2001}, where they relate the observed surface brightness and geometry of a stream to the age and mass of the progenitor and to the mass of the host galaxy. 
Later, \cite{Martinez_Delgado_et_al_2008} compared the projected geometry of the stream around NGC5907 to the predictions of N-body simulations to gather hints on the accretion history of this galaxy. \cite{Amorisco_et_al_2015} modelled the 2D surface brightness map of the stream around NGC1097 and reconstructed the accretion scenario of a progenitor dwarf galaxy on the host. The 2D map was informative enough to put constraints on the density profile of NGC1097, on the initial properties of the dwarf galaxy and its evolution in the host's potential. More recently, \cite{Pearson_2022} showed that adding one radial velocity of the progenitor helps breaking the degeneracy between the host halo mass and the stream morphology.
\cite{Nibauer_et_al_2023} showed that the curvature of the projected 2D stream track can provide information on the halo shape parameters and place limits of disk-to-halo mass ratios. 
\cite{Walder_2024} showed that streams wrapping multiple times around their host galaxies can constrain the overall radial profile and scale radius of the potential (assumed to be spherical), even without radial velocities. Similarly to \cite{Pearson_2022}, they found that adding one radial velocity measurement can break degeneracies between halo mass and progenitor velocity.  
Using numerical simulations, \cite{Chemaly_et_al_2025} recently demonstrated that the flattening of dark matter halos can be constrained at a population level, provided a sufficiently large number of streams (more than 50) are analysed and modelled. This finding highlights the need for its application to observational data. In this paper, we aim at disclosing a large population of streams in external galaxies, suitable for modelling using the method described in \cite{Chemaly_et_al_2025}.

Over the past decade, the search for stellar streams and tidal features around external galaxies has rapidly advanced. Their detection was long challenged by their intrinsically low surface brightness (LSB), but has become increasingly possible due to progress in observational strategies, CCD advancements, and improved data processing pipelines. Searches for LSB tidal features has been mostly conducted visually \citep[e.g.,][]{Schweizer_and_Seitzer_1992, Tal_et_al_2009, Kaviraj_2010, Sheen_et_al_2012, Atkinson_Abraham_Ferguson_2013, Casteels_et_al_2013, Morales_et_al_2018,Hood_et_al_2018, Muller_et_al_2019, Bilek_et_al_2020,Yoon_and_Lim_2020, Sola_et_al_2022,Sola_et_al_2025, Huang_and_Fan_2022, Jackson_et_al_2023, Martinez-Delgado_et_al_2023, Skryabina_et_al_2024, Miro-Carretero_et_al_2024, Yoon_et_al_2024,Pippert_et_al_2025}. The advent of surveys covering thousands of square degrees such as Euclid \citep{Borlaff_2022} or Rubin/LSST \citep{Martin_et_al_2022} will push even further the limit of the faint Universe. Automated methods \citep[e.g.,][]{Conselice_et_al_2003, Abraham_et_al_2003, Lotz_2004, Pawlik_et_al_2016, Kado-Fong_et_al_2018, Mantha_et_al_2019,Hendel_et_al_2019} will become necessary to deal with the huge amount of LSB-compliant images. 
Another avenue in the search for tidal features is machine learning, to quickly process millions of images, whether using supervised learning (such as convolutional neural networks) that require training sets of pre-annotated images \citep[e.g.,][]{Walmsley_et_al_2019,Pearson_et_al_2019,Bickley_et_al_2021,Walmsley_et_al_2022,Dominguez-Sanchez_2023,Gordon_et_al_2024}, semi-supervised or unsupervised learning \citep[e.g.,][]{Desmons_et_al_2024}.

One could think that the aforementioned studies provide good datasets of streams suitable for modelling, when publicly available. Yet, significant differences exist in the images depth and processing, definitions of the different types of tidal features, methods, sample selection \citep[e.g.,][]{Atkinson_Abraham_Ferguson_2013,Hood_et_al_2018}, resulting in large discrepancies in the results. Most streams cannot even be modelled, either because their geometry is inadequate (e.g., too short, too straight, too bright - i.e. rather originating from a major merger) or because the host galaxy has recently undergone major mergers which makes impossible the estimation of its potential.

Therefore, the goal of this paper is to obtain a catalogue of streams around nearby galaxies that are suitable to be modelled for the inference of the properties of dark matter haloes at a population level.
We visually inspect residual images from the Dark Energy Spectroscopic Instrument Legacy Imaging Surveys \citep[DESI-LS,][]{Dey_et_al_2019} to build and publicly release a catalogue of tidal features around nearby galaxies. From it, we select a sample of long, narrow and curved streams suitable to be modelled. Note that while \cite{Nibauer_et_al_2023} demonstrated that long streams with straight segments are highly informative, our definition of streams selected for modelling requires the presence of some curvature, rather than consisting solely of short, straight segments. The application of the method from \cite{Chemaly_et_al_2025} on this dataset will be presented in \cite{Chemaly_et_al_2025b}. 

Although other studies have also used DESI-LS to provide catalogues of tidal features and streams \citep[e.g.,][]{Walmsley_et_al_2022,Miro-Carretero_et_al_2024,Yoon_et_al_2024}, we leverage for the first time the publicly available DESI-LS residual images. Residual images \citep[e.g.,][]{Bell_et_al_2006,McIntosh_et_al_2008,Tal_et_al_2009} are obtained by building a model of every astronomical source in an image, and then subtracting it from the original image. This leaves behind asymmetries such as faint, LSB features, therefore greatly enhancing their detectability. By inspecting a combination of DESI-LS original, model and residual images of nearby galaxies, we increase our chances to find streams suitable for modelling. That search will be extended in future works where we will apply machine learning algorithms \citep[for instance, by fine-tuning the Zoobot model from][]{Walmsley_et_al_2022b} on the combination of DESI-LS images and our tidal feature labels to more distant galaxies.

The paper is organised as follows. In Section \ref{section:data} we describe the imaging data and galaxy catalogue we used. We explain our tidal feature identification method in Section \ref{section:method}. We present the results on the whole tidal feature catalogue in Section \ref{section:results}. Our STRRINGS sample of streams selected for modelling along with their properties is presented in in Section \ref{section:best_streams}. We discuss our findings in Section \ref{section:discussion} before summarising our conclusions in Section \ref{section:conclusion}.

\section{Data} \label{section:data}
\subsection{DESI Legacy Imaging Surveys}\label{sec:desi-ls}

The Dark Energy Spectroscopic Instrument (DESI) Legacy Imaging Surveys\footnote{\href{DESI-LS}{https://www.legacysurvey.org/}} \citep[DESI-LS][]{Dey_et_al_2019} are a collection of images from three ground-based surveys. The goal is to provide images to the DESI instrument, which is a multi-object spectrograph that will measure the spectra of more than 30 millions of galaxies and quasars over 14,000 $\deg^2$ \citep{DESI_2016a,DESI_2016b}. 

The DESI-LS was conducted using telescopes located at the Kitt Peak National Observatory and the Cerro Tololo Inter-American Observatory, namely the Dark Energy Camera (DECam) Legacy Survey (DECaLS), the Beijing-Arizona Sky Survey (BASS) and the Mayall $z-$band Legacy Survey (MzLS). These surveys cover $14,000 \deg^2$ in the Northern hemisphere in the $g, r, z$ optical bands. The images have been processed in a consistent way that enable to reach approximate $5\sigma$ depth of $g=24.0, r=23.4$ and $z=22.5$ AB magnitudes for faint galaxies \citep{Dey_et_al_2019}. Four mid-infrared bands from the WISE Wide-field Infrared Survey Explorer (WISE) satellite are complementing DESI-LS. Additional $g,r,i,z$ images from public DECam data have recently been added, expanding the sky coverage of the DESI-LS to more than $20,000 \deg^2$ in the data release 10 (DR10). 

In addition to images, DESI-LS provides source catalogues that are generated using an inference-based forward-modelling approach, the \texttt{Tractor} \citep{Lang_et_al_2016}. The processing pipeline, \texttt{legacypipe}\footnote{\href{legacypipe}{https://github.com/legacysurvey/legacypipe}} \citep{Dey_et_al_2019}, starts by extracting sources in the individual $g,r,z$ bands  and stacked images. Each source is then modelled by the \texttt{Tractor} as either a point source, a de Vaucouleurs profile, an exponential profile, a Sérsic profile or {a round exponential galaxy model (i.e. with zero ellipticity) with a variable half-light radius. 
The best-fit model (based on a likelihood estimation) is kept, and forced photometry is applied to the mid-infrared bands. This produces catalogues of object position, fluxes and colours measured in a consistent way across all DESI-LS images. Therefore, for each image, a model is created and a residual image is obtained by subtracting the model from the image. 

Residual images play a central role in our detection of faint tidal features. Because standard modelling algorithms are optimised for compact, high surface brightness structures, they typically fail to capture the extended, diffuse light from LSB features. As a result, these structures are not included in the model and remain visible in the residuals. In an ideal case, residual images should be flat, with any significant deviation indicating the presence of unmodelled components such as tidal features or other faint structures.
An illustration of the \texttt{legacypipe} pipeline applied to an image of a galaxy with a stream is shown in Figure \ref{figure:data-img_model_res}. For better visualisation, a black-and-white residual image can be produced by computing the median of the residual images in g, r, and z, then clipping the pixel distribution. However, any inaccuracies in the modelling of stars, galaxies, or internal galactic structures can introduce artefacts in the residual images. High residuals also commonly appear around bright stars, in the extended stellar haloes of galaxies. To mitigate these issues, tidal feature classification must be based on the simultaneous inspection of the original, model, and residual images in order to distinguish real features from artefacts. Stretching the original image would obviously make tidal feature identification easier, but they appear more clearly in the residuals.

The other key advantage of residuals lies in the possibility to segment tidal features more easily than on the original image. Segmentation is a key part of tidal features analyses and studies, as it enables to retrieve geometrical properties and automated aperture photometry. Yet, for now segmentation has been performed manually \citep[e.g.,][]{Sola_et_al_2022,Sola_et_al_2025}. Automated efforts are ongoing \citep[e.g.,][]{Richards_et_al_2022,Richards_et_al_2023} but rely on previously manually-segmented features. Although in this paper stream segmentation will be carried out manually (see Section \ref{section:method-segmentation}), the long-term goal is to make the segmentation fully automated from the residual images. 

We estimate the depth of the residual DESI-LS images following the surface brightness limit definition of Appendix A of \cite{Roman_et_al_2020} as $3\sigma, 10\arcsec \times 10\arcsec$ (in boxes of $10\arcsec \times 10\arcsec$ randomly placed on images with all sources masked, and take the median over all boxes in all images). We find a surface brightness limit of 28.5, 29 and 27.4 mag$\,$arcsec$^{-2}$ in $r$, $g$ and $z$- band, respectively. These DESI-LS depth values are consistent with the ones reported in the literature: 28.4 mag$\,$arcsec$^{-2}$ in the $r$-band according to \cite{Skryabina_et_al_2024}, and from the coadded images it ranges between 26.99-28.98, 27.44-28.84 and 25.45-27.59 mag$\,$arcsec$^{-2}$ in $g,r$ and $z$-band, respectively \citep[for the `upper limit' surface brightness,][]{Miro-Carretero_et_al_2024}.

\begin{figure*}
    \includegraphics[width=\textwidth]{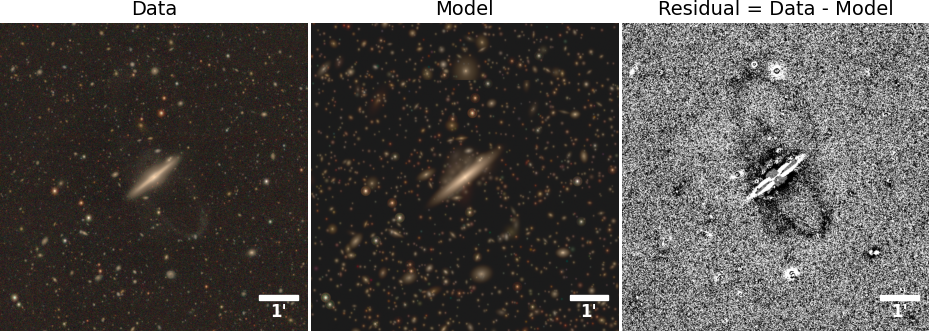}
    \caption{Illustration of the \texttt{legacypipe} processing of one image (UGC08717). A scalebar of $1'$ is indicated at the bottom of each image. North is up, East is left. \textit{Left:} original $grz$ data image. \textit{Middle:} \texttt{Tractor} $grz$ model image of all astronomical sources. \textit{Right:} residual image, obtained by subtracting the model from the original data. This black-and-white residual image is obtained by computing the median of the residual images in $g$, $r$ and $z$, then computing the 35-90\% percentile range of the pixel distribution of that median image and clipping the pixel values accordingly. White regions indicate areas where the model overestimates the light, while black regions show an excess of light in the data. The images is plotted as the median of the $g,r,z$ residuals, smoothed and stretched to enhance the visibility of faint features, that appear as black regions. Note that these images were used for photometric measurements, but a different set of images were used for visual inspection (i.e. the already available PNG montage of the $grz$ colour image, model and residuals, see Section \ref{section:data-images}).  }
    \label{figure:data-img_model_res}
\end{figure*}

\subsection{Siena Galaxy Atlas}
We select galaxies in the nearby Universe from the 2020 version of the Siena Galaxy Atlas (SGA), SGA-2020\footnote{SGA-2020, \url{https://sga.legacysurvey.org}} \citep{Moustakas_et_al_2023}. 
This atlas presents optical ($grz$) and mid-infrared (3.4–22\textmu m) imaging from the DESI-LS for 383,620 nearby galaxies, selected based on their apparent angular diameter. It includes galaxies within the $20,000 \deg^2$ footprint of DESI-LS DR9 and is 95\% complete for galaxies brighter than $r \simeq 18$ and have an angular diameter greater than $25\arcsec$ at a surface brightness level of 26 mag$\,$arcsec$^{-2}$ in the $r$-band.

In addition to the precise coordinates of the galaxies, their photometry and image model obtained with the \texttt{Tractor}, the SGA-2020 atlas provides extensive ancillary information\footnote{The description of all the columns available in the SGA-2020 atlas can be found at \url{https://www.legacysurvey.org/sga/sga2020/}} such as azimuthally average surface brightness and colour profiles, elliptical curves of growth and half-light radii. Morphological type, redshift and local environment (i.e. presence of a group of galaxies) are available and drawn from the HyperLeda\footnote{HyperLeda, \url{http://atlas.obs-hp.fr/hyperleda/}} database.

\subsection{Galaxy sample} \label{section:data-galaxy_sample}
In this paper, we restrict our study to nearby galaxies with redshifts z$\leq$0.02, corresponding to a maximum distance of approximately 85 Mpc\footnote{Cosmology calculator \citep{Wright_2006}, \url{https://astro.ucla.edu/~wright/CosmoCalc.html}, assuming $H_0=69.6$ km$\,$s$^{-1}$Mpc$^{-1}$, $\Omega_M=0.286$, $\Omega_{vac}=0.714$}. This arbitrary distance threshold allows us to build a large statistical sample; however, we plan to extend the search for streams and tidal features to more distant galaxies in future work.
Our redshift criteria limits the sample to $20,077$ individual galaxies. Some are located in galaxy groups, in which case we only consider the central, primary galaxy and remove from our sample the other galaxies. The definition of a group is based on a projected angular separation between galaxies smaller than 10$'$ \citep{Moustakas_et_al_2023}, without requiring them to be physically bounded. This results in a final sample of $19,387$ objects. The mean angular size (\texttt{D25\_LEDA}) of galaxies with $0.019\leq z \leq 0.02$ is 47$''$.

We use the morphology information from the SGA-2020 catalogue to divide our sample between early-type galaxies (ETGs, that include elliptical and lenticular galaxies), late-type galaxies (LTGs, including spiral and irregular galaxies) and galaxies with unknown morphology: 23.4\% of our sample are ETGs, 64.9\% are LTGs and 11.7\% have missing morphological information.

We estimate the stellar masses of the galaxies from the total flux in the $g$ and $r$-bands from the \texttt{Tractor} fitting. Note that this value is a proxy, the model can be wrong for some cases, and/or underestimate the real flux so the estimated stellar masses should be considered with caution. Nonetheless, they provide an order of magnitude of the galaxy mass in our sample. The fluxes are converted to magnitudes using the redshift and the extinction in the SGA-2020 catalogue. We use a mass-to-light ratio in the $r$-band $\gamma = 10^{a + (g-r)\times b}$ with $a=-0.306$ and $b=1.097$ \citep{Bell_et_al_2003}.
The distribution of the stellar masses of the galaxies has a median of $10^{9.7} M_\odot$, with 16-84\% quantiles of $10^{8.7} - 10^{10.7} M_\odot$, leading to be a low- to intermediate-mass galaxy sample. The distribution of the ratio between ETGs, LTGs and galaxies with missing morphological information as a function of the galaxy's stellar mass is shown in the top panel of Figure \ref{fig:feature_fraction_vs_mass}. LTGs dominate the sample for masses below $\sim 10^{11} M_\odot$.

\subsection{Images}\label{section:data-images}
For each galaxy, we download the already available montage PNG images \footnote{Available at \url{https://portal.nersc.gov/project/cosmo/data/sga/2020/html/}.} which are 3-channel colour composites of the original, model, and residual data produced by the \texttt{legacypipe} pipeline. Each channel is stored as integer values on a 0–255 scale. The field-of-view corresponds to three times the size of the galaxy\footnote{The galaxy size is taken in the \texttt{legacypipe} pipeline as \textit{D25}, the approximate major-axis diameter at the 25 mag$\,$arcsec$^{-2}$ optical SB isophote.}. From these colour images, we compute the median of the residuals in g, r, and z, resulting in a grayscale image. To enhance faint features, we calculate the 35th–90th percentile range of the pixel distribution in this median residual image and clip the pixel values accordingly.

These images, illustrated in Figure \ref{fig:appendix-montage_png}, are used to perform the classification through visual inspection. While their availability significantly accelerates the data processing workflow, this comes at the expense of a limited field of view of three times the galaxy size, which may reduce the number of detectable features (see Section \ref{sec:discussion-streams_literature}). This trade-off was intentionally made to enable the rapid inspection of thousands of images, as our focus is primarily on tidal features located in the immediate vicinity of the target galaxies.

A different set of images is used to derive the photometry of the STRRINGS sample of streams selected for modelling defined in Section \ref{section:best_streams}. The photometry is computed on the residual $g,r,z$ images obtained with the DESI-LS viewer cutouts, at the original pixel size of 0.262$\arcsec$. The field-of-view is 2.5 times the galaxy size\footnote{We use \textit{D26} from the SGA-2020 catalogue as the galaxy size. Some images required a larger field-of-view, such as NGC4388, NGC5907 and NGC5055.}, large enough to encompass the galaxy, the stream and the nearby environment. We found a surface brightness limit of 28.5, 29 and 27.4 mag$\,$arcsec$^{-2}$ in $r$, $g$ and $z$- band, respectively (see Section \ref{sec:desi-ls}).

\section{Method} \label{section:method}
\subsection{Visual classification}\label{sec:method_voting}
We visually classify the $19,387$ galaxy images in our sample using a custom visualization tool. This tool displays the original, model and residual montage image described in Section \ref{section:data-images}, alongside a list of mutually exclusive classification categories and an optional notes section for each galaxy. The categories include streams (from disrupted satellite galaxies during minor mergers with mass ratios around 1:100 to 1:10), shells (from radial mergers), tails and plumes (from disturbed host galaxies due to major mergers with mass ratios higher than 1:10), and disturbed, asymmetric stellar halos. Galaxies can optionally be flagged as grand-design spirals or marked as curiosities for rare or unusual morphologies. More details on the classification process and examples of each category are provided in Appendix \ref{section:appendix_classification}. We emphasise that the visual classification is based solely on projected morphology, and that some ambiguity between classes can arise in certain cases. 

Nine classifiers have contributed to the visual classification. Two of them performed the initial classification of all objects: $5,120$ objects were flagged by at least one person as displaying a feature of any type. These objects were subsequently also classified by the seven other classifiers. Galaxies were presented in a random order to prevent all users from becoming more confident on the same objects later in the classification process. We then devised a voting system to combine the individual catalogues. 

For each galaxy, we start by checking if the galaxy has a feature of any type or not: in case of ties, the galaxy is considered as having a feature. Each galaxy is assigned a single label, as the categories are mutually exclusive. 
We then remove the `No feature' from the votes and we determine majority feature, i.e. the label that occurs the most frequently. In case of ties between several labels, we always favour streams before any other label, then tidal tails, then shells, then plumes, then disturbed/asymmetric halo, then grand-design spiral, then curiosity. 
This ordering reflects both diagnostic power and clarity of identification. Streams are given the highest priority, with tidal tails ranked second due to the potential ambiguity in distinguishing them from streams. Shells are placed third, as their presence provides a strong indication of tidal interactions, most likely associated with radial mergers. Plumes are ranked next, as they are less well-defined than tails. Disturbed stellar haloes are prioritised after plumes, since they indicate recent interactions without any clearly recognisable substructure. The two optional categories, grand-design spirals and curiosity, are placed last.
Finally, we define the confidence score of the majority feature as the number of votes for that feature divided by the total number of votes. 

By doing this voting system, we bias the results towards finding more streams as this is the purpose of this study. To be comparable to other studies, we not only provide the majority label but also the individual votes and the confidence score on the majority feature, so that the majority feature can be recomputed according to another voting system.

\subsection{Feature segmentation}\label{section:method-segmentation}
In addition to the classification, we aim at providing detailed properties of a selected sub-sample of streams that are suitable to be modeled, in order to infer the DM halo properties. The selection criteria of these STRRINGS sample will be described in Section \ref{section:best_streams}. For these streams, we want to obtain their segmentation, which will enable us to recover their track, width, length and photometry. 

We use the \texttt{Jafar}\footnote{\texttt{Jafar}, \url{https://jafar.astro.unistra.fr}} annotation tool \citep{Sola_et_al_2022} to segment streams. \texttt{Jafar} is a user-friendly online tool that enables collaborators to navigate in images and draw with precision the shapes of given features, superimposed on images. Several drawing tools are available (e.g., circles, ellipses, polygons) and the shapes can be easily modified through control points. A label (i.e. feature type) is attached to each shape. The label and coordinates (in right ascension and declination) of the contours of the annotations are saved into a database. Geometrical properties (area, width, length), contour coordinates, label and additional data of each annotation can be exported as csv files. From these, masks of the annotations can be created and used to perform automated aperture photometry. 

\texttt{Jafar} requires images to be in the HiPS format \citep{Fernique_et_al_2015}. We use as images the DESI-LS cutouts of the $g,r,z$ residuals from Section \ref{section:data-images}. We keep the median value of the three bands and clip the data between 35\% and 90\% of the pixel distributions. We rescale the images at a pixel size of 0.6$\arcsec$ to enhance the stream visibility. FITS are then converted to HiPS. The already available coloured HiPS of DESI-LS DR10 and BASS DR3 are used as a complement.

From the segmentation, we derive the stream's track following \cite{Chemaly_et_al_2025}. The track fitting process is described in detail in Appendix \ref{sec:appendix-strrings_properties} and illustrated in Figure \ref{fig:illustration_track}. 
Briefly, after segmenting the stream using non-overlapping masks in azimuthal space, we divide the image into fixed angular bins and extract radial flux profiles. In each bin, we fit one or more Gaussians depending on whether the stream overlaps with itself. Because the segmentation masks correspond to distinct angular segments, this process avoids confusion from azimuthally overlapping regions and allows for an automated reconstruction of the unwrapped stream.

The stream photometry is computed from the residual $g,r,z$ cutout images from the DESI-LS viewer, with a native pixel size of 0.262$\arcsec$ (Section \ref{section:data-images}). All remaining sources (such as artefacts from bright stars of dwarf galaxies) in the residuals images are masked using SEP \citep{Barbary_et_al_2016}. We then apply the mask of the segmented stream and retrieve the flux, surface brightness and colour inside the structure.

\section{Tidal features catalogue} \label{section:results}

\begin{table*}
\caption{Classification of the $19,387$ galaxies in our sample. This table is available in electronic format in the online version. (1): identifier of the galaxy in the SGA-2020 catalogue. (2): number of votes for \textit{No feature}. (3): number of votes for \textit{Streams}. (4): number of votes for \textit{Tails/Arms}. (5): number of votes for \textit{Plumes}.  (6): number of votes for \textit{Shells}. (7): number of votes for \textit{Disturbed/asymmetric stellar halo}. (8): number of votes for \textit{Grand-design spiral}. (9): number of votes for \textit{Curiosities}. (10): feature kept as the final label using our voting system. (11):  number of votes for the majority feature divided by the total number of votes. (12):  a `*' is indicated if the galaxy has a stream that is part of our STRRINGS sample of long, narrow streams selected for modelling (see Section \ref{section:best_streams}.)}
\label{table:catalogue_galaxies} 
\begin{tabular}{cccccccccccc}
\hline
Name & No & Streams & Tails & Plumes & Shells & Disturbed & Spiral & Curiosities  & Majority feature & Score & STRRINGS \\
(1) & (2) & (3) & (4) & (5) & (6) & (7) & (8) & (9) & (10) & (11) & (12) \\
\hline
PGC2214734 & 2 & 0 & 0 & 0 & 0 & 0 & 0 & 0 & No & 1.0 &  \\
NGC0788 & 1 & 2 & 0 & 0 & 5 & 0 & 0 & 0 & Shells & 0.63 &  \\
PGC013229 & 2 & 4 & 1 & 0 & 1 & 0 & 0 & 0 & Streams & 0.5 &  \\
\dots & \dots & \dots & \dots & \dots &\dots & \dots & \dots & \dots & \dots&\dots \\

UGC05116 & 0 & 0 & 8 & 0 & 0 & 0 & 0 & 0 & Tails & 1.0 &  \\
ESO079-003\_GROUP & 1 & 8 & 0 & 0 & 0 & 0 & 0 & 0 & Streams & 0.89 & * \\
\hline
\end{tabular}
\end{table*}

We present the catalogue of the galaxies classification in Table \ref{table:catalogue_galaxies}. It summarises the individual votes about the presence of features per galaxy, the majority label kept after our voting procedure and its  confidence score. We refer to streams as  galaxies having the majority feature `stream', and likewise for the other features.
We emphasise that our categories were mutually exclusive and the voting system biases the percentages towards streams (then iteratively towards tails, shells, plumes, disturbed halo, grand-design spiral and curiosities). Therefore, the stream fractions derived here might be upper bounds, whereas other feature types might slightly be underestimated due to the bias. Depending on the science goal, another methodology could be devised, such as applying a cut on the confidence score to keep the more reliable features. This score reflects the agreement between classifiers and can vary substantially (see Appendix \ref{section:appendix_classification}), for instance 30.9\% of the streams have a score higher than 0.5, and only 6.9\% have a score higher than 0.75. In the following, we do not to apply a hard cut, to avoid excluding
genuine features that are simply challenging; however, users of our catalogue
can impose their own confidence cut as needed.

Table \ref{tab:perc_tidal_features} summarises the fraction of galaxies with a given type of feature, with their $1\sigma$ standard errors on proportions, as a function of the morphological type. The majority of the galaxies ($80.9 \pm 0.3\%$) do not display any peculiarity, while $4.4\pm 0.1\%$ have streams, $3.3 \pm 0.1\%$  have tails, $2.4 \pm 0.1 \%$ have shells, $1.8 \pm 0.1\%$ have plumes. Hence, $11.9 \pm 0.2\%$ of the galaxies host any type of detectable tidal feature (stream, tail, plume or shell), and $1.6 \pm 0.1\%$ display disturbed or asymmetric stellar haloes. An additional $5.2 \pm 0.2\%$ are flagged as being grand-design spirals and $0.4 \pm 0.0\%$ appear as curiosities. 

When considering ETGs and LTGs separately, $18.9 \pm 0.6 \%$ of ETGs host tidal features against $10.4 \pm 0.3\%$ for LTGs, which is a statistically significant difference at a 5\% confidence level according to a Z-test on proportions. Shells are more frequent around ETGs than LTGs, which is statistically confirmed by a Z-test. Tails seem to be more present around LTGs (the conditions of application of the Z-test were not verified so it could not be used\footnote{A Z-test for comparing two proportions can be applied when the samples are independent, the outcome is binomial, the data are randomly sampled, and the sample sizes are large enough for the normal approximation to be valid (typically np$\geq$5 and n(1-p)$\geq$5 for each group, where n is the sample size and p the proportion). This last condition was not met here.}). 
The fraction of galaxies with streams is similar for ETGs and LTGs for a galaxy mass below $\sim 1.5\times10^{11}M_\odot$. However, for higher masses, ETGs have a statistically significantly higher stream fraction than LTGs.

The mass of the host galaxy has a strong impact on tidal feature presence. Figure \ref{fig:feature_fraction_vs_mass} shows the evolution of the fraction of galaxies hosting tidal features as a function of the host stellar mass. The fraction of galaxies hosting any type of features increases with mass, ranging from $2.4 \pm 0.4\%$ in the lowest mass bin up to $36.5 \pm 1.2\%$ in the highest mass bin. A sharper increase is visible for masses above $2\times10^{10} M_\odot$. The trend is mostly driven by streams, but this is also due to our voting procedure favouring streams over other labels. The fraction of plumes remains constant across mass. Likewise, no noticeable increase is seen for tails, streams or shells below $2\times 10^{10} M_\odot$. In the higher mass bin, the fraction of shells and streams strongly increases, while the fraction of tails slightly decreases. This is actually related to the morphological type of the galaxies: for ETGs, the increase is driven by streams and shells while for LTGs it is driven by streams and tails (see Figure \ref{fig:appendix_fraction_features_vs_mass_morphtype}).

Finally, we examine the role of environment, quantified by group multiplicity in the SGA-2020 catalogue, on the tidal feature fraction. We emphasise that our analysis considers only the central, primary galaxy of each group and does not include tidal features around satellite members. Galaxies are divided into isolated systems (group multiplicity of one) and central galaxies in groups (multiplicity $\geq$ 2), without distinguishing between groups and clusters. Most galaxies in our sample (92.3\%) are isolated.
The overall tidal feature fraction is significantly higher for central galaxies in groups ($33.0 \pm 1.2\%$) than for isolated galaxies ($11.9 \pm 0.2\%$), a difference confirmed by a Z-test. In groups, streams and tails are found in $14.0 \pm 0.9\%$ and $9.7 \pm 0.8\%$ of galaxies, respectively, compared to only $3.6 \pm 0.1\%$ and $2.7 \pm 0.1\%$ for isolated systems. This enhancement cannot be attributed solely to stellar mass, as at fixed host mass the tidal feature fraction for group centrals remains two to three times higher than for isolated galaxies. Such elevated fractions in dense environments are expected, given the greater availability of nearby companions and the higher likelihood of mergers.
These results on the tidal feature catalogue will be discussed and compared to literature values in Section \ref{sec:discussion_tidal_features}.

\begin{figure}
    \centering
    \includegraphics[width=0.9\linewidth]{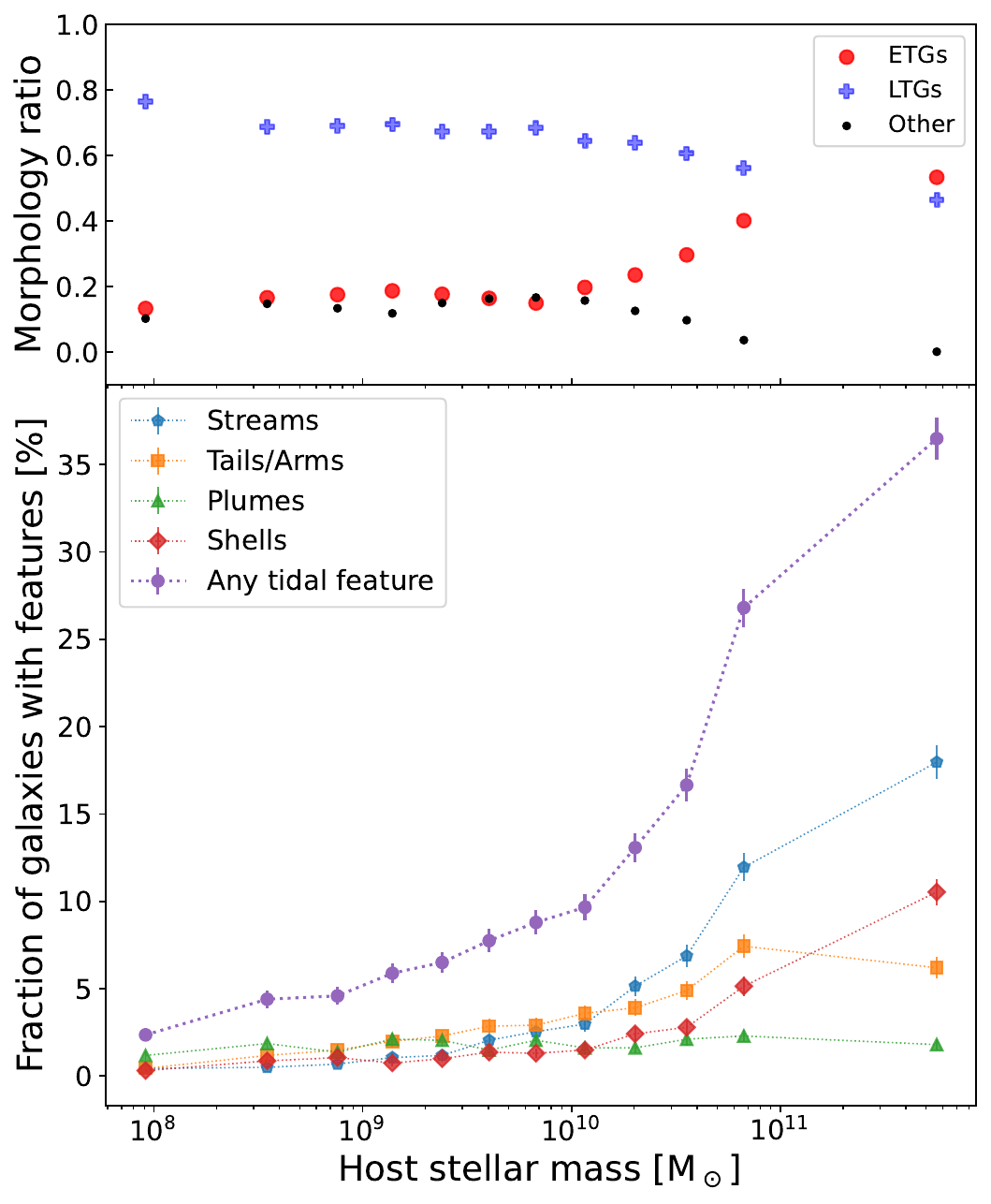}
    \caption{\textit{Top}: Morphology ratio per bin of host galaxy's stellar mass for ETGs (red circles), LTGs (blue crosses) and galaxies with unknown morphological information (black dots). The mass bins contains approximately the same number of galaxies (about $1,615$). LTGs dominate the sample for stellar masses below $10^{11} M_\odot$. \textit{Bottom}: Tidal feature fractions as a function of the host galaxy's stellar mass (in $M_\odot$). The mass bins contains approximately the same number of galaxies (about $1,615$). The errorbars represent the $1\sigma$ standard error on proportions in each bin. Streams are plotted as blue pentagons, tails as orange squares, plumes as green triangles, shells as red rhombuses and any tidal feature as purple circles. The fraction of galaxies with tidal features increases with host stellar mass. }
    \label{fig:feature_fraction_vs_mass}
\end{figure}

\begin{table}
\centering
\caption{Percentage of the LSB features fraction in our sample, computed with respect to the whole sample, to ETGs, to LTGs and to galaxy with unknown morphology. Galaxies displaying tails, streams, plumes or shells are counted in the `Any tidal feature' fraction. The errors correspond to the $1\sigma$ standard error on proportions.}
\label{tab:perc_tidal_features}
\scriptsize
\begin{tabular}{ccccc}

\hline
Feature & All & ETG & LTG & Unknown morph. \\
 & [\%] &[\%] &[\%] &[\%]\\
\hline
Streams & $4.4 \pm 0.1$ & $7.3 \pm 0.4$ & $3.9 \pm 0.2$ & $1.7 \pm 0.3$ \\
Tails & $3.3 \pm 0.1$ & $2.2 \pm 0.2$ & $3.9 \pm 0.2$ & $1.8 \pm 0.3$ \\
Plumes & $1.8 \pm 0.1$ & $2.2 \pm 0.2$ & $1.7 \pm 0.1$ & $1.4 \pm 0.2$ \\
Shells & $2.4 \pm 0.1$ & $7.2 \pm 0.4$ & $0.9 \pm 0.1$ & $1.4 \pm 0.2$ \\
Disturbed halo & $1.6\pm0.1$ & $0.9 \pm 0.1$ & $1.8 \pm 0.1$ & $1.8 \pm 0.3$ \\
\hline
Any tidal feature & $11.9 \pm 0.2$ & $18.9 \pm 0.6$ & $10.4 \pm 0.3$ & $6.3 \pm 0.5$ \\
\hline
\end{tabular}
\end{table}

\section{STRRINGS} \label{section:best_streams}
\subsection{STRRINGS sample selection}

From the catalogue of classified features, we select the stellar streams that are best suited for our modelling technique. Our method is based on comparing the radial distances of observed and modelled streams across different angular bins \citep{Chemaly_et_al_2025}, which requires curvature in the stream to meaningfully sample the angular space.  Our selection criteria hence favour streams that are relatively thin, exhibit noticeable curvature, and are sufficiently faint, which indicates low mass-ratio mergers involving small companions. Additionally, we require that the host galaxy’s central regions appear undisturbed, to avoid complications from recent major mergers that could significantly alter the gravitational potential and hinder accurate modelling. Therefore, our criteria (thin, curved, faint streams) aim to ensure that the streams are kinematically simple minor mergers, so the dynamical assumptions made for stream modelling are physically reasonable.

We select the 861 galaxies where the dominant feature is classified as `streams'. Four classifiers then re-examined these images using the same classification tool, now with three refined options: (1) `Not streams’ (the feature is not a stream or is overshadowed by multiple other features), (2) `Stream not selected for modelling’ (due to insufficient curvature for our method, excessive brightness, or central galaxy perturbation), and (3) `Stream selected for modelling’. We emphasise that this classification is designed for our specific modelling goals, other studies may adopt different criteria depending on their scientific objectives or modelling frameworks.

These selection criteria inevitably introduce biases in the statistical properties of the galaxies found to host tidal streams. By excluding systems with clearly disturbed morphology, we preferentially select dynamically quiet galaxies and avoid recent or ongoing major mergers. Our focus on thin, curved, and faint streams further biases the sample toward minor mergers with low mass ratios, as these produce streams that are best suited for our modelling assumptions. In addition, stream detectability is enhanced in galaxies viewed closer to edge-on, where the disc can be more easily separated from surrounding substructure, which may favour higher inclination hosts. As a result, the STRRINGS sample is not representative of the full population of galaxies with streams, but rather of those hosting streams that can be robustly modelled within our framework.

Our final STRRINGS sample includes only objects where all classifiers agreed on a stream selected for modelling, resulting in 35 galaxies and their streams, which are marked with a (*) in Table \ref{table:catalogue_galaxies}. We then annotate these streams using \texttt{Jafar}, carefully delineating their contours primarily from the residual image while simultaneously checking with the original image.

\subsection{STRRINGS stream properties}
We retrieve quantitative measurements of the segmented streams from the database, and we compute the coordinates, length and width of the track. Using the coordinates of the contours of the annotations, we create masks of the streams and apply them to the  $g, r, z$ residual cutout images from the DESI-LS viewer to derive their surface brightness (SB) and colours. An illustration of the stream segmentation and track is presented in Figure \ref{fig:results-segmentated_stream_image_example}, while Figure \ref{fig:streams_all} shows the Cartesian projection of
all the streams, centred around their respective host galaxy. Several of our STRRINGS streams in residual images are illustrated in Figure \ref{fig:only_best_streams}. 
The images and segmentations for all the STRRINGS sample is shown in Figure \ref{fig:appendix-best_stream_segmentation_all-part1}. We provide in Appendix \ref{section:appendix_streams_segmentation} the summary of the geometric and photometric properties of the streams in Table \ref{tab:best_streams} and the coordinates of the contours of the annotations in a format readable by SAOImageDS9 and Aladin. We checked that the photometry estimated from the residuals is comparable to the one obtained from the original images in Appendix \ref{section:appendix-photometry_coadd_vs_residual}.

\begin{figure*}
    \includegraphics[width=\linewidth]{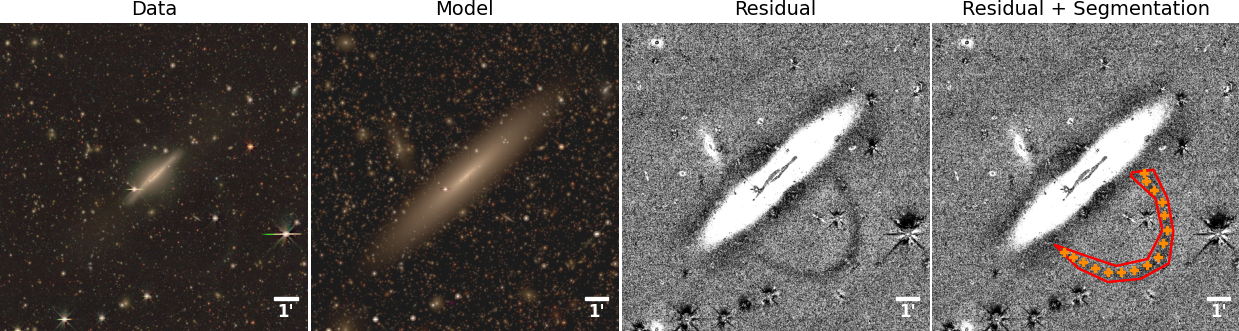} 
    \caption{Illustration of the original and model images, segmentation and track of a stream for one galaxy (ESO079-003\_GROUP). A scalebar of $1'$ is indicated at the bottom of each panel. North is up, East is left. \textit{Left}:  original $grz$ data image. \textit{Middle, left}: \texttt{Tractor} $grz$ model of all astronomical sources. \textit{Middle, right}: residual image (obtained by subtracting the model from the data), plotted as the median of the $r,g,z$ residuals, smoothed and stretched to enhance faint features. White areas correspond to an excess of light in the model, black areas show an excess of light in the data.  \textit{Right}: residual image with the stream segmentation (red contours) from \texttt{Jafar} and the stream track (orange crosses) as described in Section \ref{section:method-segmentation}.}
    \label{fig:results-segmentated_stream_image_example}
\end{figure*}

\begin{figure}
    \centering
    \includegraphics[width=\linewidth]{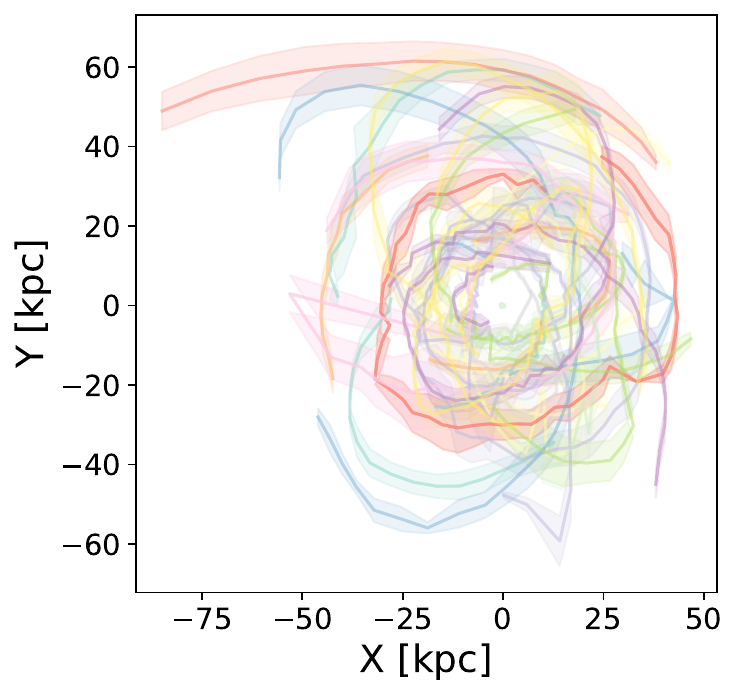}
    \caption{Cartesian projection (in kpc) of the track (solid lines) and width (shaded regions) of all the individual STRRINGS streams, centred around their respective host galaxy (at X=0,Y=0). The majority of streams are located within 30 kpc of their host galaxy.}
    \label{fig:streams_all}
\end{figure}

\begin{figure*}
    \centering
    \includegraphics[width=\linewidth]{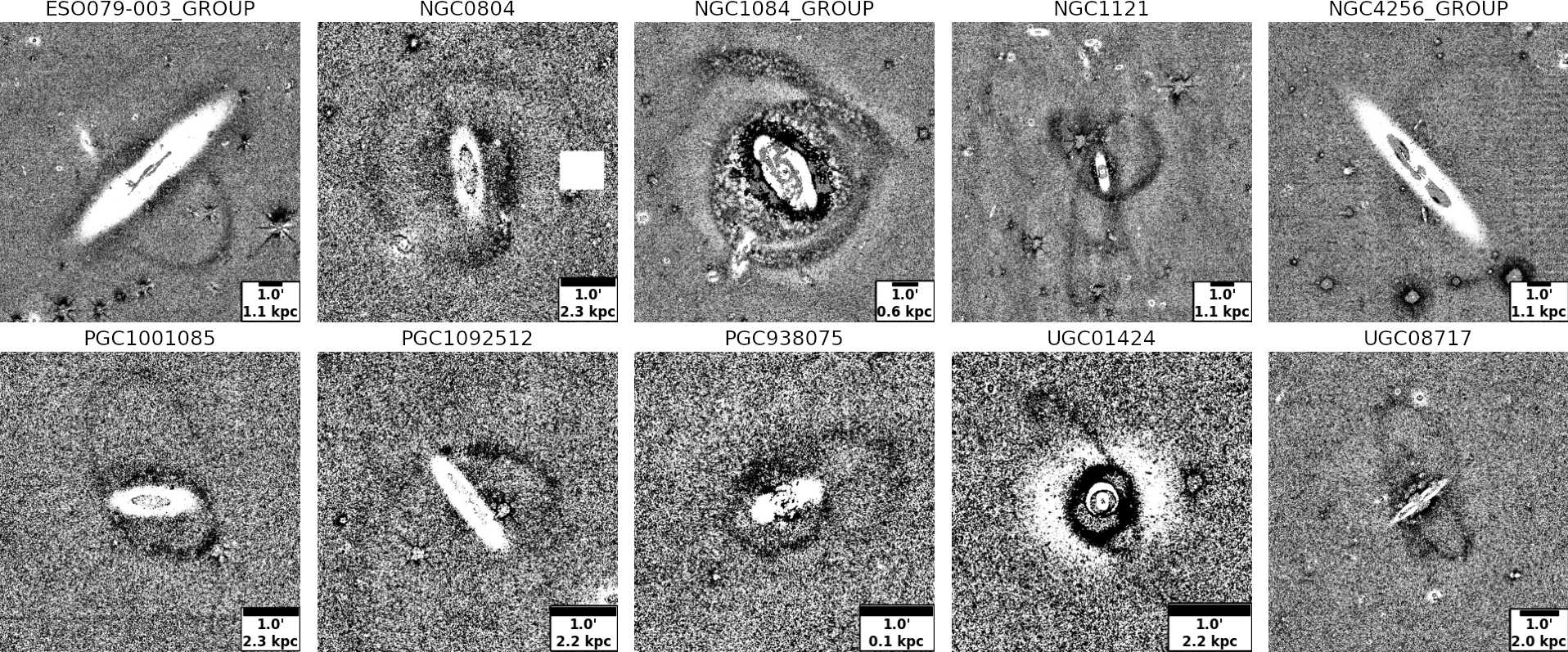}
    \caption{ Illustration of some streams from our STRRINGS sample. The residual images are displayed, and represent the median of $g,r,z$ residual images, smoothed and stretched to enhance faint features. Black regions correspond to light excess in the data, white regions to areas where the model overestimates the light. Streams clearly appear as black regions in the residuals. A scalebar of $1'$ and the corresponding physical length (in kpc) is indicated at the bottom of each panel, the galaxy names are at the top. North is up, East is left.  }
    \label{fig:only_best_streams}
\end{figure*}

\subsubsection{Surface brightness and colours}\label{sec:sb_colors_streams}
The surface brightness of each stream is not uniform but varies along its length. Additional variability arises from the measurement method, whether calculated per pixel or within small aperture boxes. In this study, we report the median SB values, as they provide the most representative estimate and best reflect the central tendency of the distribution.
Figure \ref{fig:results-stream_properties} presents the kernel density estimates\footnote{In this work, we plot kernel density estimates instead of histograms because of their smoother representation of distributions. We use the \texttt{scipy.stats.gaussian\_kde} function, which applies Gaussian kernels, with the bandwidth determined using the default `scott' method. } of the distributions of the streams median SB and colours. The streams have a median surface brightness of 26.4, 26.8 and 25.1 mag$\,$arcsec$^{-2}$ in the $r,g$ and $z$-band, respectively. The faintest feature reaches median SB of 27.4, 27.5 and 25.4  mag$\,$arcsec$^{-2}$ in $r$, $g$ and $z$-band, respectively (where the SB limits are 28.5, 29 and 27.4 mag$\,$arcsec$^{-2}$, respectively). 

The distribution of the median $r$ and $g$-band SB values are similar to the mean SB values reported by \cite{Martinez-Delgado_et_al_2023} (24 streams), \cite{Miro-Carretero_et_al_2023} (22 streams) and \cite{Miro-Carretero_et_al_2024} (63 streams) who utilised DESI-LS images reprocessed to avoid oversubtraction of the background, leading to a few fainter streams. A thorough comparison of the streams in common between STRRINGS and these literature works will be presented in Sections \ref{sec:discussion-streams_literature} and  \ref{sec:comparison_streams_catalogues}. The $r$-band are similar to the ones reported by \cite{Skryabina_et_al_2024} for their 9 streams (measured on the SDSS Stripe 82 images, with a depth of 28.6 mag$\,$arcsec$^{-2}$)\footnote{\href{http://research.iac.es/proyecto/stripe82/}{http://research.iac.es/proyecto/stripe82/}}. The fainter 100 streams in \cite{Sola_et_al_2025} are found in deeper CFHT images (with a depth of about 29 mag$\,$arcsec$^{-2}$) for galaxies with a maximum distance of 45 Mpc. The 0.8 mag discrepancy in the $z$-band between our values and those of \cite{Martinez-Delgado_et_al_2023} and \cite{Miro-Carretero_et_al_2024} could be due to their more refined background re-estimation. This implies that our $g-z$ and $r-z$ colours are systematically offset from theirs and not reliable. The streams median $g-r$ colour of 0.5 is in agreement with the literature.

The stream around NGC4632 has a negative $g-r$ colour. An inspection of the image revealed an instrumental artefact in the $g$-band, which artificially boosted the flux and resulted in bluer colours. However, even after masking the affected region, the stream remained blue. \cite{Deg_et_al_2023} and \cite{Mosenkov_et_al_2024} identified NGC4632 as a potential HI-rich ring around a polar ring galaxy, accompanied by a faint stellar component. The stream is clearly visible in the deeper $grz$ HSC image shown in Fig. 5 of \cite{Deg_et_al_2023}, and appears to lie within the HI ring. The blue colours could be explained by ongoing star formation within the HI structure that encompasses the stream.

It is worth noting the difference between the faintest median SB values we measure (27.4, 27.5 and 25.4 mag$\,$arcsec$^{-2}$ for $r$, $g$, and $z$ bands) and the nominal survey depths (28.5, 29 and 27.4 mag$\,$arcsec$^{-2}$ for $r$, $g$, and $z$ bands), with  differences of 1.1, 1.5, and 2 mag$\,$arcsec$^{-2}$ in the $r$, $g$, and $z$ bands, respectively.
Several factors likely contribute to this gap \citep[e.g.,][]{Sola_et_al_2022, Sola_et_al_2025, Martinez-Delgado_et_al_2023, Skryabina_et_al_2024, Miro-Carretero_et_al_2024}. First, there is an important distinction between identifying individual pixels above the noise and confidently attributing them to a coherent tidal feature, meaning the faintest outskirts of these structures may be missed. Moreover, survey depth is typically measured using noise on $10\arcsec \times 10\arcsec$ scales, while tidal features are often far more extended and exhibit surface brightness variations as previously discussed. The overall depth of the survey also plays a role, with deeper surveys revealing fainter features \citep[e.g.,][]{Martin_et_al_2022, Vera-Casanova_et_al_2022}. Additionally, this discrepancy may reflect the short-lived visibility of tidal features, as we may preferentially detect younger structures that have not yet phase-mixed \citep{Sola_et_al_2022}.

\begin{figure*}
    \centering
    \includegraphics[width=\linewidth]{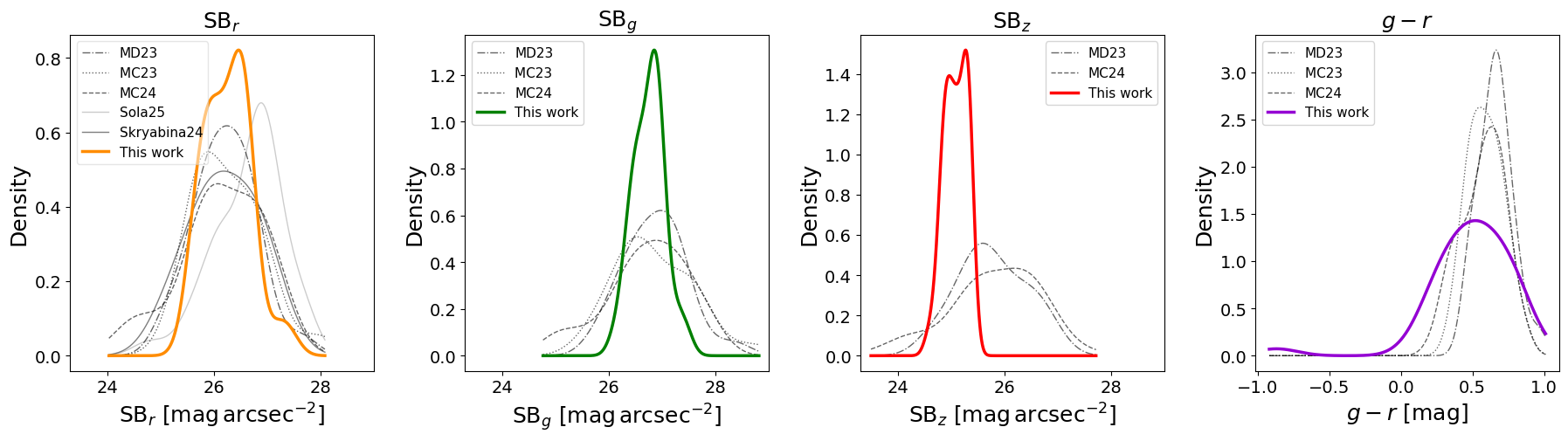}
    \caption{Kernel density estimates of the distributions of surface brightness and colours of the STRRINGS sample (thick, solid, coloured lines), compared to literature values (black lines). We compare to works based on DESI-LS images: \protect\cite{Martinez-Delgado_et_al_2023} (MD23), \protect\cite{Miro-Carretero_et_al_2023} (MC23), \protect\cite{Miro-Carretero_et_al_2024} (MC24), on CFHT images: \protect\cite{Sola_et_al_2025} and on SDSS Stripe 82 images: \protect\cite{Skryabina_et_al_2024}. From left to right: median surface brightness (mag$\,$arcsec$^{-2}$) of the streams in $r$, $g$, and $z$-band, and of the median $g-r$ colours (mag). Our STRRINGS streams have $g-$,$r-$band surface brightnesses and $g-r$ colours comparable to the literature that uses DESI-LS images, but we find brighter streams in the shallower $z-$band. }
    \label{fig:results-stream_properties}
\end{figure*}

\subsubsection{Stellar mass estimates}
We retrieve the total flux in $g$- and $r$-bands inside the stream segmentation from the residual images. The negative values, due to the model over-prediction, are removed as they can lead to negative total fluxes. As for the estimate of the stellar mass of the galaxies in our sample (see Section \ref{section:data-galaxy_sample}), we convert them to magnitudes, taking into account the extinction, and we assume a mass-to-light ratio varying depending on the median $g-r$ colour of the stream. These values likely underestimate the stream mass as the photometry is limited to the manual segmentation, which may miss the faintest outskirts of the streams, but it is still a reasonable estimate. The distributions of the computed stream stellar masses and mass ratios between the stream and the host galaxy are presented in Figure \ref{fig:stream_mass}, and the values are listed in Table \ref{tab:best_streams}. The stellar mass of the host is derived from the total flux given by the \texttt{Tractor} model. We caution that the stream-to-host mass ratio likely underestimates the true merger mass ratio, as it does not account for the full mass of the disrupted progenitor, so our values should be considered lower bounds.

\begin{figure}
    \centering
    \includegraphics[width=\linewidth]{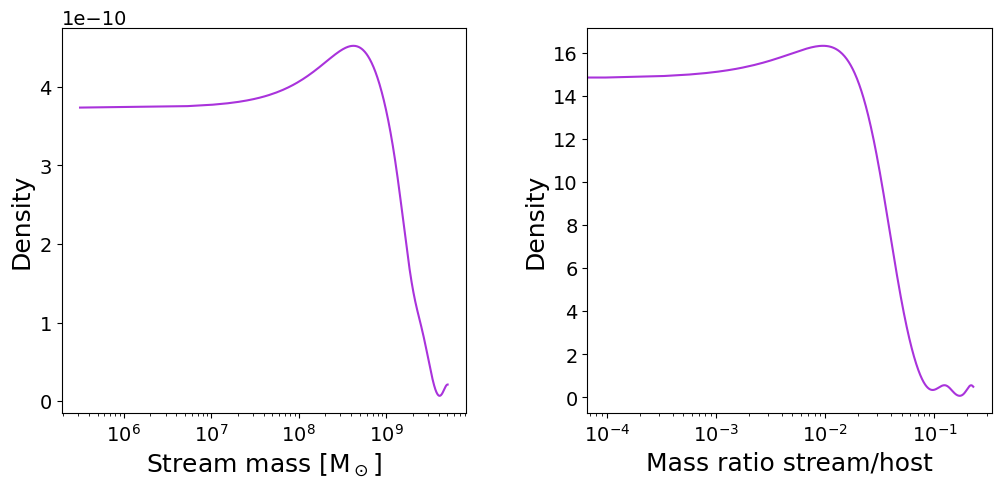}
    \caption{\textit{Left}: Kernel density estimate of the stellar mass of the streams (in M$_\odot$). \textit{Right}: Kernel density estimate of the stellar mass ratio between the stream and the host galaxy. The stream-to-host mass ratios are consistent with minor mergers (1:100 to 1:10).}
    \label{fig:stream_mass}
\end{figure}

The median and mean mass ratios between the stream and the host galaxy are 0.007 and 0.02 respectively, with only four streams with a mass ratio higher than 0.05.
These values are consistent with the expected mass ratios for minor mergers, which typically range from about 1:100 to 1:10, supporting the validity of our visual selection approach for our goal of modelling faint, curved streams originating from low-mass progenitors.
Our streams are relatively massive, with a median of is $6.2 \times 10^{8} M_\odot$. This value is much higher than the mass of most streams around the Milky Way \citep[e.g.,][]{Bonaca_and_Prince-Whelan_2025} (but such faint, low-mass streams would not be detectable at large distances and in relatively shallow surveys such as DESI-LS). Thus, our sample is biased towards the more massive end of the stream mass function, on the order of $10^8-10^9 M_\odot$, which is comparable to the predicted mass of progenitors of the Sagittarius dwarf galaxy \citep[e.g.,][]{Law_and_Majewski_2010}, M31's Giant Southern Stream \citep[e.g.,][]{Fardal_et_al_2006}, other extragalactic streams \citep[e.g.,][]{Johnston_et_al_2001,Martinez-Delgado_et_al_2009} or numerical simulations \citep[e.g.,][]{Shipp_et_al_2022}.

The left panel of Figure \ref{fig:stream_mass_vs_properties} presents the relation between host galaxy and stream stellar masses. To test for the presence of a correlation, we performed both a linear fit and a Spearman rank test between the logarithms of the stream mass and of the host mass. The linear fit gives $y = 0.74x + 4.32$ with an R$^2$ correlation coefficient of 0.53, indicating a weak positive correlation. Spearman’s test yields a correlation coefficient of 0.49 with a p-value of 0.0019, demonstrating a statistically significant monotonic trend at the 5\% level. Hence, we find a weak but statistically significant correlation between host galaxy mass and stream stellar mass, irrespective of the morphological type, with the data broadly following a $\sim$1:100 mass ratio line despite substantial scatter. More massive galaxies should accrete higher numbers of low-mass companions, with higher mass-ratio events being less frequent \citep[e.g.,][]{Lotz+2011,Ownsworth+2014,Conselice+2022}. Given the survey depth and small-number statistics, our measurements are likely tracing the upper envelope of the stream-to-host mass relation rather than the full distribution.  

We caution that the stream around PGC938075, with the lowest stellar mass, might be in the prolongation of some very faint spiral arms.

\begin{figure*}
    \centering
    \includegraphics[width=\linewidth]{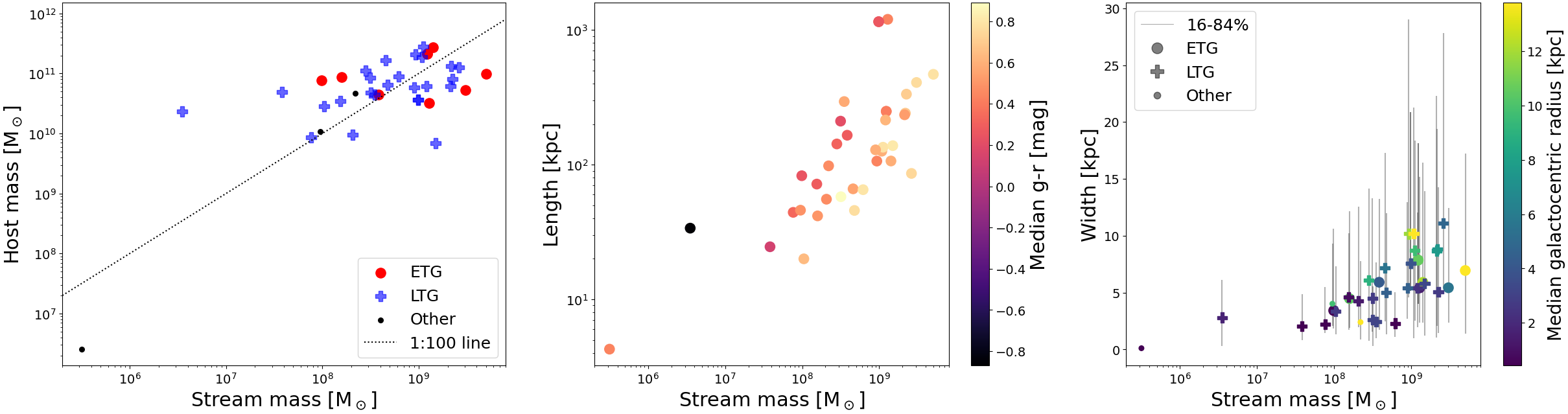}
    \caption{Correlation between the stellar mass of the stream, its length and width and the host galaxy mass for our STRRINGS sample. \textit{Left}: Host galaxy stellar mass (M$_\odot$) as a function of stream stellar mass (M$_\odot$). A 1:100 mass ratio is plotted as a dotted line. ETGs are plotted as red circles, LTGs as blue crosses and galaxies with missing morphological information as black dots. There is a weak trend that more massive galaxies host more massive streams, irrespective of the morphological type.  \textit{Middle}: Stream length (kpc) as a function of the stream mass (M$_\odot$), and colour-coded by the stream median $g-r$ (mag). The stream length correlates with the stream mass, with a hint of more massive streams being redder. \textit{Right}: Stream width (kpc) as a function of the stream mass (M$_\odot$), colour-coded by the median distance to the host galaxy (kpc). The errorbars correspond to the 16\% and 84\% percentiles of the width distribution along the track for each stream. The markers' shapes represent the morphological type of the host galaxy (with ETGs as circles, LTGs as crosses and galaxies with missing morphological information as dots). There is a slight correlation between the stream width and its mass.}
    \label{fig:stream_mass_vs_properties}
\end{figure*}

\subsubsection{Geometry}
The geometry of the streams provides complementary information about the properties of the progenitor. Indeed, the width is related to the kinematic state of the progenitor, as higher velocity dispersions will result in broader streams \citep[e.g.,][]{Johnston_et_al_1996,Hendel_and_Johnston_2015}. The broadening is also a function of the streams dynamical age \citep[e.g.,][]{Erkal_et_al_2016}. Additional constrains can be placed on the potential of the host galaxy and on the orbit of the progenitor \citep[e.g.,][]{Ibata_2001,Helmi_2004b, Johnston_et_al_2005}. \cite{Erkal_et_al_2016} showed that the physical width of the stream is expected to increase for flattened potentials, and progenitors on polar orbits will produce broader streams. The physical width along the stream is also supposed to evolve in a complex, non-linear manner, but overall it should increase further away from the progenitor. They showed the width is also expected to increase with the distance to the host galaxy. \cite{Erkal_et_al_2016} also showed that the angular width of the stream, defined as the ratio of its physical width to the galactocentric radius, should slowly decrease with the radius of the progenitor.

We start by exploring the distributions of width and length of streams in Figure \ref{fig:stream_width_length}. The median and mean width are 5.6 kpc, with 5-95\% percentiles of 2.3-9.9 kpc. For the length, the median is 124 kpc, the mean is 224 kpc, with the 5-95\% percentiles of 24-1151 kpc. The longest streams are actually the ones looping on themselves, i.e. with an angular extent larger than $2\pi$. The distribution of the width is comparable to the ones from \cite{Sola_et_al_2025} and \cite{Miro-Carretero_et_al_2024}. \cite{Pippert_et_al_2025} developed an automated method to model and measure the properties of tails and streams and they report 8 streams with a width between 3-16 kpc, for a length between 25 - 80 kpc, closer to our values. Our streams are overall longer than the literature, but this is due to our targeting of long, curved streams as they are the one that are more suitable to be modelable. The stream around PGC938075 is located very close to the galaxy, it has a small width and small length, but this is related to the small size of the host galaxy. 
\begin{figure}
    \centering
    \includegraphics[width=\linewidth]{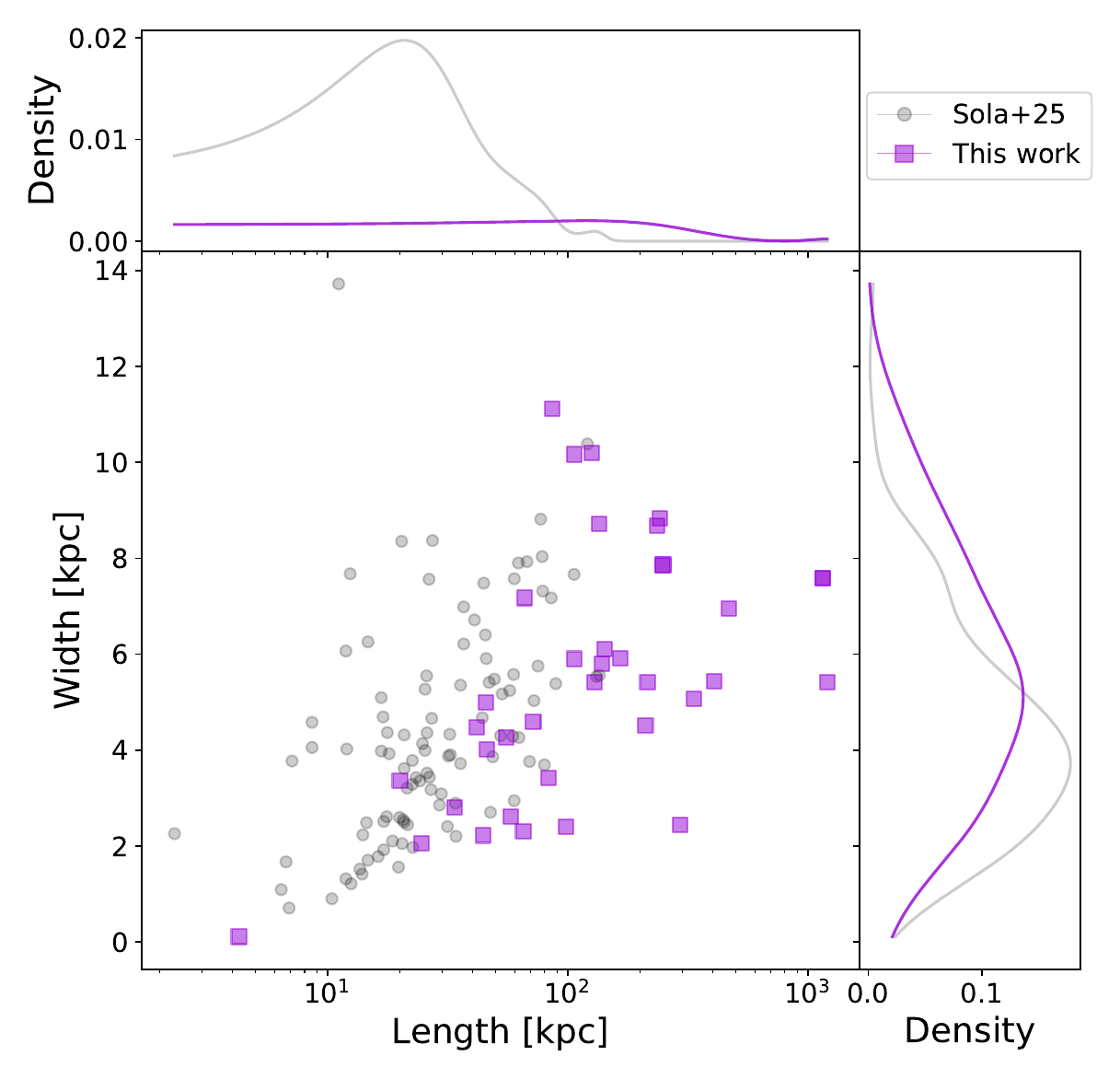}
    \caption{Scatter plot and kernel density estimates of the marginal histograms of the width (in kpc) and length (in kpc) of our STRRINGS streams (purple squares) compared to literature values (grey circles). Due to our selection method, our streams are longer than those in the literature. }
    \label{fig:stream_width_length}
\end{figure}

From the middle panel of Figure \ref{fig:stream_mass_vs_properties}, we find that the length of the stream correlates with its mass.
There is also a hint that more massive streams are redder, although the interpretation of this colour trend is not straightforward. Since streams preferentially trace stars stripped from the progenitor outskirts, metallicity gradients must be taken into account \citep[e.g.][]{Liao_et_al_2023}, as well as the total amount of stripped stars, which depends on the progenitor mass. Moreover, our sample spans a relatively narrow mass range, and thus a correspondingly limited metallicity range, making the interpretation of this trend in terms of a mass–metallicity relation uncertain and could instead reflect age effects or measurement uncertainties.

The right panel of Figure \ref{fig:stream_mass_vs_properties} shows a slight correlation between the stream width and the mass. More massive galaxies have higher velocity dispersions \citep[e.g.,][]{Tully_and_Fisher_1977,McGaugh_et_al_2000,Karachentsev_et_al_2017}, which in turns leads to wider streams \citep[e.g.,][]{Johnston_et_al_1996,Johnston_et_al_1998, Hendel_and_Johnston_2015}. 
In addition, (physically) wider streams tend to be located further away from the galaxy, as seen in the left panel of Figure \ref{fig:width_vs_radius}, consistent with expectations \citep[e.g.,][]{Johnston_et_al_1998,Erkal_et_al_2016}.  However, we caution that we might be biased towards large radii as it is observationally more difficult to detect wide streams close to their host galaxy's centre due to surface brightness contrast.
The angular width of the stream appears to decrease slightly with increasing radius, as shown in the right panel of Figure \ref{fig:width_vs_radius}, which is consistent with the mild trend from \cite{Erkal_et_al_2016} where streams further out would be slightly narrower due to the larger enclosed host galaxy's mass. This expectation relies on the assumption that the observer is in the stream plane, otherwise the measured width is affected by the spread of the debris within the stream plane \citep{Erkal_et_al_2016}. A linear fit to the median values in the right panel of Figure \ref{fig:width_vs_radius} yields a statistically significant non-zero slope (p-value = 0.03 from Wald's test), indicating a trend. However, the low coefficient of determination R$^2$ suggests that a linear model does not capture the full complexity of the data. A fit to all individual data points reveals a more pronounced non-linear relationship.

\begin{figure*}
    \centering
    \includegraphics[width=0.35\linewidth]{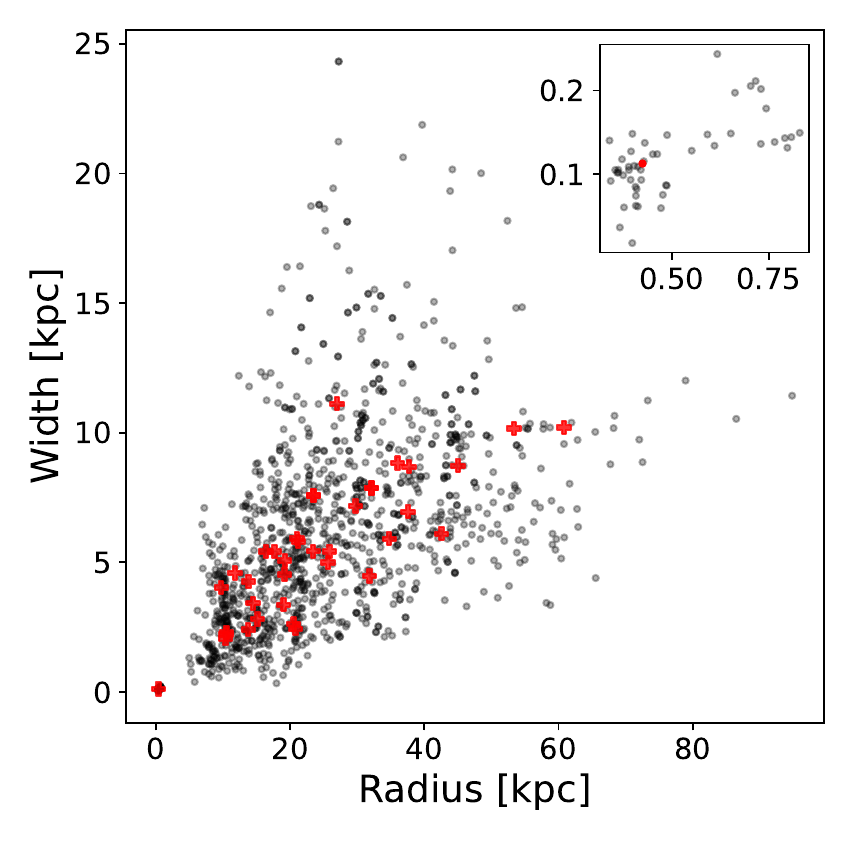}
    \includegraphics[width=0.36\linewidth]{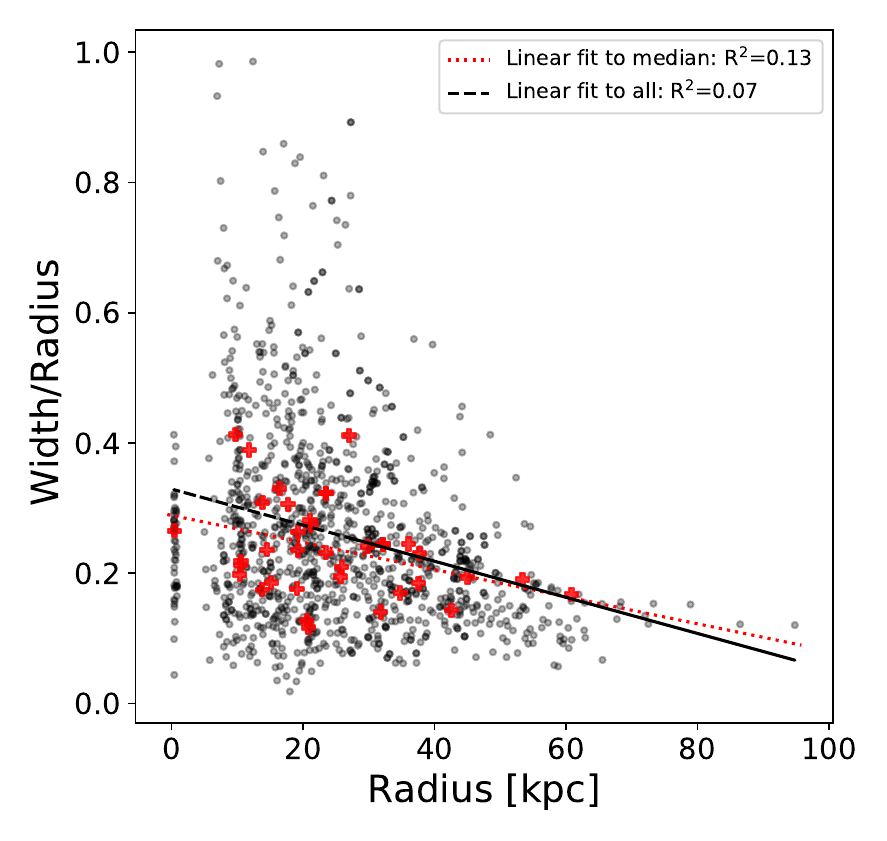}
    \caption{Physical and angular widths of the STRRINGS streams as a function of galactocentric radius. The black dots correspond to individual measurements along the tracks of the streams, while the red crosses are the median values for each stream. \textit{Left}: Physical width of the streams (kpc) as a function of the galactocentric radius (kpc).  The inset is a zoom-in on the stream around PGC938075, with small radius and width. Wider streams are found at higher galactocentric distances.
    \textit{Right}: Angular width of the stream, taken as the ratio between the physical width and the galactocentric radius, as a function of the galactocentric radius (kpc). A linear fit to the median values (dotted red) gives $y=-0.0021x + 0.29$ and R$^2=0.13$, and a linear fit to all values (dashed black) gives $y=-0.0028x + 0.33$ and R$^2=0.07$. There is a slight decrease of the angular width with increasing galactocentric distance, but the low coefficient of determination R$^2$ suggest a non-linear relationship.} 
    \label{fig:width_vs_radius}
\end{figure*}

Figure \ref{fig:streams_all} shows the Cartesian projection of all the streams, centred around their respective host galaxy. Most streams are contained within a projected distance 30 kpc from the galaxy centre, with a few extending up to 40-60 kpc. This implies that the streams can be used to probe the dark matter distribution close to the galaxy centre, not the furthest outskirts. In comparison, the Sagittarius stream around the Milky way is located further away. Its debris extend from 20-100 kpc, with apocenters for the trailing and leading arms of about 90 kpc and 47 kpc, respectively \citep[e.g.,][]{Belokurov_et_al_2014,Sesar_et_al_2017,Ramos_et_al_2020}. For the streams that have a remaining potential dwarf galaxy progenitor (see Section \ref{sec:progenitors}), we could not find an obvious trend between the width and the progenitor's location, which is partly due to our manual stream segmentation. Additionally, depending on the dwarf galaxy's SB, it may have been modelled and removed in the residual image. 

For the few streams that display a loop, we compute and show the pericenter, apocenter and projected ellipticity in Figure \ref{fig:streams_width_radius}. The ellipticities varies significantly, from 0.4 to 0.8, and no obvious correlation is found between ellipticity and length of the track.  Streams are expected to have more circular than radial orbits, as radial mergers lead to the formation of shells rather than long, thin streams \citep[e.g.,][]{Johnston_et_al_1998,Amorisco_2015,Karademir_2019}. Here, we do not have ellipticity values for shells which makes a comparison difficult, and we do not correct for the projection effect, so the projected morphology is degenerate with respect to viewing angles. More looping streams are needed to get more robust statistics.
\begin{figure}
    \centering
    \includegraphics[width=0.9\linewidth]{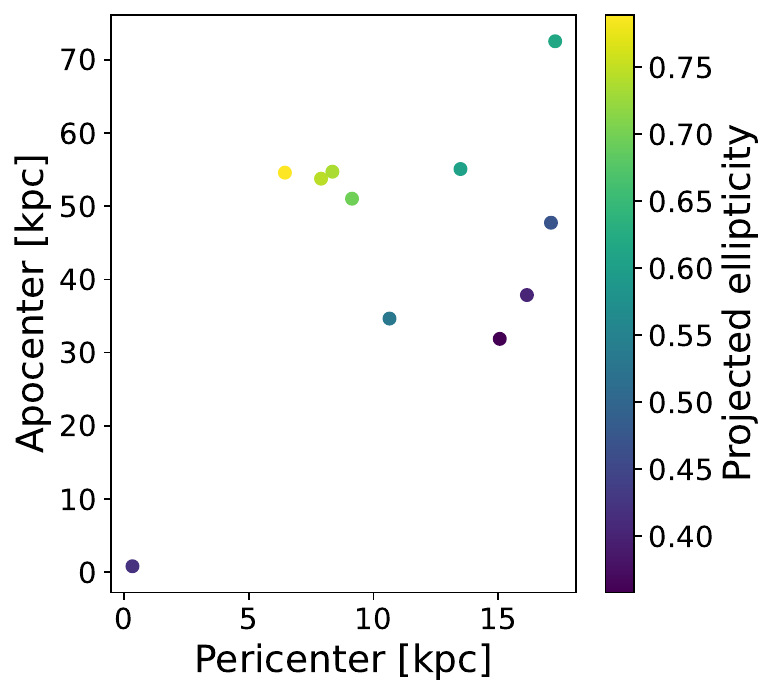}
    \caption{For the streams that present at least one loop, stream apocenter (in kpc) as a function of the pericenter (in kpc), colour-coded by the projected ellipticity. The galaxy with small pericenter and apocenter is PGC93807. The projected ellipticities vary significantly, without obvious correlations.}
    \label{fig:streams_width_radius}
\end{figure}

To summarise, due to our selection criteria we find that our streams are longer than the literature values, but with a similar width. Physically wider streams are located further away from their host galaxy. The angular width of streams shows a slight decrease with increasing galactocentric radius. More massive streams are both slightly redder (potentially from a more massive, metal-rich and/or older progenitor) and are wider (more massive progenitors have more velocity dispersion). Our streams are mostly concentrated within 30 kpc, with only a few extending more than 60 kpc from the galaxy centre, making them probes of the inner dark matter distribution. 

A thorough comparison of the stream properties in observations to the predictions of numerical simulations is out of scope in this paper, but will be the focus of future works. 
Here, we only briefly compare these results to the prediction of some numerical simulations, such as the \texttt{StreamGen} semi-analytic galaxy formation code from \cite{Dropulic_et_al_2024}. They find their streams to be associated with satellites on high-eccentricity orbits with pericenters mostly in the 0-20 kpc range and apocenters in the 0-120 kpc range, along with a high halo-to-halo variability. The pericenter values agree well with ours, but we do not find the most extended apocenters, as our values lie in the 30-60 kpc range. \cite{Shipp_et_al_2022} carried out a detailed analysis of the properties of streams in the FIRE-2 simulation of Milky Way analogues and of observed streams around our Galaxy. Although the number and stellar mass distributions of streams are consistent between simulations and observations, they find larger pericenters and apocenters than in the MW or in \cite{Dropulic_et_al_2024}. \cite{Shipp_et_al_2024} studied streams in the Auriga simulation and found streams on orbits consistent with the MW streams, i.e. with pericenters lower than 20 kpc and apocenters lower than 50 kpc -- along with a large halo-to-halo variance and some streams extending much further than those around the MW. These values are also consistent with our findings. The origin of the discrepancies between these three simulations is not fully understood and could be linked to differences in the feedback prescription, host properties, set of assumptions for semi-analytic modeling or numerical effects. They could also result from the large halo-to-halo variance as the number of halos studied is small. Conversely, if these numerical predictions are correct, they imply a much higher disrupted satellite population around the MW, whose tidal disruption signatures could be fainter than our current detection limits. The implications may extend to galaxies in the nearby Universe, as our results seem to suggest, assuming similar analyses are applied to larger samples. However, meaningful comparisons with numerical simulations will require a careful assessment of detection biases, projection effects, and mock images generation (including the size of field-of-view as well as instrumental and astrophysical noise sources).

\subsection{Candidate satellite dwarf galaxy progenitors} \label{sec:progenitors}
During the manual segmentation, we identified six streams that feature a candidate satellite dwarf galaxy progenitor, visible as a compact object embedded in the stream, for the galaxies: ESO356-012, PGC1092512, NGC0804, PGC430221, NGC5513\_GROUP and UGC09239. We search in the literature for the redshifts of these satellite candidates\footnote{We note that some of these satellites have stellar masses exceeding $10^{9}M_\odot$ (Table \ref{tab:table_progenitors}), which is above the conventional threshold for dwarf galaxies; however, we retain the usage of the term `dwarf' in this section for consistency.}. We find from the NED database that the dwarf galaxy around NGC5513\_GROUP is located at the same redshift ($z=0.0167$) as the host, and similarly for the dwarf around NGC0804 ($z=0.0175$), which was also reported in \cite{Miro-Carretero_et_al_2024}. The dwarf galaxy around ESO356-012 was reported in \cite{Tanoglidis_et_al_2021} (\texttt{coadd-oid=321855008}) but without redshift estimate. To our knowledge, the dwarf galaxy candidates around PGC1092512 and PGC430221 have not been previously reported, such that no redshift information is available. For UGC09239, the dwarf candidate appears in NED as \texttt{SDSS J142508.77+134511.4}, without reliable redshift\footnote{A warning about potentially unreliable photometry is raised in SDSS DR18 for this object. In SDSS DR12 a photometric redshift of 0.167 is indicated but seems unlikely given the physical appearance of the object.}.

We compute the photometry from the original images (as the dwarf candidates are partly modelled, the residuals would not provide accurate measurements). We perform ellipse fitting using \texttt{Autoprof} \citep{Stone_et_al_2021}, by providing an estimation of the centre of the dwarf. We trace the $g,r$ and $g-r$ radial profiles, and sum the flux inside every isophote to retrieve the total magnitude. We estimate the colour as the median value of the $g-r$ profiles and from that we estimate the stellar mass. Table \ref{tab:table_progenitors} summarises the values and compares it to the stream properties, while the images are shown in Figure \ref{fig:progenitors_zoom}. Caution should be taken when comparing the colours of the satellite candidates (obtained from ellipse-fitting on the original images), of the host (taken from the \texttt{Tractor} model fitting from the SGA-2020 catalogue) and of the streams (measured on residuals from the segmentation masks). The stream-to-dwarf mass ratio is an indicator of the degree of the disruption process, with higher ratios indicating later disruption stages.
\begin{figure*}
    \centering
    \includegraphics[width=0.9\linewidth]{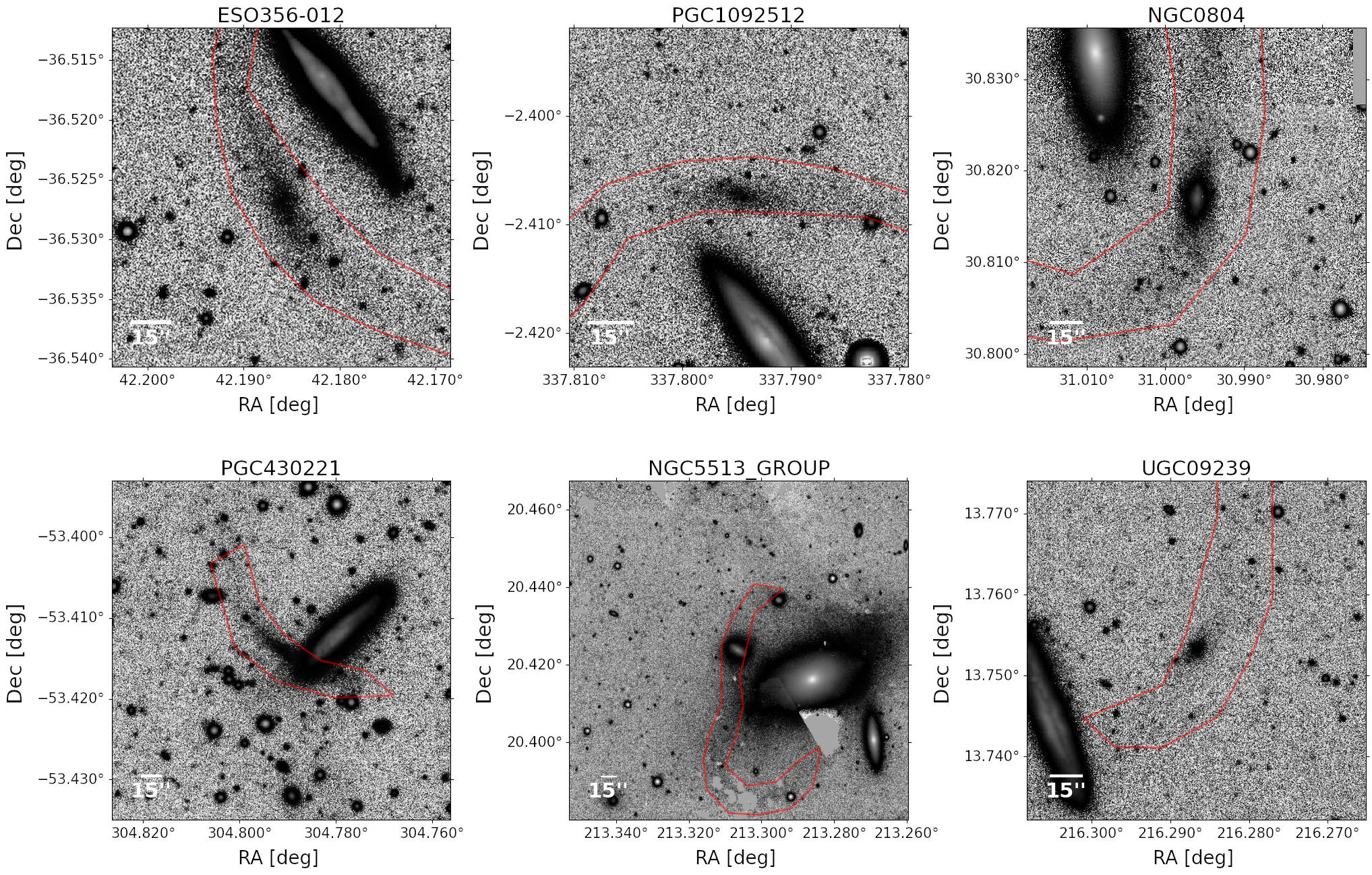}
    \caption{Zoom on the $g$-band images centred on the candidate dwarf galaxy progenitors of the streams. A different stretch is used to visualise both the bright inner regions and the faint outskirts of the galaxies. A scalebar of $15\arcsec$ is indicated at the bottom of each panel. North is up, East is left. The faint red lines outline the contours of the segmented stream. For NGC5513\_GROUP, the identification of the dwarf galaxy as the progenitor of the stream remains uncertain, whereas for the other five cases, the association is more confidently established. }
    \label{fig:progenitors_zoom}
\end{figure*}

In all cases, the satellite candidates have bluer colours than the host galaxy, which may reflect both an age and metallicity effect in the minor merger scenario.
NGC0804's dwarf candidate (top, right panel of Figure \ref{fig:progenitors_zoom}) is very likely to be the progenitor of the stream, as it has the same redshift as the host galaxy, it is aligned with the stream and has similar colour. The stream-to-dwarf mass ratio is 1.85 which means that the dwarf galaxy has already been stripped of a lot of its stellar mass, which makes sense when considering the extent of the stream. A smaller and fainter second overdensity is visible along the stream, North of the dwarf, but it is not reported in the literature. Modelling would be needed to determine its origin. 

NGC5513\_GROUP's satellite galaxy (bottom, middle panel) has a redshift similar to the host's and the stream is redder than the dwarf. The stream-to-dwarf mass ratio is 0.07, the satellite is massive (above the usual stellar mass limit of $10^{9}M_\odot$)  and does not show prominent tidal disruption. Other galaxies are located next to NGC5513\_GROUP and may contribute to a fraction of the light of the stream. The association between the dwarf and the stream can be doubted. 

For ESO356-012 (top, left panel), even though no redshift information is available, the alignment of the dwarf with the stream, the similar colours and the mass ratio of 0.8 give us confidence in asserting that this dwarf is the progenitor of the stream.

PGC1092512's candidate dwarf galaxy (top, middle panel) has similar colours and it is aligned with the stream. The stream-to-dwarf mass ratio of 6.4 is not so surprising when comparing the large extent of the stream. We are relatively confident in saying that this dwarf is the progenitor of the stream. 

PGC430221's dwarf galaxy candidate (bottom, left panel) is six times more massive than the stream and 0.12 mag redder. The photometry should be considered with caution as the stream and the dwarf are overlapping with the host galaxy. The dwarf candidate shows clear signs of tidal disruption and is aligned with the stream, so we can consider it as the progenitor.

UGC09239's faint dwarf galaxy candidate (bottom, right panel) is 8.4 times less massive than the stream, which is not surprising when considering the large spatial extent and brightness of the stream. The dwarf candidate is aligned with the stream and it is slightly redder by 0.08 mag. Although distance information would be needed, we can assume that this is the stream progenitor.

These examples illustrate diverse phases of tidal disruption and enable a better understanding of the current merger. Adding the positions of these five potential progenitors will provide additional constraints when modelling dark matter haloes in \cite{Chemaly_et_al_2025b}.

\begin{table*}
\caption{Properties of the dwarf galaxy candidates that are potential progenitors of the streams in which they are embedded. \textit{(1)}: Name of the host galaxy; \textit{(2)}: Right ascension of the dwarf candidate (deg), \textit{(3)}: Declination of the dwarf candidate (deg);  \textit{(4)}: Stellar mass of the stream (M$_\odot$); \textit{(5)}: Stellar mass of the dwarf candidate (M$_\odot$); \textit{(6)}: $g-r$ colour of the host galaxy, computed from the \texttt{Tractor} model fluxes from the SGA-2020 catalogue; \textit{(7)}: Median $g-r$ colour of the stream; \textit{(8)}: Median $g-r$ colour of the dwarf candidate, \textit{(9)}: Redshift of the host galaxy (from the SGA-2020 catalogue); \textit{(10)}: Redshift of the dwarf candidate (from NED). } 
\label{tab:table_progenitors}
\begin{tabular}{cccccccccc}
\hline
Host & RA$_\texttt{dwarf}$ & Dec$_\texttt{dwarf}$ &   $M_{\star,\texttt{stream}}$ &  $M_{\star,\texttt{dwarf}}$ & $g-r_{\texttt{Host}}$ & $g-r_{\texttt{stream}}$ & $g-r_{\texttt{dwarf}}$ & $z_{\texttt{host}}$ &$z_{\texttt{dwarf}}$\\ 
(1) & (2) & (3) & (4) & (5) & (6) & (7) & (8) & (9) & (10) \\
\hline
ESO356-012 & 42.186 & -36.526 & 2.1E+08 & 2.6E+08 & 0.60 & 0.47 $\pm$ 0.23 & 0.47 $\pm$ 0.07 & 0.0171 & -- \\
NGC0804 & 30.996 & 30.817 & 5.0E+09 & 2.7E+09     & 0.87 & 0.78 $\pm$ 0.07 & 0.71 $\pm$ 0.02 & 0.0176& 0.0175 \\
NGC5513\_GROUP & 213.306 & 20.424 & 1.4E+09 & 1.9E+10& 0.80 & 0.59 $\pm$ 0.10 & 0.75 $\pm$ 0.01& 0.0167 & 0.0167  \\
PGC1092512 & 337.795 & -2.407 &  9.0E+08 & 1.4E+08 & 0.91& 0.59 $\pm$ 0.14 & 0.6 $\pm$ 0.12 & 0.0173 & -- \\
PGC430221 & 304.791 & -53.414 &9.5E+07 & 5.9E+08& 0.66 & 0.49 $\pm$ 0.38 & 0.61 $\pm$ 0.07 & 0.0177 & -- \\
UGC09239 & 216.287 & 13.753 & 2.1E+09 & 2.5E+08& 0.91 & 0.53 $\pm$0.08 & 0.61 $\pm$ 0.06 & 0.0186 & -- \\

\hline
\end{tabular}
\end{table*}

\section{Discussion}\label{section:discussion}
\subsection{Tidal features census} \label{sec:discussion_tidal_features}

The frequency of tidal features is closely linked to the late assembly history of galaxies and can be used to constrain models of galaxy formation and evolution through comparisons with numerical simulations. Several factors influence the observed fraction of tidal features, both observational (survey depth, image processing techniques) and astrophysical (galaxy mass, morphological type, distance, environment). In this section, we compare our results with the literature.

We remind here that our galaxy sample was selected on a redshift criteria ($z\leq0.02$) from the SGA-2020 catalogue of galaxies above a given apparent angular diameter \citep{Moustakas_et_al_2023}. Our sample of $19,387$ galaxies is composed of 64.9\% LTGs, 23.4\% ETGs and 11.7\% have missing morphological information. Galaxies have a median stellar mass of $10^{9.7} M_\odot$ (16\%-84\% quantiles of $10^{8.7} - 10^{10.7} M_\odot$). They are located in different environments, ranging from the field to galaxy groups and clusters (e.g., Virgo and Coma clusters), but the majority (92.3\%) of galaxies are isolated.  

Our tidal feature fractions are reported for an approximate DESI-LS depth of 28.5 and 29 mag$\,$arcsec$^{-2}$ in the $r$ and $g$-band (we do not consider the $z$-band for detection as it is too shallow). We find $11.9 \pm 0.2\%$ of our galaxies to host detectable tidal features, with $4.4 \pm 0.1\%$ hosting streams, $3.3 \pm 0.1\%$ tails, $2.4 \pm 0.1\%$ shells and $1.8 \pm 0.1\%$ plumes. These fractions increase with host galaxy stellar mass and vary depending on the morphological type. 

\subsubsection{Impact of the morphological type}
In Section \ref{section:results}, we found a higher incidence of tidal tails around LTGs compared to ETGs. This can be attributed to the fact that prominent, extended tails are more readily produced by progenitors with ordered rotational motions, such as disc galaxies \citep[e.g.,][]{Duc_and_Renaud_2013}. In contrast, tidal features originating from ETGs tend to exhibit a more plume-like morphology.
Our category `Tails/Arms' can include some spiral arms when the origin is ambiguous. We also found an higher incidence of shells around ETGs compared to LTGs. Shells are thought to originate from intermediate-mass radial mergers \citep[e.g.,][]{Prieur_1990,Ebrova_2013,Duc_et_al_2015,Pop_2018} that would destroy the disc of LTGs, hence higher fraction of shells are expected around ETGs. This is also found in other observations \citep[e.g.,][]{Kado-Fong_et_al_2018} and in simulations \citep[e.g.,][]{Valenzuela_and_Remus_2024}. Streams are expected to form around both morphological types without preference (for a fixed galaxy mass) as they originate from a disrupted minor companion \citep[e.g.,][]{Sola_et_al_2025, Valenzuela_and_Remus_2024}, which is what we observe for galaxies with stellar masses below $\sim 1.5\times10^{11}M_\odot$. The higher stream fraction around the most massive ETGs compared to the most massive LTGs is consistent with the hierarchical models in which ETGs are more evolved objects that have undergone more (including minor) mergers events.

\subsubsection{Comparison to observations and simulations}

A comparison of the overall tidal feature fraction to the literature should take into account the fact that different works used different definitions, depths, methods or galaxy sample. This leads to a wide range of reported tidal feature fractions: from a few to about 10\% \citep[e.g.,][]{Kado-Fong_et_al_2018,Skryabina_et_al_2024,Miro-Carretero_et_al_2024}; between 10-20\% \citep[e.g.,][]{Malin_and_Carter_1983,Kaviraj_2010, Miskolczi_et_al_2011, Kim_et_al_2012,Atkinson_Abraham_Ferguson_2013,Hood_et_al_2018,Morales_et_al_2018,Vazquez-Mata_et_al_2022}, 20-40\% \citep[e.g.,][]{Bilek_et_al_2020,Huang_and_Fan_2022, Desmons_et_al_2023,Yoon_et_al_2023,Rutherford_et_al_2024,Sola_et_al_2025}, 40-50\% \citep[e.g.,][]{Sheen_et_al_2012,Jackson_et_al_2023}, and up to 50-70\% \citep[e.g.,][]{van_Dokkum_2005,Tal_et_al_2009,Duc_et_al_2015}.
Our fraction falls at the lower end of reported values, and when considering streams alone, it is notably lower than those found in the literature.
An important point to consider is the morphological type of the host galaxy and its environment. While most of the aforementioned studies focus on ETGs often in groups or clusters, our sample is dominated by LTGs located in different environments including the field, with a large majority of isolated galaxies, where tidal feature fractions are lower. Indeed, we find 31\% of the galaxies in groups to host tidal features, while it is of 10.3\% for isolated galaxies. Our results are more comparable to similarly LTG-dominated study by \cite{Skryabina_et_al_2024}, who found 11\% and 5.8\% of their edge-on galaxies to host LSB features and streams, respectively.

Observational studies must be compared to numerical simulations that provide insights on the tidal feature fraction using simulated mock images. \cite{Martin_et_al_2022} studied the tidal feature detectability from visual classification of mock images as a function of depth, galaxy mass and environment. They report increased tidal feature flux fractions with depth and galaxy mass. This fraction ranges from a zero to few percents for a depth of 27 mag$\,$arcsec$^{-2}$ and increases up to almost 100\% for a depth of 33 mag$\,$arcsec$^{-2}$, the values showing a strong dependence on galaxy mass. They estimate 60\% of the tidal feature flux would be detectable at 29.5 mag$\,$arcsec$^{-2}$.
\cite{Vera-Casanova_et_al_2022} find similar trends through visual inspection of SB maps, with a significant increase of tidal feature detection fraction between 28-29 mag$\,$arcsec$^{-2}$. Their fraction ranges from a few percent below 28 mag$\,$arcsec$^{-2}$, up to 87\% for 31 mag$\,$arcsec$^{-2}$. Similar trends are reported in \cite{Miro-Carretero_et_al_2024_simu}, where the fraction of visually detected streams in mock images reaches 70\% for a depth of 32 mag$\,$arcsec$^{-2}$. 
\cite{Valenzuela_and_Remus_2024} found that 10-30\% of their massive (M$_\star > 10^{11}M_\odot$) galaxies in the Magneticum Pathfinder simulation host tidal features, and 29\% hosting streams. Overall, with their method of visually inspecting a three-dimensional rendering of the simulated galaxies, about 18\% of all the galaxies in their sample host any type of tidal feature. \cite{Khalid_et_al_2024} visually inspect mock images of four different simulations and find an overall tidal feature fraction of  37\% for EAGLE, 32\% for TNG, 32\% for Magneticum and 40\% for NewHorizon.
These findings highlight a large variability of the feature fraction is seen for simulations, which relates both to the methods (preparation of mock images and tidal feature identification lead to different percentages while using the same simulation), as well as the simulation type and properties.

An important source of differences between simulations lies in their underlying physical assumptions. 
For example, \citet{Zhu+2016} showed that baryonic processes such as adiabatic contraction, tidal disruption, and reionization strongly influence the distribution of dark matter, while \citet{Sawala+2017} demonstrated that the inclusion of baryons in the APOSTLE simulation leads to enhanced tidal disruption and a reduced number of surviving subhaloes compared to dark matter–only runs.
Since galaxy properties are tightly connected to their dark matter haloes, variations in halo characteristics or in the stellar-to-halo mass relation can significantly affect stellar stripping. For example, haloes with higher concentrations tend to strip stars more efficiently \citep{Contini+2023,Martin_et_al_2024}.  
While \cite{Khalid_et_al_2024} found overall tidal feature fractions consistent across several large cosmological simulations, suggesting they are primarily driven by gravitational interactions and relatively insensitive to subgrid physics, they also emphasised that differences in the stellar–halo mass relation can increase the amount of stellar material stripped during mergers (such as NewHorizon producing more stars at fixed halo mass and lacking the most massive haloes). Similarly, \cite{Kelly+2022} showed with the AURIGA and APOSTLE simulations that, while subgrid physics models can reproduce realistic stellar components of L$^{*}$ galaxies, they result in different gas properties, which in turn can influence the characteristics of tidal features.
Resolution effects must also be taken into account, as \cite{Martin_et_al_2024} demonstrated that the numerical resolution of simulations strongly affects stellar stripping rates: low-resolution dark matter haloes tend to develop cored profiles, enhancing stripping, and the radius from which stars are stripped depends on both stellar and dark matter resolution. Similarly, \cite{Khalid_et_al_2024} reported that higher stellar mass resolution increases the detectability of tidal features.

Then, we explore potential reasons for our low tidal feature fraction compared to other observational works, such as differences in tidal feature definitions, identification methods, surface brightness limit, or data reduction techniques.  Techniques that enhance image contrast beyond simple stretching, such as unsharp masking \citep[e.g.,][]{Colbert_et_al_2001,Kim_et_al_2012,Giri_et_al_2023} or residual imaging \citep[e.g.,][]{Bell_et_al_2006,McIntosh_et_al_2008,Tal_et_al_2009}, should facilitate the detection of features by reducing galaxy contamination. However, our fraction remains notably low, suggesting additional contributing factors. 
The experience of the classifiers is an important factor, as more expert users tend to identify fainter and more extended structures \citep[e.g.,][]{Bilek_et_al_2020}. The classification differences between experts and citizen scientists was also discussed in \cite{Gordon_et_al_2024}.For this reason, our two most experienced classifiers inspected the entire galaxy sample during the first visual inspection round. The remaining seven classifiers were trained beforehand, and to mitigate differences in expertise, we combined all votes and assigned a confidence score to each feature, reflecting the level of agreement in its classification (see Figure \ref{fig:appendix-results_pie_chart_scores}). We still see that our two most experienced classifiers found higher combined fractions of tails, streams, shells and plumes than the other classifiers, further motivating the voting system to mitigate for the experience bias.  We do not account for self-variability within classifiers \citep[e.g.,][]{Bridge_et_al_2010} that is another source of classification variability.

Image depth is another key factor, as deeper imaging reveals more tidal features, but differences in how depth is estimated \citep[e.g.,][]{Roman_et_al_2020} can also affect the comparison between studies. In this context, numerical simulations provide predictions for the expected tidal feature fraction as a function of imaging depth. For a depth of about 28.5 mag$\,$arcsec$^{-2}$, \cite{Martin_et_al_2022} predicted from the NewHorizon simulation that, depending on the galaxy mass, between 0-50\% of the flux and 0-15\% of the area of tidal features will be detected. For our median mass of $10^{9.7} M_\odot$, these fractions are very low, below a few percent.
\cite{Vera-Casanova_et_al_2022} predicted for Milky Way mass haloes from the Auriga simulation that 20-30\% of the galaxies should host streams at a SB limit of 28.5 mag$\,$arcsec$^{-2}$. Using a different method to generate mock images from simulations, \cite{Miro-Carretero_et_al_2024_simu} found for a depth of 28.5 mag$\,$arcsec$^{-2}$ an average percentage of galaxies hosting streams of 6\% for the TNG50 galaxies, 9\% for Auriga and a much higher fraction of 30\% for COCO.  Additionally, the quality of the images, including the presence of artifacts, Galactic cirrus, the differences in data processing pipelines with the problem of oversubtraction of the background, all lead to discrepancies within observational studies and between observations and simulations (either idealised or that include only some of the contaminants). Oversubtraction of the background is visible in the DESI-LS images, so we are less likely to recover faint tidal features close to the galaxy. This is not an issue for our search of long, curved streams that extend further away from the galaxy, but still impacts the overall tidal feature fraction and can lead to lower fractions compared to surveys with an LSB-optimised data reduction pipeline.
The size of the images is another factor to consider, as discussed in Section \ref{sec:discussion-streams_literature}. The images used in this study have a predefined field of view of three times to the galaxy size, that can lead to non-detection of tidal features further away from the galaxy. 
One could also wonder whether the less accurate modelling of LTGs, compared to ETGs, could result in more artefacts or higher non-detection rates. However, we mitigate this concern by jointly inspecting the original, modelled, and residual images to ensure robust feature identification.
Finally, projection effects and intrinsic alignment between tidal features and galaxies must also be considered \citep[e.g.,][]{Bournaud_et_al_2004, Lotz_et_al_2008,Mancillas_2019} as they can significantly impact feature detectability.

\subsubsection{Impact of the host galaxy mass}
The host galaxy stellar mass has a strong influence of the tidal feature fraction (Figure \ref{fig:feature_fraction_vs_mass}): it ranges from $2.4 \pm 0.4\%$ at $\sim10^8 M_\odot$ to $36.5 \pm 1.2\%$ at $\sim 5\times10^{11} M_\odot$. This mass dependency is in line with the literature.
Indeed, many observational works report and discuss this trend \citep[e.g.,][]{Atkinson_Abraham_Ferguson_2013,Duc_et_al_2015,Kado-Fong_et_al_2018, Yoon_and_Lim_2020,Bilek_et_al_2020,  Vazquez-Mata_et_al_2022,Desmons_et_al_2023, Skryabina_et_al_2024, Sola_et_al_2025}, which is also supported by numerical simulations \citep[e.g.,][]{Martin_et_al_2022, Khalid_et_al_2024,Valenzuela_and_Remus_2024}. 
These results align with hierarchical models of galaxy formation, where more massive galaxies (both in stellar and halo mass) have undergone more merger events, resulting in a higher frequency of tidal features. 
These trends can also be understood in the context of the stellar mass – halo mass relation \citep[e.g.,][]{Wechsler_and_Tinker_2018,Girelli_et_al_2020}. While the dark matter distribution is approximately self-similar across a wide range of halo masses, lower-mass subhaloes tend to host significantly fewer stars, or in some cases none at all. As a result, the tidal features they produce are fainter and more difficult to detect, particularly at the low-mass end, which would bias completeness due to limited survey depth.

Therefore, the combination of our relatively low-mass (median stellar mass $10^{9.7} M_\odot$), LTG-dominated galaxy sample and observational effects (depth, data reduction, visual classification) could drive the relatively low tidal feature fraction we observe.

\subsection{STRRINGS and literature stream catalogues} \label{sec:discussion-streams_literature}

We compare our tidal feature classification to existing stream catalogues in the literature that rely on DESI-LS images, namely \cite{Martinez-Delgado_et_al_2023, Miro-Carretero_et_al_2023, Miro-Carretero_et_al_2024} (hereafter \citetalias{Martinez-Delgado_et_al_2023, Miro-Carretero_et_al_2023,Miro-Carretero_et_al_2024}, respectively). While other stream catalogues exist \citep[e.g.,][]{Skryabina_et_al_2024,Pippert_et_al_2025,Sola_et_al_2025}, we limit our comparison to those based on the same imaging dataset.
We acknowledge that the streams in our STRRINGS sample and in our broader tidal feature catalogue may have been previously identified in other studies. Our aim is not to conduct an exhaustive comparison with the existing literature, but rather to assess whether we may have missed some of the galaxies hosting streams suitable for modelling (i.e.,  long, curved, faint features around otherwise unperturbed hosts) that have already identified in other studies using the same images.

Our classification was designed with a specific objective in mind, and the categories used were mutually exclusive. As a result, features that did not match our definition may have been classified differently in other studies.  We emphasise that the distinction between tidal feature classes is often ambiguous (sometimes leading to different categories by our classifiers hence lower confidence scores), and definitions vary across studies. In the absence of a ground truth, the classification of such features inevitably involves a degree of subjectivity. 

With this in mind, we cross-match our tidal feature catalogue to the stream catalogues in the literature and find 75 galaxies in common (including 20 galaxies with \citetalias{Martinez-Delgado_et_al_2023}, 22 with \citetalias{Miro-Carretero_et_al_2023} and 36 with \citetalias{Miro-Carretero_et_al_2024}). The individual comparison of the classification is presented in Table \ref{tab:appendix-compare_streams}. The classifications agree on 28 streams.
Only 3 out of our 35 STRRINGS streams are in common with these three literature works. Among the 75 galaxies with literature-reported streams that are also in our sample, we classified 6 galaxies as having tails, 4 with plumes, and 14 with shells. We find 9 galaxies to exhibit spiral features, one is classified as curiosity and we marked 13 as having no feature. 

We investigate the origin of these discrepancies by visualising the original, model and residual images. We re-download the images $g,r,z$-band images with a fixed size of 1500 pixels and a pixel scale of 0.4 arcsecond. We combine and stretch the images in the same manner as for Figure \ref{figure:data-img_model_res}. It is important to note that these are not the images we used for the visual classification, as we used the montage PNG already available from the DESI LS website\footnote{Montage PNG images available at \url{https://portal.nersc.gov/project/cosmo/data/sga/2020/html/}.} that have a field-of-view three times larger than the galaxy size, so the scaling and field of view are different. Some examples are shown in Figure \ref{fig:comparison_literature_streams}.

The ambiguity between streams, tails and plumes can be quite high and some cases can easily be put in one category or another. A good example is UGC04132 with a majority vote for tails but with a mix of streams and plumes votes. 
The galaxies we identified as having shells generally display prominent shells and/or disturbed morphologies with several features. We did not consider umbrellas (e.g., NGC0922, NGC5971) as streams as the host is quite perturbed and this would prevent a proper modelling of the dark matter halo. PGC597851 does have an extended stream but also has inner shells, in which case it was classified as shells as it may not be usable for modelling. Another ambiguous case is ESO476-004 that got three votes for shells and two for streams, but the host is slightly perturbed.

ESO485-011 is interacting with a dwarf galaxy but we did not considered it disrupted enough to be classified as stream, instead we labeled it as a curiosity. For the galaxies we identified as grand-design spirals, in two cases there is a stream but we did not reach a consensus vote: NGC2543 (2 votes for streams, 2 for tails, 3 for spirals) and NGC3614 (2 votes for streams, 4 for spirals). These streams, with visible progenitors, are relatively short and would not be suitable to be modelled. However, we missed the long stream of NGC0171. This is actually due to the size of the field-of-view between these images and the ones we used for the classification. We only could see a region close to the galaxy, while the stream extends much further away. In addition, a bad \texttt{Tractor} modelling with strong and sharp artefact partially hid the stream in the residual image. We do not include it in our STRRINGS sample, to be consistent with the method used for the whole classification.
The same issue of the too small field-of-view holds for ESO242-007, NGC4799 and NGC7400: the streams are located further away from the galaxy than observable with our images. For PGC13411, there is one vote for shell, one for stream, one for disturbed halo and five votes for no feature, but the stream is in any case too short to be modelled. The short, straight stream of IC1904 we overlooked could also not have been modelled.

Therefore, the only galaxies (NGC0171 and maybe NGC4799) with streams suitable for modelling in these DESI-LS literature works that are not in our catalogue are due to the too small field-of-view of the images used during the classification.

\section{Conclusions} \label{section:conclusion}

Tidal features can be used as probes of the late galactic assembly and of the dark matter haloes surrounding galaxies. Putting constraints on the properties of dark matter haloes at a population level requires large samples of long, faint streams around unperturbed host galaxies.

In this paper, we provide a census of tidal features around $19,387$ nearby galaxies ($z\leq0.02$) from the SGA-2020 catalogue. We utilise the combined information from the original, model and residual images from the DESI Legacy Imaging Surveys to disclose the faintest features. We compile and publicly release a catalogue of LSB features, including streams, tails, shells, plumes, and disturbed haloes. We find that $11.9 \pm 0.2\%$ of the galaxies host detectable tidal features: $4.4 \pm 0.1\%$ have streams, $3.3 \pm 0.1\%$ have tails, $2.4 \pm 0.1\%$ have shells and $1.8 \pm 0.1\%$ have plumes. The fraction of galaxies with features increases with stellar mass, from about $2.4 \pm 0.4\%$ up to $36.5 \pm 1.2\%$ for the most massive galaxies. Early-type galaxies host more tidal features than late-type, especially streams and shells. The trends and values are broadly consistent with previous observational studies and with the predictions from numerical simulations, given the depth of our data.

From the tidal feature catalogue, we select the STRRINGS sample, a subset of 35 long, faint and curved enough streams to be modelled within our modelling framework. We manually segment them and extract their geometrical properties, tracks, surface brightnesses, and colours, from which the stellar mass is computed. The median surface brightness is 26.4, 26.8 and 25.1 mag$\,$arcsec$^{-2}$ in $r$, $g$ and $z$-band respectively, although SB variations are present along each stream. The mass ratio between the stream and the host galaxy (mostly between 0.007 and 0.02, with only four higher than 0.05) are consistent with minor mergers, which makes them excellent tracers of the underlying gravitational potential. 

Our STRRINGS streams have a median width of 5.6 kpc and length of 124 kpc with an extended distribution, which makes them longer than the literature values, but they were selected to be long. These streams, being long, curved, minor-merger remnants, are ideal testbeds for modelling halo potentials. We find a positive correlation between the mass of the stream and its length, as well as with redder colours. A slight trend with the width is observed. This is consistent with the idea that more massive galaxies have higher velocity dispersions (hence wider and longer streams).  Wider streams are located at higher galactocentric radii, as expected from theoretical predictions. 

Furthermore, we identify six candidate satellite dwarf galaxy progenitors of the streams. We performed ellipse-fitting to retrieve their photometry and colours, and compare them to the stream values. Five of these candidates are highly likely to be the progenitors (one has a confirmed redshift), while more doubt remains for the sixth. 

Our STRRINGS sample, and the positions of the candidate dwarf galaxy progenitors, will be used in \cite{Chemaly_et_al_2025b} to constrain the flattening of the dark matter haloes at a population level. The STRRINGS catalogue will be extended in future works to more distant galaxy, using machine learning tools.

The methods presented in this series of papers serve as a proof of concept based on a relatively small catalogue. With the advent of large-scale surveys optimised for low surface brightness science, such as Rubin/LSST and Euclid, the number of streams suitable for modelling is expected to increase dramatically. Applying a similar approach to these forthcoming datasets holds great promise for advancing our understanding of galaxy assembly and dark matter halo structures.

\section*{Acknowledgements}
We thank the referee, Garreth Martin, for constructive comments that helped improve the manuscript. 
We thank Kyle Adams and Maria Skryabina for sharing their SB measurements of streams from their paper \citep{Skryabina_et_al_2024}.
E.S. and V.B. are grateful for the support from the Leverhulme Trust Research Project Grant RPG-2021-205 “The Faint Universe Made Visible with Machine Learning”.
OM is grateful to the Swiss National Science Foundation
for financial support under the grant number PZ00P2\_202104. AAA acknowledges support from the Herchel Smith Fellowship at the University of Cambridge and a Fitzwilliam College research fellowship supported by the Isaac Newton Trust. EYD thanks the Science and Technology Facilities Council (STFC) for a PhD studentship (UKRI grant number 2605433). HZ thanks the Science and Technology Facilities Council (STFC) for a PhD studentship. DC acknowledges funding from the Harding Distinguished Postgraduate Scholars Program. JLM acknowledges the support of a fellowship from the “la Caixa” Foundation (ID 100010434), the fellowship code is LCF/BQ/EU24/12060091. 

The Siena Galaxy Atlas was made possible by funding support from the U.S. Department of Energy, Office of Science, Office of High Energy Physics under Award Number DE-SC0020086 and from the National Science Foundation under grant AST-1616414.

The Legacy Surveys consist of three individual and complementary projects: the Dark Energy Camera Legacy Survey (DECaLS; Proposal ID \#2014B-0404; PIs: David Schlegel and Arjun Dey), the Beijing-Arizona Sky Survey (BASS; NOAO Prop. ID \#2015A-0801; PIs: Zhou Xu and Xiaohui Fan), and the Mayall z-band Legacy Survey (MzLS; Prop. ID \#2016A-0453; PI: Arjun Dey). DECaLS, BASS and MzLS together include data obtained, respectively, at the Blanco telescope, Cerro Tololo Inter-American Observatory, NSF’s NOIRLab; the Bok telescope, Steward Observatory, University of Arizona; and the Mayall telescope, Kitt Peak National Observatory, NOIRLab. Pipeline processing and analyses of the data were supported by NOIRLab and the Lawrence Berkeley National Laboratory (LBNL). The Legacy Surveys project is honored to be permitted to conduct astronomical research on Iolkam Du’ag (Kitt Peak), a mountain with particular significance to the Tohono O’odham Nation.

NOIRLab is operated by the Association of Universities for Research in Astronomy (AURA) under a cooperative agreement with the National Science Foundation. LBNL is managed by the Regents of the University of California under contract to the U.S. Department of Energy.

This project used data obtained with the Dark Energy Camera (DECam), which was constructed by the Dark Energy Survey (DES) collaboration. Funding for the DES Projects has been provided by the U.S. Department of Energy, the U.S. National Science Foundation, the Ministry of Science and Education of Spain, the Science and Technology Facilities Council of the United Kingdom, the Higher Education Funding Council for England, the National Center for Supercomputing Applications at the University of Illinois at Urbana-Champaign, the Kavli Institute of Cosmological Physics at the University of Chicago, Center for Cosmology and Astro-Particle Physics at the Ohio State University, the Mitchell Institute for Fundamental Physics and Astronomy at Texas A\&M University, Financiadora de Estudos e Projetos, Fundacao Carlos Chagas Filho de Amparo, Financiadora de Estudos e Projetos, Fundacao Carlos Chagas Filho de Amparo a Pesquisa do Estado do Rio de Janeiro, Conselho Nacional de Desenvolvimento Cientifico e Tecnologico and the Ministerio da Ciencia, Tecnologia e Inovacao, the Deutsche Forschungsgemeinschaft and the Collaborating Institutions in the Dark Energy Survey. The Collaborating Institutions are Argonne National Laboratory, the University of California at Santa Cruz, the University of Cambridge, Centro de Investigaciones Energeticas, Medioambientales y Tecnologicas-Madrid, the University of Chicago, University College London, the DES-Brazil Consortium, the University of Edinburgh, the Eidgenossische Technische Hochschule (ETH) Zurich, Fermi National Accelerator Laboratory, the University of Illinois at Urbana-Champaign, the Institut de Ciencies de l’Espai (IEEC/CSIC), the Institut de Fisica d’Altes Energies, Lawrence Berkeley National Laboratory, the Ludwig Maximilians Universitat Munchen and the associated Excellence Cluster Universe, the University of Michigan, NSF’s NOIRLab, the University of Nottingham, the Ohio State University, the University of Pennsylvania, the University of Portsmouth, SLAC National Accelerator Laboratory, Stanford University, the University of Sussex, and Texas A\&M University.

BASS is a key project of the Telescope Access Program (TAP), which has been funded by the National Astronomical Observatories of China, the Chinese Academy of Sciences (the Strategic Priority Research Program “The Emergence of Cosmological Structures” Grant \# XDB09000000), and the Special Fund for Astronomy from the Ministry of Finance. The BASS is also supported by the External Cooperation Program of Chinese Academy of Sciences (Grant \# 114A11KYSB20160057), and Chinese National Natural Science Foundation (Grant \# 12120101003, \# 11433005).

The Legacy Survey team makes use of data products from the Near-Earth Object Wide-field Infrared Survey Explorer (NEOWISE), which is a project of the Jet Propulsion Laboratory/California Institute of Technology. NEOWISE is funded by the National Aeronautics and Space Administration.

The Legacy Surveys imaging of the DESI footprint is supported by the Director, Office of Science, Office of High Energy Physics of the U.S. Department of Energy under Contract No. DE-AC02-05CH1123, by the National Energy Research Scientific Computing Center, a DOE Office of Science User Facility under the same contract; and by the U.S. National Science Foundation, Division of Astronomical Sciences under Contract No. AST-0950945 to NOAO.

\section*{Data Availability}
Our catalogue of tidal feature classification for the $19,387$ galaxies (Table \ref{table:catalogue_galaxies}), the properties of the STRRINGS sample (Table \ref{tab:best_streams}), the coordinates of the segmented streams, and the comparison between our catalogue and literature streams (Table \ref{tab:appendix-compare_streams}) are available in the online version of this article.



\bibliographystyle{mnras}
\bibliography{bibliography} 

\begin{thebibliography}{}
\makeatletter
\relax
\def\mn@urlcharsother{\let\do\@makeother \do\$\do\&\do\#\do\^\do\_\do\%\do\~}
\def\mn@doi{\begingroup\mn@urlcharsother \@ifnextchar [ {\mn@doi@} {\mn@doi@[]}}
\def\mn@doi@[#1]#2{\def\@tempa{#1}\ifx\@tempa\@empty \href {http://dx.doi.org/#2} {doi:#2}\else \href {http://dx.doi.org/#2} {#1}\fi \endgroup}
\def\mn@eprint#1#2{\mn@eprint@#1:#2::\@nil}
\def\mn@eprint@arXiv#1{\href {http://arxiv.org/abs/#1} {{\tt arXiv:#1}}}
\def\mn@eprint@dblp#1{\href {http://dblp.uni-trier.de/rec/bibtex/#1.xml} {dblp:#1}}
\def\mn@eprint@#1:#2:#3:#4\@nil{\def\@tempa {#1}\def\@tempb {#2}\def\@tempc {#3}\ifx \@tempc \@empty \let \@tempc \@tempb \let \@tempb \@tempa \fi \ifx \@tempb \@empty \def\@tempb {arXiv}\fi \@ifundefined {mn@eprint@\@tempb}{\@tempb:\@tempc}{\expandafter \expandafter \csname mn@eprint@\@tempb\endcsname \expandafter{\@tempc}}}

\bibitem[\protect\citeauthoryear{{Abraham}, {van den Bergh}  \& {Nair}}{{Abraham} et~al.}{2003}]{Abraham_et_al_2003}
{Abraham} R.~G.,  {van den Bergh} S.,   {Nair} P.,  2003, \mn@doi [\apj] {10.1086/373919}, \href {https://ui.adsabs.harvard.edu/abs/2003ApJ...588..218A} {588, 218}

\bibitem[\protect\citeauthoryear{{Akhlaghi} \& {Ichikawa}}{{Akhlaghi} \& {Ichikawa}}{2015}]{Akhlaghi_and_Ichikawa_2015}
{Akhlaghi} M.,  {Ichikawa} T.,  2015, \mn@doi [\apjs] {10.1088/0067-0049/220/1/1}, \href {https://ui.adsabs.harvard.edu/abs/2015ApJS..220....1A} {220, 1}

\bibitem[\protect\citeauthoryear{{Amorisco}}{{Amorisco}}{2015}]{Amorisco_2015}
{Amorisco} N.~C.,  2015, \mn@doi [\mnras] {10.1093/mnras/stv648}, \href {https://ui.adsabs.harvard.edu/abs/2015MNRAS.450..575A} {450, 575}

\bibitem[\protect\citeauthoryear{{Amorisco}, {Martinez-Delgado}  \& {Schedler}}{{Amorisco} et~al.}{2015}]{Amorisco_et_al_2015}
{Amorisco} N.~C.,  {Martinez-Delgado} D.,   {Schedler} J.,  2015, \mn@doi [arXiv e-prints] {10.48550/arXiv.1504.03697}, \href {https://ui.adsabs.harvard.edu/abs/2015arXiv150403697A} {p. arXiv:1504.03697}

\bibitem[\protect\citeauthoryear{{Arp}}{{Arp}}{1966}]{Arp_1966}
{Arp} H.,  1966, \mn@doi [\apjs] {10.1086/190147}, \href {https://ui.adsabs.harvard.edu/abs/1966ApJS...14....1A} {14, 1}

\bibitem[\protect\citeauthoryear{{Atkinson}, {Abraham}  \& {Ferguson}}{{Atkinson} et~al.}{2013}]{Atkinson_Abraham_Ferguson_2013}
{Atkinson} A.~M.,  {Abraham} R.~G.,   {Ferguson} A. M.~N.,  2013, \mn@doi [\apj] {10.1088/0004-637X/765/1/28}, \href {https://ui.adsabs.harvard.edu/abs/2013ApJ...765...28A} {765, 28}

\bibitem[\protect\citeauthoryear{{Balcells} \& {Quinn}}{{Balcells} \& {Quinn}}{1990}]{Balcells_and_Quinn_1990}
{Balcells} M.,  {Quinn} P.~J.,  1990, \mn@doi [\apj] {10.1086/169204}, \href {https://ui.adsabs.harvard.edu/abs/1990ApJ...361..381B} {361, 381}

\bibitem[\protect\citeauthoryear{Barbary}{Barbary}{2016}]{Barbary_et_al_2016}
Barbary K.,  2016, \mn@doi [Journal of Open Source Software] {10.21105/joss.00058}, 1, 58

\bibitem[\protect\citeauthoryear{{Barnes}}{{Barnes}}{1988}]{Barnes_1988}
{Barnes} J.~E.,  1988, \mn@doi [\apj] {10.1086/166593}, \href {https://ui.adsabs.harvard.edu/abs/1988ApJ...331..699B} {331, 699}

\bibitem[\protect\citeauthoryear{{Barnes}}{{Barnes}}{2004}]{Barnes_2004}
{Barnes} J.~E.,  2004, \mn@doi [\mnras] {10.1111/j.1365-2966.2004.07725.x}, \href {https://ui.adsabs.harvard.edu/abs/2004MNRAS.350..798B} {350, 798}

\bibitem[\protect\citeauthoryear{{Bell}, {McIntosh}, {Katz}  \& {Weinberg}}{{Bell} et~al.}{2003}]{Bell_et_al_2003}
{Bell} E.~F.,  {McIntosh} D.~H.,  {Katz} N.,   {Weinberg} M.~D.,  2003, \mn@doi [\apjs] {10.1086/378847}, \href {https://ui.adsabs.harvard.edu/abs/2003ApJS..149..289B} {149, 289}

\bibitem[\protect\citeauthoryear{{Bell} et~al.,}{{Bell} et~al.}{2006}]{Bell_et_al_2006}
{Bell} E.~F.,  et~al., 2006, \mn@doi [\apj] {10.1086/499931}, \href {https://ui.adsabs.harvard.edu/abs/2006ApJ...640..241B} {640, 241}

\bibitem[\protect\citeauthoryear{{Belokurov} et~al.,}{{Belokurov} et~al.}{2006}]{Belokurov_2006}
{Belokurov} V.,  et~al., 2006, \mn@doi [\apjl] {10.1086/504797}, \href {https://ui.adsabs.harvard.edu/abs/2006ApJ...642L.137B} {642, L137}

\bibitem[\protect\citeauthoryear{{Belokurov} et~al.,}{{Belokurov} et~al.}{2014}]{Belokurov_et_al_2014}
{Belokurov} V.,  et~al., 2014, \mn@doi [\mnras] {10.1093/mnras/stt1862}, \href {https://ui.adsabs.harvard.edu/abs/2014MNRAS.437..116B} {437, 116}

\bibitem[\protect\citeauthoryear{{Bertin} \& {Arnouts}}{{Bertin} \& {Arnouts}}{1996}]{Bertin_and_Arnouts_1996}
{Bertin} E.,  {Arnouts} S.,  1996, \mn@doi [\aaps] {10.1051/aas:1996164}, \href {https://ui.adsabs.harvard.edu/abs/1996A&AS..117..393B} {117, 393}

\bibitem[\protect\citeauthoryear{{Bickley} et~al.,}{{Bickley} et~al.}{2021}]{Bickley_et_al_2021}
{Bickley} R.~W.,  et~al., 2021, \mn@doi [\mnras] {10.1093/mnras/stab806}, \href {https://ui.adsabs.harvard.edu/abs/2021MNRAS.504..372B} {504, 372}

\bibitem[\protect\citeauthoryear{{B{\'\i}lek} et~al.,}{{B{\'\i}lek} et~al.}{2020}]{Bilek_et_al_2020}
{B{\'\i}lek} M.,  et~al., 2020, \mn@doi [\mnras] {10.1093/mnras/staa2248}, \href {https://ui.adsabs.harvard.edu/abs/2020MNRAS.498.2138B} {498, 2138}

\bibitem[\protect\citeauthoryear{{B{\'\i}lek}, {Duc}  \& {Sola}}{{B{\'\i}lek} et~al.}{2023}]{Bilek_2023}
{B{\'\i}lek} M.,  {Duc} P.~A.,   {Sola} E.,  2023, \mn@doi [\aap] {10.1051/0004-6361/202244749}, \href {https://ui.adsabs.harvard.edu/abs/2023A&A...672A..27B} {672, A27}

\bibitem[\protect\citeauthoryear{{Binney}}{{Binney}}{2008}]{Binney_2008}
{Binney} J.,  2008, \mn@doi [\mnras] {10.1111/j.1745-3933.2008.00458.x}, \href {https://ui.adsabs.harvard.edu/abs/2008MNRAS.386L..47B} {386, L47}

\bibitem[\protect\citeauthoryear{{Blumenthal} \& {Barnes}}{{Blumenthal} \& {Barnes}}{2018}]{Blumenthal_and_Barnes_2018}
{Blumenthal} K.~A.,  {Barnes} J.~E.,  2018, \mn@doi [\mnras] {10.1093/mnras/sty1605}, \href {https://ui.adsabs.harvard.edu/abs/2018MNRAS.479.3952B} {479, 3952}

\bibitem[\protect\citeauthoryear{{Bois} et~al.,}{{Bois} et~al.}{2011}]{Bois_2011}
{Bois} M.,  et~al., 2011, \mn@doi [\mnras] {10.1111/j.1365-2966.2011.19113.x}, \href {https://ui.adsabs.harvard.edu/abs/2011MNRAS.416.1654B} {416, 1654}

\bibitem[\protect\citeauthoryear{{Bonaca} \& {Hogg}}{{Bonaca} \& {Hogg}}{2018}]{Bonaca_2018}
{Bonaca} A.,  {Hogg} D.~W.,  2018, \mn@doi [\apj] {10.3847/1538-4357/aae4da}, \href {https://ui.adsabs.harvard.edu/abs/2018ApJ...867..101B} {867, 101}

\bibitem[\protect\citeauthoryear{{Bonaca} \& {Price-Whelan}}{{Bonaca} \& {Price-Whelan}}{2025}]{Bonaca_and_Prince-Whelan_2025}
{Bonaca} A.,  {Price-Whelan} A.~M.,  2025, \mn@doi [\nar] {10.1016/j.newar.2024.101713}, \href {https://ui.adsabs.harvard.edu/abs/2025NewAR.10001713B} {100, 101713}

\bibitem[\protect\citeauthoryear{{Bournaud}, {Duc}, {Amram}, {Combes}  \& {Gach}}{{Bournaud} et~al.}{2004}]{Bournaud_et_al_2004}
{Bournaud} F.,  {Duc} P.~A.,  {Amram} P.,  {Combes} F.,   {Gach} J.~L.,  2004, \mn@doi [\aap] {10.1051/0004-6361:20040394}, \href {https://ui.adsabs.harvard.edu/abs/2004A&A...425..813B} {425, 813}

\bibitem[\protect\citeauthoryear{{Bovy}, {Bahmanyar}, {Fritz}  \& {Kallivayalil}}{{Bovy} et~al.}{2016}]{Bovy_2016}
{Bovy} J.,  {Bahmanyar} A.,  {Fritz} T.~K.,   {Kallivayalil} N.,  2016, \mn@doi [\apj] {10.3847/1538-4357/833/1/31}, \href {https://ui.adsabs.harvard.edu/abs/2016ApJ...833...31B} {833, 31}

\bibitem[\protect\citeauthoryear{{Bridge}, {Carlberg}  \& {Sullivan}}{{Bridge} et~al.}{2010}]{Bridge_et_al_2010}
{Bridge} C.~R.,  {Carlberg} R.~G.,   {Sullivan} M.,  2010, \mn@doi [\apj] {10.1088/0004-637X/709/2/1067}, \href {https://ui.adsabs.harvard.edu/abs/2010ApJ...709.1067B} {709, 1067}

\bibitem[\protect\citeauthoryear{{Bullock} \& {Johnston}}{{Bullock} \& {Johnston}}{2005}]{Bullock_and_Johnston_2005}
{Bullock} J.~S.,  {Johnston} K.~V.,  2005, \mn@doi [\apj] {10.1086/497422}, \href {https://ui.adsabs.harvard.edu/abs/2005ApJ...635..931B} {635, 931}

\bibitem[\protect\citeauthoryear{{Casteels} et~al.,}{{Casteels} et~al.}{2013}]{Casteels_et_al_2013}
{Casteels} K. R.~V.,  et~al., 2013, \mn@doi [\mnras] {10.1093/mnras/sts391}, \href {https://ui.adsabs.harvard.edu/abs/2013MNRAS.429.1051C} {429, 1051}

\bibitem[\protect\citeauthoryear{{Chemaly}, {Sola}, {Belokurov}  et~al.}{{Chemaly} et~al.}{2025a}]{Chemaly_et_al_2025}
{Chemaly} D.,  {Sola} E.,  {Belokurov} V.,   et~al., in prep, 2025a

\bibitem[\protect\citeauthoryear{{Chemaly}, {Belokurov}, {Sola}  et~al.}{{Chemaly} et~al.}{2025b}]{Chemaly_et_al_2025b}
{Chemaly} D.,  {Belokurov} V.,  {Sola} E.,   et~al., in prep, 2025b

\bibitem[\protect\citeauthoryear{{Colbert}, {Mulchaey}  \& {Zabludoff}}{{Colbert} et~al.}{2001}]{Colbert_et_al_2001}
{Colbert} J.~W.,  {Mulchaey} J.~S.,   {Zabludoff} A.~I.,  2001, \mn@doi [\aj] {10.1086/318758}, \href {https://ui.adsabs.harvard.edu/abs/2001AJ....121..808C} {121, 808}

\bibitem[\protect\citeauthoryear{{Cole}, {Lacey}, {Baugh}  \& {Frenk}}{{Cole} et~al.}{2000}]{Cole_2000}
{Cole} S.,  {Lacey} C.~G.,  {Baugh} C.~M.,   {Frenk} C.~S.,  2000, \mn@doi [\mnras] {10.1046/j.1365-8711.2000.03879.x}, \href {https://ui.adsabs.harvard.edu/abs/2000MNRAS.319..168C} {319, 168}

\bibitem[\protect\citeauthoryear{{Comerford} et~al.,}{{Comerford} et~al.}{2024}]{Comerford_2024}
{Comerford} J.~M.,  et~al., 2024, \mn@doi [\apj] {10.3847/1538-4357/ad1a15}, \href {https://ui.adsabs.harvard.edu/abs/2024ApJ...963...53C} {963, 53}

\bibitem[\protect\citeauthoryear{{Conselice}}{{Conselice}}{2014}]{Conselice_2014}
{Conselice} C.~J.,  2014, \mn@doi [\araa] {10.1146/annurev-astro-081913-040037}, \href {https://ui.adsabs.harvard.edu/abs/2014ARA&A..52..291C} {52, 291}

\bibitem[\protect\citeauthoryear{{Conselice}, {Bershady}, {Dickinson}  \& {Papovich}}{{Conselice} et~al.}{2003}]{Conselice_et_al_2003}
{Conselice} C.~J.,  {Bershady} M.~A.,  {Dickinson} M.,   {Papovich} C.,  2003, \mn@doi [\aj] {10.1086/377318}, \href {https://ui.adsabs.harvard.edu/abs/2003AJ....126.1183C} {126, 1183}

\bibitem[\protect\citeauthoryear{{Conselice}, {Mundy}, {Ferreira}  \& {Duncan}}{{Conselice} et~al.}{2022}]{Conselice+2022}
{Conselice} C.~J.,  {Mundy} C.~J.,  {Ferreira} L.,   {Duncan} K.,  2022, \mn@doi [\apj] {10.3847/1538-4357/ac9b1a}, \href {https://ui.adsabs.harvard.edu/abs/2022ApJ...940..168C} {940, 168}

\bibitem[\protect\citeauthoryear{{Contini}, {Jeon}, {Rhee}, {Han}  \& {Yi}}{{Contini} et~al.}{2023}]{Contini+2023}
{Contini} E.,  {Jeon} S.,  {Rhee} J.,  {Han} S.,   {Yi} S.~K.,  2023, \mn@doi [\apj] {10.3847/1538-4357/acfd25}, \href {https://ui.adsabs.harvard.edu/abs/2023ApJ...958...72C} {958, 72}

\bibitem[\protect\citeauthoryear{{Cooper} et~al.,}{{Cooper} et~al.}{2010}]{Cooper_2010}
{Cooper} A.~P.,  et~al., 2010, \mn@doi [\mnras] {10.1111/j.1365-2966.2010.16740.x}, \href {https://ui.adsabs.harvard.edu/abs/2010MNRAS.406..744C} {406, 744}

\bibitem[\protect\citeauthoryear{{Cui} et~al.,}{{Cui} et~al.}{2012}]{Cui_2012}
{Cui} X.-Q.,  et~al., 2012, \mn@doi [Research in Astronomy and Astrophysics] {10.1088/1674-4527/12/9/003}, \href {https://ui.adsabs.harvard.edu/abs/2012RAA....12.1197C} {12, 1197}

\bibitem[\protect\citeauthoryear{{DESI Collaboration} et~al.,}{{DESI Collaboration} et~al.}{2016a}]{DESI_2016a}
{DESI Collaboration} et~al., 2016a, \mn@doi [arXiv e-prints] {10.48550/arXiv.1611.00036}, \href {https://ui.adsabs.harvard.edu/abs/2016arXiv161100036D} {p. arXiv:1611.00036}

\bibitem[\protect\citeauthoryear{{DESI Collaboration} et~al.,}{{DESI Collaboration} et~al.}{2016b}]{DESI_2016b}
{DESI Collaboration} et~al., 2016b, \mn@doi [arXiv e-prints] {10.48550/arXiv.1611.00037}, \href {https://ui.adsabs.harvard.edu/abs/2016arXiv161100037D} {p. arXiv:1611.00037}

\bibitem[\protect\citeauthoryear{{Dark Energy Survey Collaboration} et~al.,}{{Dark Energy Survey Collaboration} et~al.}{2016}]{DES_2016}
{Dark Energy Survey Collaboration} et~al., 2016, \mn@doi [\mnras] {10.1093/mnras/stw641}, \href {https://ui.adsabs.harvard.edu/abs/2016MNRAS.460.1270D} {460, 1270}

\bibitem[\protect\citeauthoryear{{Davis}, {Efstathiou}, {Frenk}  \& {White}}{{Davis} et~al.}{1985}]{Davis_1985}
{Davis} M.,  {Efstathiou} G.,  {Frenk} C.~S.,   {White} S.~D.~M.,  1985, \mn@doi [\apj] {10.1086/163168}, \href {https://ui.adsabs.harvard.edu/abs/1985ApJ...292..371D} {292, 371}

\bibitem[\protect\citeauthoryear{{Deg} et~al.,}{{Deg} et~al.}{2023}]{Deg_et_al_2023}
{Deg} N.,  et~al., 2023, \mn@doi [\mnras] {10.1093/mnras/stad2312}, \href {https://ui.adsabs.harvard.edu/abs/2023MNRAS.525.4663D} {525, 4663}

\bibitem[\protect\citeauthoryear{{Desmons}, {Brough}, {Mart{\'\i}nez-Lombilla}, {De Propris}, {Holwerda}  \& {L{\'o}pez-S{\'a}nchez}}{{Desmons} et~al.}{2023}]{Desmons_et_al_2023}
{Desmons} A.,  {Brough} S.,  {Mart{\'\i}nez-Lombilla} C.,  {De Propris} R.,  {Holwerda} B.,   {L{\'o}pez-S{\'a}nchez} {\'A}.~R.,  2023, \mn@doi [\mnras] {10.1093/mnras/stad1639}, \href {https://ui.adsabs.harvard.edu/abs/2023MNRAS.523.4381D} {523, 4381}

\bibitem[\protect\citeauthoryear{{Desmons}, {Brough}  \& {Lanusse}}{{Desmons} et~al.}{2024}]{Desmons_et_al_2024}
{Desmons} A.,  {Brough} S.,   {Lanusse} F.,  2024, \mn@doi [\mnras] {10.1093/mnras/stae1402}, \href {https://ui.adsabs.harvard.edu/abs/2024MNRAS.531.4070D} {531, 4070}

\bibitem[\protect\citeauthoryear{{Dey} et~al.,}{{Dey} et~al.}{2019}]{Dey_et_al_2019}
{Dey} A.,  et~al., 2019, \mn@doi [\aj] {10.3847/1538-3881/ab089d}, \href {https://ui.adsabs.harvard.edu/abs/2019AJ....157..168D} {157, 168}

\bibitem[\protect\citeauthoryear{{Di Matteo}, {Jog}, {Lehnert}, {Combes}  \& {Semelin}}{{Di Matteo} et~al.}{2009}]{DiMatteo_2009}
{Di Matteo} P.,  {Jog} C.~J.,  {Lehnert} M.~D.,  {Combes} F.,   {Semelin} B.,  2009, \mn@doi [\aap] {10.1051/0004-6361/200912354}, \href {https://ui.adsabs.harvard.edu/abs/2009A&A...501L...9D} {501, L9}

\bibitem[\protect\citeauthoryear{{Dodd}, {Helmi}  \& {Koppelman}}{{Dodd} et~al.}{2022}]{Dodd_2022}
{Dodd} E.,  {Helmi} A.,   {Koppelman} H.~H.,  2022, \mn@doi [\aap] {10.1051/0004-6361/202141354}, \href {https://ui.adsabs.harvard.edu/abs/2022A&A...659A..61D} {659, A61}

\bibitem[\protect\citeauthoryear{{Dom{\'\i}nguez S{\'a}nchez} et~al.,}{{Dom{\'\i}nguez S{\'a}nchez} et~al.}{2023}]{Dominguez-Sanchez_2023}
{Dom{\'\i}nguez S{\'a}nchez} H.,  et~al., 2023, \mn@doi [\mnras] {10.1093/mnras/stad750}, \href {https://ui.adsabs.harvard.edu/abs/2023MNRAS.521.3861D} {521, 3861}

\bibitem[\protect\citeauthoryear{{Dropulic}, {Shipp}, {Kim}, {Mezghanni}, {Necib}  \& {Lisanti}}{{Dropulic} et~al.}{2024}]{Dropulic_et_al_2024}
{Dropulic} A.,  {Shipp} N.,  {Kim} S.,  {Mezghanni} Z.,  {Necib} L.,   {Lisanti} M.,  2024, \mn@doi [arXiv e-prints] {10.48550/arXiv.2409.13810}, \href {https://ui.adsabs.harvard.edu/abs/2024arXiv240913810D} {p. arXiv:2409.13810}

\bibitem[\protect\citeauthoryear{{Duc} \& {Renaud}}{{Duc} \& {Renaud}}{2013}]{Duc_and_Renaud_2013}
{Duc} P.-A.,  {Renaud} F.,  2013, in {Souchay} J.,  {Mathis} S.,   {Tokieda} T.,  eds, , Vol.~861, Lecture Notes in Physics, Berlin Springer Verlag.
p.~327, \mn@doi{10.1007/978-3-642-32961-6_9}

\bibitem[\protect\citeauthoryear{{Duc} et~al.,}{{Duc} et~al.}{2015}]{Duc_et_al_2015}
{Duc} P.-A.,  et~al., 2015, \mn@doi [\mnras] {10.1093/mnras/stu2019}, \href {https://ui.adsabs.harvard.edu/abs/2015MNRAS.446..120D} {446, 120}

\bibitem[\protect\citeauthoryear{{Ebrova}}{{Ebrova}}{2013}]{Ebrova_2013}
{Ebrova} I.,  2013, arXiv e-prints, \href {https://ui.adsabs.harvard.edu/abs/2013arXiv1312.1643E} {p. arXiv:1312.1643}

\bibitem[\protect\citeauthoryear{{Erkal}, {Sanders}  \& {Belokurov}}{{Erkal} et~al.}{2016}]{Erkal_et_al_2016}
{Erkal} D.,  {Sanders} J.~L.,   {Belokurov} V.,  2016, \mn@doi [\mnras] {10.1093/mnras/stw1400}, \href {https://ui.adsabs.harvard.edu/abs/2016MNRAS.461.1590E} {461, 1590}

\bibitem[\protect\citeauthoryear{{Erkal} et~al.,}{{Erkal} et~al.}{2019}]{Erkal_et_al_2019}
{Erkal} D.,  et~al., 2019, \mn@doi [\mnras] {10.1093/mnras/stz1371}, \href {https://ui.adsabs.harvard.edu/abs/2019MNRAS.487.2685E} {487, 2685}

\bibitem[\protect\citeauthoryear{{Euclid Collaboration} et~al.,}{{Euclid Collaboration} et~al.}{2022}]{Borlaff_2022}
{Euclid Collaboration} et~al., 2022, \mn@doi [\aap] {10.1051/0004-6361/202141935}, \href {https://ui.adsabs.harvard.edu/abs/2022A&A...657A..92E} {657, A92}

\bibitem[\protect\citeauthoryear{{Eyre} \& {Binney}}{{Eyre} \& {Binney}}{2009}]{Eyre_and_Binney_2009}
{Eyre} A.,  {Binney} J.,  2009, \mn@doi [\mnras] {10.1111/j.1365-2966.2009.15494.x}, \href {https://ui.adsabs.harvard.edu/abs/2009MNRAS.400..548E} {400, 548}

\bibitem[\protect\citeauthoryear{{Fardal}, {Babul}, {Geehan}  \& {Guhathakurta}}{{Fardal} et~al.}{2006}]{Fardal_et_al_2006}
{Fardal} M.~A.,  {Babul} A.,  {Geehan} J.~J.,   {Guhathakurta} P.,  2006, \mn@doi [\mnras] {10.1111/j.1365-2966.2005.09864.x}, \href {https://ui.adsabs.harvard.edu/abs/2006MNRAS.366.1012F} {366, 1012}

\bibitem[\protect\citeauthoryear{{Fardal} et~al.,}{{Fardal} et~al.}{2013}]{Fardal_et_al_2013}
{Fardal} M.~A.,  et~al., 2013, \mn@doi [\mnras] {10.1093/mnras/stt1121}, \href {https://ui.adsabs.harvard.edu/abs/2013MNRAS.434.2779F} {434, 2779}

\bibitem[\protect\citeauthoryear{{Fellhauer} et~al.,}{{Fellhauer} et~al.}{2006}]{Fellhauser_2006}
{Fellhauer} M.,  et~al., 2006, \mn@doi [\apj] {10.1086/507128}, \href {https://ui.adsabs.harvard.edu/abs/2006ApJ...651..167F} {651, 167}

\bibitem[\protect\citeauthoryear{{Fernique} et~al.,}{{Fernique} et~al.}{2015}]{Fernique_et_al_2015}
{Fernique} P.,  et~al., 2015, \mn@doi [\aap] {10.1051/0004-6361/201526075}, \href {https://ui.adsabs.harvard.edu/abs/2015A&A...578A.114F} {578, A114}

\bibitem[\protect\citeauthoryear{{Gaia Collaboration} et~al.,}{{Gaia Collaboration} et~al.}{2016}]{Gaia_2016}
{Gaia Collaboration} et~al., 2016, \mn@doi [\aap] {10.1051/0004-6361/201629272}, \href {https://ui.adsabs.harvard.edu/abs/2016A&A...595A...1G} {595, A1}

\bibitem[\protect\citeauthoryear{{Girelli}, {Pozzetti}, {Bolzonella}, {Giocoli}, {Marulli}  \& {Baldi}}{{Girelli} et~al.}{2020}]{Girelli_et_al_2020}
{Girelli} G.,  {Pozzetti} L.,  {Bolzonella} M.,  {Giocoli} C.,  {Marulli} F.,   {Baldi} M.,  2020, \mn@doi [\aap] {10.1051/0004-6361/201936329}, \href {https://ui.adsabs.harvard.edu/abs/2020A&A...634A.135G} {634, A135}

\bibitem[\protect\citeauthoryear{{Giri}, {Barway}  \& {Raychaudhury}}{{Giri} et~al.}{2023}]{Giri_et_al_2023}
{Giri} G.,  {Barway} S.,   {Raychaudhury} S.,  2023, \mn@doi [\mnras] {10.1093/mnras/stad474}, \href {https://ui.adsabs.harvard.edu/abs/2023MNRAS.520.5870G} {520, 5870}

\bibitem[\protect\citeauthoryear{{Gordon}, {Ferguson}  \& {Mann}}{{Gordon} et~al.}{2024}]{Gordon_et_al_2024}
{Gordon} A.~J.,  {Ferguson} A. M.~N.,   {Mann} R.~G.,  2024, \mn@doi [\mnras] {10.1093/mnras/stae2169}, \href {https://ui.adsabs.harvard.edu/abs/2024MNRAS.534.1459G} {534, 1459}

\bibitem[\protect\citeauthoryear{{Helmi}}{{Helmi}}{2004a}]{Helmi_2004b}
{Helmi} A.,  2004a, \mn@doi [\mnras] {10.1111/j.1365-2966.2004.07812.x}, \href {https://ui.adsabs.harvard.edu/abs/2004MNRAS.351..643H} {351, 643}

\bibitem[\protect\citeauthoryear{{Helmi}}{{Helmi}}{2004b}]{Helmi_2004}
{Helmi} A.,  2004b, \mn@doi [\apjl] {10.1086/423340}, \href {https://ui.adsabs.harvard.edu/abs/2004ApJ...610L..97H} {610, L97}

\bibitem[\protect\citeauthoryear{{Hendel} \& {Johnston}}{{Hendel} \& {Johnston}}{2015}]{Hendel_and_Johnston_2015}
{Hendel} D.,  {Johnston} K.~V.,  2015, \mn@doi [\mnras] {10.1093/mnras/stv2035}, \href {https://ui.adsabs.harvard.edu/abs/2015MNRAS.454.2472H} {454, 2472}

\bibitem[\protect\citeauthoryear{{Hendel}, {Johnston}, {Patra}  \& {Sen}}{{Hendel} et~al.}{2019}]{Hendel_et_al_2019}
{Hendel} D.,  {Johnston} K.~V.,  {Patra} R.~K.,   {Sen} B.,  2019, \mn@doi [\mnras] {10.1093/mnras/stz1107}, \href {https://ui.adsabs.harvard.edu/abs/2019MNRAS.486.3604H} {486, 3604}

\bibitem[\protect\citeauthoryear{Hood, Kannappan, Stark, Dell’Antonio, Moffett, Eckert, Norris  \& Hendel}{Hood et~al.}{2018}]{Hood_et_al_2018}
Hood C.~E.,  Kannappan S.~J.,  Stark D.~V.,  Dell’Antonio I.~P.,  Moffett A.~J.,  Eckert K.~D.,  Norris M.~A.,   Hendel D.,  2018, \mn@doi [The Astrophysical Journal] {10.3847/1538-4357/aab719}, 857, 144

\bibitem[\protect\citeauthoryear{{Huang} \& {Fan}}{{Huang} \& {Fan}}{2022}]{Huang_and_Fan_2022}
{Huang} Q.,  {Fan} L.,  2022, \mn@doi [\apjs] {10.3847/1538-4365/ac85b1}, \href {https://ui.adsabs.harvard.edu/abs/2022ApJS..262...39H} {262, 39}

\bibitem[\protect\citeauthoryear{{Ibata}, {Lewis}, {Irwin}, {Totten}  \& {Quinn}}{{Ibata} et~al.}{2001}]{Ibata_2001}
{Ibata} R.,  {Lewis} G.~F.,  {Irwin} M.,  {Totten} E.,   {Quinn} T.,  2001, \mn@doi [\apj] {10.1086/320060}, \href {https://ui.adsabs.harvard.edu/abs/2001ApJ...551..294I} {551, 294}

\bibitem[\protect\citeauthoryear{{Ibata}, {Irwin}, {Lewis}, {Ferguson}  \& {Tanvir}}{{Ibata} et~al.}{2003}]{Ibata_2003}
{Ibata} R.~A.,  {Irwin} M.~J.,  {Lewis} G.~F.,  {Ferguson} A.~M.~N.,   {Tanvir} N.,  2003, \mn@doi [\mnras] {10.1046/j.1365-8711.2003.06545.x}, \href {https://ui.adsabs.harvard.edu/abs/2003MNRAS.340L..21I} {340, L21}

\bibitem[\protect\citeauthoryear{{Jackson}, {Pasquali}, {La Barbera}, {More}  \& {Grebel}}{{Jackson} et~al.}{2023}]{Jackson_et_al_2023}
{Jackson} T.~M.,  {Pasquali} A.,  {La Barbera} F.,  {More} S.,   {Grebel} E.~K.,  2023, \mn@doi [\mnras] {10.1093/mnras/stad131}, \href {https://ui.adsabs.harvard.edu/abs/2023MNRAS.520.1155J} {520, 1155}

\bibitem[\protect\citeauthoryear{{Johnston}}{{Johnston}}{1998}]{Johnston_et_al_1998}
{Johnston} K.~V.,  1998, \mn@doi [\apj] {10.1086/305273}, \href {https://ui.adsabs.harvard.edu/abs/1998ApJ...495..297J} {495, 297}

\bibitem[\protect\citeauthoryear{{Johnston}, {Hernquist}  \& {Bolte}}{{Johnston} et~al.}{1996}]{Johnston_et_al_1996}
{Johnston} K.~V.,  {Hernquist} L.,   {Bolte} M.,  1996, \mn@doi [\apj] {10.1086/177418}, \href {https://ui.adsabs.harvard.edu/abs/1996ApJ...465..278J} {465, 278}

\bibitem[\protect\citeauthoryear{{Johnston}, {Zhao}, {Spergel}  \& {Hernquist}}{{Johnston} et~al.}{1999}]{Johnston_1999}
{Johnston} K.~V.,  {Zhao} H.,  {Spergel} D.~N.,   {Hernquist} L.,  1999, \mn@doi [\apjl] {10.1086/311876}, \href {https://ui.adsabs.harvard.edu/abs/1999ApJ...512L.109J} {512, L109}

\bibitem[\protect\citeauthoryear{{Johnston}, {Sackett}  \& {Bullock}}{{Johnston} et~al.}{2001}]{Johnston_et_al_2001}
{Johnston} K.~V.,  {Sackett} P.~D.,   {Bullock} J.~S.,  2001, \mn@doi [\apj] {10.1086/321644}, \href {https://ui.adsabs.harvard.edu/abs/2001ApJ...557..137J} {557, 137}

\bibitem[\protect\citeauthoryear{{Johnston}, {Law}  \& {Majewski}}{{Johnston} et~al.}{2005}]{Johnston_et_al_2005}
{Johnston} K.~V.,  {Law} D.~R.,   {Majewski} S.~R.,  2005, \mn@doi [\apj] {10.1086/426777}, \href {https://ui.adsabs.harvard.edu/abs/2005ApJ...619..800J} {619, 800}

\bibitem[\protect\citeauthoryear{{Johnston}, {Bullock}, {Sharma}, {Font}, {Robertson}  \& {Leitner}}{{Johnston} et~al.}{2008}]{Johnston_2008}
{Johnston} K.~V.,  {Bullock} J.~S.,  {Sharma} S.,  {Font} A.,  {Robertson} B.~E.,   {Leitner} S.~N.,  2008, \mn@doi [\apj] {10.1086/592228}, \href {https://ui.adsabs.harvard.edu/abs/2008ApJ...689..936J} {689, 936}

\bibitem[\protect\citeauthoryear{{Kado-Fong} et~al.,}{{Kado-Fong} et~al.}{2018}]{Kado-Fong_et_al_2018}
{Kado-Fong} E.,  et~al., 2018, \mn@doi [\apj] {10.3847/1538-4357/aae0f0}, \href {https://ui.adsabs.harvard.edu/abs/2018ApJ...866..103K} {866, 103}

\bibitem[\protect\citeauthoryear{{Karachentsev}, {Kaisina}  \& {Kashibadze Nasonova}}{{Karachentsev} et~al.}{2017}]{Karachentsev_et_al_2017}
{Karachentsev} I.~D.,  {Kaisina} E.~I.,   {Kashibadze Nasonova} O.~G.,  2017, \mn@doi [\aj] {10.3847/1538-3881/153/1/6}, \href {https://ui.adsabs.harvard.edu/abs/2017AJ....153....6K} {153, 6}

\bibitem[\protect\citeauthoryear{{Karademir}, {Remus}, {Burkert}, {Dolag}, {Hoffmann}, {Moster}, {Steinwandel}  \& {Zhang}}{{Karademir} et~al.}{2019}]{Karademir_2019}
{Karademir} G.~S.,  {Remus} R.-S.,  {Burkert} A.,  {Dolag} K.,  {Hoffmann} T.~L.,  {Moster} B.~P.,  {Steinwandel} U.~P.,   {Zhang} J.,  2019, \mn@doi [\mnras] {10.1093/mnras/stz1251}, \href {https://ui.adsabs.harvard.edu/abs/2019MNRAS.487..318K} {487, 318}

\bibitem[\protect\citeauthoryear{{Kaviraj}}{{Kaviraj}}{2010}]{Kaviraj_2010}
{Kaviraj} S.,  2010, \mn@doi [\mnras] {10.1111/j.1365-2966.2010.16714.x}, \href {https://ui.adsabs.harvard.edu/abs/2010MNRAS.406..382K} {406, 382}

\bibitem[\protect\citeauthoryear{{Kelly}, {Jenkins}, {Deason}, {Fattahi}, {Grand}, {Pakmor}, {Springel}  \& {Frenk}}{{Kelly} et~al.}{2022}]{Kelly+2022}
{Kelly} A.~J.,  {Jenkins} A.,  {Deason} A.,  {Fattahi} A.,  {Grand} R. J.~J.,  {Pakmor} R.,  {Springel} V.,   {Frenk} C.~S.,  2022, \mn@doi [\mnras] {10.1093/mnras/stac1019}, \href {https://ui.adsabs.harvard.edu/abs/2022MNRAS.514.3113K} {514, 3113}

\bibitem[\protect\citeauthoryear{{Khalid}, {Brough}, {Martin}, {Kimmig}, {Lagos}, {Remus}  \& {Martinez-Lombilla}}{{Khalid} et~al.}{2024}]{Khalid_et_al_2024}
{Khalid} A.,  {Brough} S.,  {Martin} G.,  {Kimmig} L.~C.,  {Lagos} C.~D.~P.,  {Remus} R.~S.,   {Martinez-Lombilla} C.,  2024, \mn@doi [\mnras] {10.1093/mnras/stae1064}, \href {https://ui.adsabs.harvard.edu/abs/2024MNRAS.530.4422K} {530, 4422}

\bibitem[\protect\citeauthoryear{{Kim} et~al.,}{{Kim} et~al.}{2012}]{Kim_et_al_2012}
{Kim} T.,  et~al., 2012, \mn@doi [\apj] {10.1088/0004-637X/753/1/43}, \href {https://ui.adsabs.harvard.edu/abs/2012ApJ...753...43K} {753, 43}

\bibitem[\protect\citeauthoryear{{Knapen}, {Cisternas}  \& {Querejeta}}{{Knapen} et~al.}{2015}]{Knapen_2015}
{Knapen} J.~H.,  {Cisternas} M.,   {Querejeta} M.,  2015, \mn@doi [\mnras] {10.1093/mnras/stv2135}, \href {https://ui.adsabs.harvard.edu/abs/2015MNRAS.454.1742K} {454, 1742}

\bibitem[\protect\citeauthoryear{{Koposov}, {Rix}  \& {Hogg}}{{Koposov} et~al.}{2010}]{Koposov_2010}
{Koposov} S.~E.,  {Rix} H.-W.,   {Hogg} D.~W.,  2010, \mn@doi [\apj] {10.1088/0004-637X/712/1/260}, \href {https://ui.adsabs.harvard.edu/abs/2010ApJ...712..260K} {712, 260}

\bibitem[\protect\citeauthoryear{{Koposov} et~al.,}{{Koposov} et~al.}{2023}]{Koposov_2023}
{Koposov} S.~E.,  et~al., 2023, \mn@doi [\mnras] {10.1093/mnras/stad551}, \href {https://ui.adsabs.harvard.edu/abs/2023MNRAS.521.4936K} {521, 4936}

\bibitem[\protect\citeauthoryear{{Lang}, {Hogg}  \& {Mykytyn}}{{Lang} et~al.}{2016}]{Lang_et_al_2016}
{Lang} D.,  {Hogg} D.~W.,   {Mykytyn} D.,  2016, {The Tractor: Probabilistic astronomical source detection and measurement}, Astrophysics Source Code Library, record ascl:1604.008

\bibitem[\protect\citeauthoryear{{Law} \& {Majewski}}{{Law} \& {Majewski}}{2010}]{Law_and_Majewski_2010}
{Law} D.~R.,  {Majewski} S.~R.,  2010, \mn@doi [\apj] {10.1088/0004-637X/714/1/229}, \href {https://ui.adsabs.harvard.edu/abs/2010ApJ...714..229L} {714, 229}

\bibitem[\protect\citeauthoryear{{Law}, {Johnston}  \& {Majewski}}{{Law} et~al.}{2005}]{Law_2005}
{Law} D.~R.,  {Johnston} K.~V.,   {Majewski} S.~R.,  2005, \mn@doi [\apj] {10.1086/426779}, \href {https://ui.adsabs.harvard.edu/abs/2005ApJ...619..807L} {619, 807}

\bibitem[\protect\citeauthoryear{{Liao} \& {Cooper}}{{Liao} \& {Cooper}}{2023}]{Liao_et_al_2023}
{Liao} L.-W.,  {Cooper} A.~P.,  2023, \mn@doi [\mnras] {10.1093/mnras/stac3327}, \href {https://ui.adsabs.harvard.edu/abs/2023MNRAS.518.3999L} {518, 3999}

\bibitem[\protect\citeauthoryear{{Lotz}, {Primack}  \& {Madau}}{{Lotz} et~al.}{2004}]{Lotz_2004}
{Lotz} J.~M.,  {Primack} J.,   {Madau} P.,  2004, \mn@doi [\aj] {10.1086/421849}, \href {https://ui.adsabs.harvard.edu/abs/2004AJ....128..163L} {128, 163}

\bibitem[\protect\citeauthoryear{{Lotz}, {Jonsson}, {Cox}  \& {Primack}}{{Lotz} et~al.}{2008}]{Lotz_et_al_2008}
{Lotz} J.~M.,  {Jonsson} P.,  {Cox} T.~J.,   {Primack} J.~R.,  2008, \mn@doi [\mnras] {10.1111/j.1365-2966.2008.14004.x}, \href {https://ui.adsabs.harvard.edu/abs/2008MNRAS.391.1137L} {391, 1137}

\bibitem[\protect\citeauthoryear{{Lotz}, {Jonsson}, {Cox}, {Croton}, {Primack}, {Somerville}  \& {Stewart}}{{Lotz} et~al.}{2011}]{Lotz+2011}
{Lotz} J.~M.,  {Jonsson} P.,  {Cox} T.~J.,  {Croton} D.,  {Primack} J.~R.,  {Somerville} R.~S.,   {Stewart} K.,  2011, \mn@doi [\apj] {10.1088/0004-637X/742/2/103}, \href {https://ui.adsabs.harvard.edu/abs/2011ApJ...742..103L} {742, 103}

\bibitem[\protect\citeauthoryear{{Lux}, {Read}, {Lake}  \& {Johnston}}{{Lux} et~al.}{2013}]{Lux_2013}
{Lux} H.,  {Read} J.~I.,  {Lake} G.,   {Johnston} K.~V.,  2013, \mn@doi [\mnras] {10.1093/mnras/stt1744}, \href {https://ui.adsabs.harvard.edu/abs/2013MNRAS.436.2386L} {436, 2386}

\bibitem[\protect\citeauthoryear{{Lynden-Bell} \& {Lynden-Bell}}{{Lynden-Bell} \& {Lynden-Bell}}{1995}]{Lynden-Bell_1995}
{Lynden-Bell} D.,  {Lynden-Bell} R.~M.,  1995, \mn@doi [\mnras] {10.1093/mnras/275.2.429}, \href {https://ui.adsabs.harvard.edu/abs/1995MNRAS.275..429L} {275, 429}

\bibitem[\protect\citeauthoryear{{Majewski} et~al.,}{{Majewski} et~al.}{2017}]{Majewski_2017}
{Majewski} S.~R.,  et~al., 2017, \mn@doi [\aj] {10.3847/1538-3881/aa784d}, \href {https://ui.adsabs.harvard.edu/abs/2017AJ....154...94M} {154, 94}

\bibitem[\protect\citeauthoryear{{Malhan} \& {Ibata}}{{Malhan} \& {Ibata}}{2019}]{Malhan_and_Ibata_2019}
{Malhan} K.,  {Ibata} R.~A.,  2019, \mn@doi [\mnras] {10.1093/mnras/stz1035}, \href {https://ui.adsabs.harvard.edu/abs/2019MNRAS.486.2995M} {486, 2995}

\bibitem[\protect\citeauthoryear{{Malin} \& {Carter}}{{Malin} \& {Carter}}{1983}]{Malin_and_Carter_1983}
{Malin} D.~F.,  {Carter} D.,  1983, \mn@doi [\apj] {10.1086/161467}, \href {https://ui.adsabs.harvard.edu/abs/1983ApJ...274..534M} {274, 534}

\bibitem[\protect\citeauthoryear{{Mancillas}, {Duc}, {Combes}, {Bournaud}, {Emsellem}, {Martig}  \& {Michel-Dansac}}{{Mancillas} et~al.}{2019}]{Mancillas_2019}
{Mancillas} B.,  {Duc} P.-A.,  {Combes} F.,  {Bournaud} F.,  {Emsellem} E.,  {Martig} M.,   {Michel-Dansac} L.,  2019, \mn@doi [\aap] {10.1051/0004-6361/201936320}, \href {https://ui.adsabs.harvard.edu/abs/2019A&A...632A.122M} {632, A122}

\bibitem[\protect\citeauthoryear{{Mantha} et~al.,}{{Mantha} et~al.}{2019}]{Mantha_et_al_2019}
{Mantha} K.~B.,  et~al., 2019, \mn@doi [\mnras] {10.1093/mnras/stz872}, \href {https://ui.adsabs.harvard.edu/abs/2019MNRAS.486.2643M} {486, 2643}

\bibitem[\protect\citeauthoryear{{Martin} et~al.,}{{Martin} et~al.}{2022}]{Martin_et_al_2022}
{Martin} G.,  et~al., 2022, \mn@doi [\mnras] {10.1093/mnras/stac1003}, \href {https://ui.adsabs.harvard.edu/abs/2022MNRAS.513.1459M} {513, 1459}

\bibitem[\protect\citeauthoryear{{Martin}, {Pearce}, {Hatch}, {Contreras-Santos}, {Knebe}  \& {Cui}}{{Martin} et~al.}{2024}]{Martin_et_al_2024}
{Martin} G.,  {Pearce} F.~R.,  {Hatch} N.~A.,  {Contreras-Santos} A.,  {Knebe} A.,   {Cui} W.,  2024, \mn@doi [\mnras] {10.1093/mnras/stae2488}, \href {https://ui.adsabs.harvard.edu/abs/2024MNRAS.535.2375M} {535, 2375}

\bibitem[\protect\citeauthoryear{{Mart{\'\i}nez-Delgado}, {Pe{\~n}arrubia}, {Gabany}, {Trujillo}, {Majewski}  \& {Pohlen}}{{Mart{\'\i}nez-Delgado} et~al.}{2008}]{Martinez_Delgado_et_al_2008}
{Mart{\'\i}nez-Delgado} D.,  {Pe{\~n}arrubia} J.,  {Gabany} R.~J.,  {Trujillo} I.,  {Majewski} S.~R.,   {Pohlen} M.,  2008, \mn@doi [\apj] {10.1086/592555}, \href {https://ui.adsabs.harvard.edu/abs/2008ApJ...689..184M} {689, 184}

\bibitem[\protect\citeauthoryear{{Mart{\'\i}nez-Delgado}, {Pohlen}, {Gabany}, {Majewski}, {Pe{\~n}arrubia}  \& {Palma}}{{Mart{\'\i}nez-Delgado} et~al.}{2009}]{Martinez-Delgado_et_al_2009}
{Mart{\'\i}nez-Delgado} D.,  {Pohlen} M.,  {Gabany} R.~J.,  {Majewski} S.~R.,  {Pe{\~n}arrubia} J.,   {Palma} C.,  2009, \mn@doi [\apj] {10.1088/0004-637X/692/2/955}, \href {https://ui.adsabs.harvard.edu/abs/2009ApJ...692..955M} {692, 955}

\bibitem[\protect\citeauthoryear{{Mart{\'\i}nez-Delgado} et~al.,}{{Mart{\'\i}nez-Delgado} et~al.}{2010}]{Martinez-Delgado_2010}
{Mart{\'\i}nez-Delgado} D.,  et~al., 2010, \mn@doi [\aj] {10.1088/0004-6256/140/4/962}, \href {https://ui.adsabs.harvard.edu/abs/2010AJ....140..962M} {140, 962}

\bibitem[\protect\citeauthoryear{{Mart{\'\i}nez-Delgado} et~al.,}{{Mart{\'\i}nez-Delgado} et~al.}{2023}]{Martinez-Delgado_et_al_2023}
{Mart{\'\i}nez-Delgado} D.,  et~al., 2023, \mn@doi [\aap] {10.1051/0004-6361/202245011}, \href {https://ui.adsabs.harvard.edu/abs/2023A&A...671A.141M} {671, A141}

\bibitem[\protect\citeauthoryear{{McGaugh}, {Schombert}, {Bothun}  \& {de Blok}}{{McGaugh} et~al.}{2000}]{McGaugh_et_al_2000}
{McGaugh} S.~S.,  {Schombert} J.~M.,  {Bothun} G.~D.,   {de Blok} W.~J.~G.,  2000, \mn@doi [\apjl] {10.1086/312628}, \href {https://ui.adsabs.harvard.edu/abs/2000ApJ...533L..99M} {533, L99}

\bibitem[\protect\citeauthoryear{{McIntosh}, {Guo}, {Hertzberg}, {Katz}, {Mo}, {van den Bosch}  \& {Yang}}{{McIntosh} et~al.}{2008}]{McIntosh_et_al_2008}
{McIntosh} D.~H.,  {Guo} Y.,  {Hertzberg} J.,  {Katz} N.,  {Mo} H.~J.,  {van den Bosch} F.~C.,   {Yang} X.,  2008, \mn@doi [\mnras] {10.1111/j.1365-2966.2008.13531.x}, \href {https://ui.adsabs.harvard.edu/abs/2008MNRAS.388.1537M} {388, 1537}

\bibitem[\protect\citeauthoryear{{Mir{\'o}-Carretero} et~al.,}{{Mir{\'o}-Carretero} et~al.}{2023}]{Miro-Carretero_et_al_2023}
{Mir{\'o}-Carretero} J.,  et~al., 2023, \mn@doi [\aap] {10.1051/0004-6361/202245003}, \href {https://ui.adsabs.harvard.edu/abs/2023A&A...669L..13M} {669, L13}

\bibitem[\protect\citeauthoryear{{Miro-Carretero} et~al.,}{{Miro-Carretero} et~al.}{2024a}]{Miro-Carretero_et_al_2024_simu}
{Miro-Carretero} J.,  et~al., 2024a, \mn@doi [arXiv e-prints] {10.48550/arXiv.2409.03585}, \href {https://ui.adsabs.harvard.edu/abs/2024arXiv240903585M} {p. arXiv:2409.03585}

\bibitem[\protect\citeauthoryear{{Mir{\'o}-Carretero} et~al.,}{{Mir{\'o}-Carretero} et~al.}{2024b}]{Miro-Carretero_et_al_2024}
{Mir{\'o}-Carretero} J.,  et~al., 2024b, \mn@doi [\aap] {10.1051/0004-6361/202451685}, \href {https://ui.adsabs.harvard.edu/abs/2024A&A...691A.196M} {691, A196}

\bibitem[\protect\citeauthoryear{{Miskolczi}, {Bomans}  \& {Dettmar}}{{Miskolczi} et~al.}{2011}]{Miskolczi_et_al_2011}
{Miskolczi} A.,  {Bomans} D.~J.,   {Dettmar} R.~J.,  2011, \mn@doi [A\&A] {10.1051/0004-6361/201116716}, \href {https://ui.adsabs.harvard.edu/abs/2011A&A...536A..66M} {536, A66}

\bibitem[\protect\citeauthoryear{{Moore}, {Ghigna}, {Governato}, {Lake}, {Quinn}, {Stadel}  \& {Tozzi}}{{Moore} et~al.}{1999}]{Moore_et_al_1999}
{Moore} B.,  {Ghigna} S.,  {Governato} F.,  {Lake} G.,  {Quinn} T.,  {Stadel} J.,   {Tozzi} P.,  1999, \mn@doi [\apjl] {10.1086/312287}, \href {https://ui.adsabs.harvard.edu/abs/1999ApJ...524L..19M} {524, L19}

\bibitem[\protect\citeauthoryear{{Morales}, {Mart{\'\i}nez-Delgado}, {Grebel}, {Cooper}, {Javanmardi}  \& {Miskolczi}}{{Morales} et~al.}{2018}]{Morales_et_al_2018}
{Morales} G.,  {Mart{\'\i}nez-Delgado} D.,  {Grebel} E.~K.,  {Cooper} A.~P.,  {Javanmardi} B.,   {Miskolczi} A.,  2018, \mn@doi [\aap] {10.1051/0004-6361/201732271}, \href {https://ui.adsabs.harvard.edu/abs/2018A&A...614A.143M} {614, A143}

\bibitem[\protect\citeauthoryear{{Mosenkov}, {Bahr}, {Reshetnikov}, {Shakespear}  \& {Smirnov}}{{Mosenkov} et~al.}{2024}]{Mosenkov_et_al_2024}
{Mosenkov} A.~V.,  {Bahr} S.~K.~H.,  {Reshetnikov} V.~P.,  {Shakespear} Z.,   {Smirnov} D.~V.,  2024, \mn@doi [\aap] {10.1051/0004-6361/202348494}, \href {https://ui.adsabs.harvard.edu/abs/2024A&A...681L..15M} {681, L15}

\bibitem[\protect\citeauthoryear{{Moustakas} et~al.,}{{Moustakas} et~al.}{2023}]{Moustakas_et_al_2023}
{Moustakas} J.,  et~al., 2023, \mn@doi [\apjs] {10.3847/1538-4365/acfaa2}, \href {https://ui.adsabs.harvard.edu/abs/2023ApJS..269....3M} {269, 3}

\bibitem[\protect\citeauthoryear{{M{\"u}ller} et~al.,}{{M{\"u}ller} et~al.}{2019}]{Muller_et_al_2019}
{M{\"u}ller} O.,  et~al., 2019, \mn@doi [\aap] {10.1051/0004-6361/201935463}, \href {https://ui.adsabs.harvard.edu/abs/2019A&A...624L...6M} {624, L6}

\bibitem[\protect\citeauthoryear{{Naab} \& {Ostriker}}{{Naab} \& {Ostriker}}{2017}]{Naab_and_Ostriker_2017}
{Naab} T.,  {Ostriker} J.~P.,  2017, \mn@doi [\araa] {10.1146/annurev-astro-081913-040019}, \href {https://ui.adsabs.harvard.edu/abs/2017ARA&A..55...59N} {55, 59}

\bibitem[\protect\citeauthoryear{{Newberg} \& {Carlin}}{{Newberg} \& {Carlin}}{2016}]{Newberg_and_Carlin_2016}
{Newberg} H.~J.,  {Carlin} J.~L.,  eds, 2016, {Tidal Streams in the Local Group and Beyond}  Astrophysics and Space Science Library Vol. 420, \mn@doi{10.1007/978-3-319-19336-6.
}

\bibitem[\protect\citeauthoryear{{Newberg}, {Willett}, {Yanny}  \& {Xu}}{{Newberg} et~al.}{2010}]{Newberg_2010}
{Newberg} H.~J.,  {Willett} B.~A.,  {Yanny} B.,   {Xu} Y.,  2010, \mn@doi [\apj] {10.1088/0004-637X/711/1/32}, \href {https://ui.adsabs.harvard.edu/abs/2010ApJ...711...32N} {711, 32}

\bibitem[\protect\citeauthoryear{{Nibauer}, {Bonaca}  \& {Johnston}}{{Nibauer} et~al.}{2023}]{Nibauer_et_al_2023}
{Nibauer} J.,  {Bonaca} A.,   {Johnston} K.~V.,  2023, \mn@doi [\apj] {10.3847/1538-4357/ace9bc}, \href {https://ui.adsabs.harvard.edu/abs/2023ApJ...954..195N} {954, 195}

\bibitem[\protect\citeauthoryear{{Ownsworth}, {Conselice}, {Mortlock}, {Hartley}, {Almaini}, {Duncan}  \& {Mundy}}{{Ownsworth} et~al.}{2014}]{Ownsworth+2014}
{Ownsworth} J.~R.,  {Conselice} C.~J.,  {Mortlock} A.,  {Hartley} W.~G.,  {Almaini} O.,  {Duncan} K.,   {Mundy} C.~J.,  2014, \mn@doi [\mnras] {10.1093/mnras/stu1802}, \href {https://ui.adsabs.harvard.edu/abs/2014MNRAS.445.2198O} {445, 2198}

\bibitem[\protect\citeauthoryear{{Pawlik}, {Wild}, {Walcher}, {Johansson}, {Villforth}, {Rowlands}, {Mendez-Abreu}  \& {Hewlett}}{{Pawlik} et~al.}{2016}]{Pawlik_et_al_2016}
{Pawlik} M.~M.,  {Wild} V.,  {Walcher} C.~J.,  {Johansson} P.~H.,  {Villforth} C.,  {Rowlands} K.,  {Mendez-Abreu} J.,   {Hewlett} T.,  2016, \mn@doi [\mnras] {10.1093/mnras/stv2878}, \href {https://ui.adsabs.harvard.edu/abs/2016MNRAS.456.3032P} {456, 3032}

\bibitem[\protect\citeauthoryear{{Pearson}, {Wang}, {Trayford}, {Petrillo}  \& {van der Tak}}{{Pearson} et~al.}{2019a}]{Pearson_et_al_2019}
{Pearson} W.~J.,  {Wang} L.,  {Trayford} J.~W.,  {Petrillo} C.~E.,   {van der Tak} F.~F.~S.,  2019a, \mn@doi [\aap] {10.1051/0004-6361/201935355}, \href {https://ui.adsabs.harvard.edu/abs/2019A&A...626A..49P} {626, A49}

\bibitem[\protect\citeauthoryear{{Pearson} et~al.,}{{Pearson} et~al.}{2019b}]{Pearson_2019}
{Pearson} W.~J.,  et~al., 2019b, \mn@doi [\aap] {10.1051/0004-6361/201936337}, \href {https://ui.adsabs.harvard.edu/abs/2019A&A...631A..51P} {631, A51}

\bibitem[\protect\citeauthoryear{{Pearson}, {Price-Whelan}, {Hogg}, {Seth}, {Sand}, {Hunt}  \& {Crnojevi{\'c}}}{{Pearson} et~al.}{2022}]{Pearson_2022}
{Pearson} S.,  {Price-Whelan} A.~M.,  {Hogg} D.~W.,  {Seth} A.~C.,  {Sand} D.~J.,  {Hunt} J. A.~S.,   {Crnojevi{\'c}} D.,  2022, \mn@doi [\apj] {10.3847/1538-4357/ac9bfb}, \href {https://ui.adsabs.harvard.edu/abs/2022ApJ...941...19P} {941, 19}

\bibitem[\protect\citeauthoryear{{Pillepich}, {Madau}  \& {Mayer}}{{Pillepich} et~al.}{2015}]{Pillepich_2015}
{Pillepich} A.,  {Madau} P.,   {Mayer} L.,  2015, \mn@doi [\apj] {10.1088/0004-637X/799/2/184}, \href {https://ui.adsabs.harvard.edu/abs/2015ApJ...799..184P} {799, 184}

\bibitem[\protect\citeauthoryear{{Pippert}, {Kluge}  \& {Bender}}{{Pippert} et~al.}{2025}]{Pippert_et_al_2025}
{Pippert} J.-N.,  {Kluge} M.,   {Bender} R.,  2025, \mn@doi [\apj] {10.3847/1538-4357/adabda}, \href {https://ui.adsabs.harvard.edu/abs/2025ApJ...980..244P} {980, 244}

\bibitem[\protect\citeauthoryear{{Pop}, {Pillepich}, {Amorisco}  \& {Hernquist}}{{Pop} et~al.}{2018}]{Pop_2018}
{Pop} A.-R.,  {Pillepich} A.,  {Amorisco} N.~C.,   {Hernquist} L.,  2018, \mn@doi [\mnras] {10.1093/mnras/sty1932}, \href {https://ui.adsabs.harvard.edu/abs/2018MNRAS.480.1715P} {480, 1715}

\bibitem[\protect\citeauthoryear{{Prieur}}{{Prieur}}{1990}]{Prieur_1990}
{Prieur} J.~L.,  1990, {Status of shell galaxies.}.
pp 72--83

\bibitem[\protect\citeauthoryear{{Quinn}}{{Quinn}}{1984}]{Quinn_1984}
{Quinn} P.~J.,  1984, \mn@doi [\apj] {10.1086/161924}, \href {https://ui.adsabs.harvard.edu/abs/1984ApJ...279..596Q} {279, 596}

\bibitem[\protect\citeauthoryear{{Ramos}, {Mateu}, {Antoja}, {Helmi}, {Castro-Ginard}, {Balbinot}  \& {Carrasco}}{{Ramos} et~al.}{2020}]{Ramos_et_al_2020}
{Ramos} P.,  {Mateu} C.,  {Antoja} T.,  {Helmi} A.,  {Castro-Ginard} A.,  {Balbinot} E.,   {Carrasco} J.~M.,  2020, \mn@doi [\aap] {10.1051/0004-6361/202037819}, \href {https://ui.adsabs.harvard.edu/abs/2020A&A...638A.104R} {638, A104}

\bibitem[\protect\citeauthoryear{{Richards}, {Sola}, {Paiement}, {Xie}  \& {Duc}}{{Richards} et~al.}{2022}]{Richards_et_al_2022}
{Richards} F.,  {Sola} E.,  {Paiement} A.,  {Xie} X.,   {Duc} P.-A.,  2022, IEEE International Conference on Image Processing (ICIP)

\bibitem[\protect\citeauthoryear{{Richards}, {Paiement}, {Xie}, {Sola}  \& {Duc}}{{Richards} et~al.}{2023}]{Richards_et_al_2023}
{Richards} F.,  {Paiement} A.,  {Xie} X.,  {Sola} E.,   {Duc} P.-A.,  2023, 18th International Conference on Machine Vision Applications, Jul 2023, Hamamatsu, Shizuoka, Japan, hal-04129549

\bibitem[\protect\citeauthoryear{{Rodriguez-Gomez} et~al.,}{{Rodriguez-Gomez} et~al.}{2017}]{Rodriguez-Gomez_2017}
{Rodriguez-Gomez} V.,  et~al., 2017, \mn@doi [\mnras] {10.1093/mnras/stx305}, \href {https://ui.adsabs.harvard.edu/abs/2017MNRAS.467.3083R} {467, 3083}

\bibitem[\protect\citeauthoryear{{Rom{\'a}n}, {Trujillo}  \& {Montes}}{{Rom{\'a}n} et~al.}{2020}]{Roman_et_al_2020}
{Rom{\'a}n} J.,  {Trujillo} I.,   {Montes} M.,  2020, \mn@doi [\aap] {10.1051/0004-6361/201936111}, \href {https://ui.adsabs.harvard.edu/abs/2020A&A...644A..42R} {644, A42}

\bibitem[\protect\citeauthoryear{{Rutherford} et~al.,}{{Rutherford} et~al.}{2024}]{Rutherford_et_al_2024}
{Rutherford} T.~H.,  et~al., 2024, \mn@doi [\mnras] {10.1093/mnras/stae398}, \href {https://ui.adsabs.harvard.edu/abs/2024MNRAS.tmp..477R} {}

\bibitem[\protect\citeauthoryear{{Sanders} \& {Binney}}{{Sanders} \& {Binney}}{2013}]{Sanders_and_Binney_2013}
{Sanders} J.~L.,  {Binney} J.,  2013, \mn@doi [\mnras] {10.1093/mnras/stt806}, \href {https://ui.adsabs.harvard.edu/abs/2013MNRAS.433.1813S} {433, 1813}

\bibitem[\protect\citeauthoryear{{Sawala}, {Pihajoki}, {Johansson}, {Frenk}, {Navarro}, {Oman}  \& {White}}{{Sawala} et~al.}{2017}]{Sawala+2017}
{Sawala} T.,  {Pihajoki} P.,  {Johansson} P.~H.,  {Frenk} C.~S.,  {Navarro} J.~F.,  {Oman} K.~A.,   {White} S. D.~M.,  2017, \mn@doi [\mnras] {10.1093/mnras/stx360}, \href {https://ui.adsabs.harvard.edu/abs/2017MNRAS.467.4383S} {467, 4383}

\bibitem[\protect\citeauthoryear{{Schweizer} \& {Seitzer}}{{Schweizer} \& {Seitzer}}{1992}]{Schweizer_and_Seitzer_1992}
{Schweizer} F.,  {Seitzer} P.,  1992, \mn@doi [\aj] {10.1086/116296}, \href {https://ui.adsabs.harvard.edu/abs/1992AJ....104.1039S} {104, 1039}

\bibitem[\protect\citeauthoryear{{Sesar}, {Hernitschek}, {Dierickx}, {Fardal}  \& {Rix}}{{Sesar} et~al.}{2017}]{Sesar_et_al_2017}
{Sesar} B.,  {Hernitschek} N.,  {Dierickx} M. I.~P.,  {Fardal} M.~A.,   {Rix} H.-W.,  2017, \mn@doi [\apjl] {10.3847/2041-8213/aa7c61}, \href {https://ui.adsabs.harvard.edu/abs/2017ApJ...844L...4S} {844, L4}

\bibitem[\protect\citeauthoryear{{Sheen}, {Yi}, {Ree}  \& {Lee}}{{Sheen} et~al.}{2012}]{Sheen_et_al_2012}
{Sheen} Y.-K.,  {Yi} S.~K.,  {Ree} C.~H.,   {Lee} J.,  2012, \mn@doi [\apjs] {10.1088/0067-0049/202/1/8}, \href {https://ui.adsabs.harvard.edu/abs/2012ApJS..202....8S} {202, 8}

\bibitem[\protect\citeauthoryear{{Shipp} et~al.,}{{Shipp} et~al.}{2023}]{Shipp_et_al_2022}
{Shipp} N.,  et~al., 2023, \mn@doi [\apj] {10.3847/1538-4357/acc582}, \href {https://ui.adsabs.harvard.edu/abs/2023ApJ...949...44S} {949, 44}

\bibitem[\protect\citeauthoryear{{Shipp} et~al.,}{{Shipp} et~al.}{2024}]{Shipp_et_al_2024}
{Shipp} N.,  et~al., 2024, \mn@doi [arXiv e-prints] {10.48550/arXiv.2410.09143}, \href {https://ui.adsabs.harvard.edu/abs/2024arXiv241009143S} {p. arXiv:2410.09143}

\bibitem[\protect\citeauthoryear{{Skrutskie} et~al.,}{{Skrutskie} et~al.}{2006}]{Skrutskie_2006}
{Skrutskie} M.~F.,  et~al., 2006, \mn@doi [\aj] {10.1086/498708}, \href {https://ui.adsabs.harvard.edu/abs/2006AJ....131.1163S} {131, 1163}

\bibitem[\protect\citeauthoryear{{Skryabina}, {Adams}  \& {Mosenkov}}{{Skryabina} et~al.}{2024}]{Skryabina_et_al_2024}
{Skryabina} M.~N.,  {Adams} K.~R.,   {Mosenkov} A.~V.,  2024, \mn@doi [\mnras] {10.1093/mnras/stae1502}, \href {https://ui.adsabs.harvard.edu/abs/2024MNRAS.532..883S} {532, 883}

\bibitem[\protect\citeauthoryear{{Sola} et~al.,}{{Sola} et~al.}{2022}]{Sola_et_al_2022}
{Sola} E.,  et~al., 2022, \mn@doi [\aap] {10.1051/0004-6361/202142675}, \href {https://ui.adsabs.harvard.edu/abs/2022A&A...662A.124S} {662, A124}

\bibitem[\protect\citeauthoryear{{Sola} et~al.,}{{Sola} et~al.}{2025}]{Sola_et_al_2025}
{Sola} E.,  et~al., 2025, \mn@doi [\mnras] {10.1093/mnras/staf1139}, \href {https://ui.adsabs.harvard.edu/abs/2025MNRAS.tmp.1100S} {}

\bibitem[\protect\citeauthoryear{{Somerville} \& {Dav{\'e}}}{{Somerville} \& {Dav{\'e}}}{2015}]{Somerville_and_Davé_2015}
{Somerville} R.~S.,  {Dav{\'e}} R.,  2015, \mn@doi [\araa] {10.1146/annurev-astro-082812-140951}, \href {https://ui.adsabs.harvard.edu/abs/2015ARA&A..53...51S} {53, 51}

\bibitem[\protect\citeauthoryear{{Springel}, {Frenk}  \& {White}}{{Springel} et~al.}{2006}]{Springel_2006}
{Springel} V.,  {Frenk} C.~S.,   {White} S. D.~M.,  2006, \mn@doi [\nat] {10.1038/nature04805}, \href {https://ui.adsabs.harvard.edu/abs/2006Natur.440.1137S} {440, 1137}

\bibitem[\protect\citeauthoryear{{Stone}, {Arora}, {Courteau}  \& {Cuillandre}}{{Stone} et~al.}{2021}]{Stone_et_al_2021}
{Stone} C.~J.,  {Arora} N.,  {Courteau} S.,   {Cuillandre} J.-C.,  2021, \mn@doi [\mnras] {10.1093/mnras/stab2709}, \href {https://ui.adsabs.harvard.edu/abs/2021MNRAS.508.1870S} {508, 1870}

\bibitem[\protect\citeauthoryear{{Tal}, {van Dokkum}, {Nelan}  \& {Bezanson}}{{Tal} et~al.}{2009}]{Tal_et_al_2009}
{Tal} T.,  {van Dokkum} P.~G.,  {Nelan} J.,   {Bezanson} R.,  2009, \mn@doi [\aj] {10.1088/0004-6256/138/5/1417}, \href {https://ui.adsabs.harvard.edu/abs/2009AJ....138.1417T} {138, 1417}

\bibitem[\protect\citeauthoryear{{Tanoglidis} et~al.,}{{Tanoglidis} et~al.}{2021}]{Tanoglidis_et_al_2021}
{Tanoglidis} D.,  et~al., 2021, \mn@doi [\apjs] {10.3847/1538-4365/abca89}, \href {https://ui.adsabs.harvard.edu/abs/2021ApJS..252...18T} {252, 18}

\bibitem[\protect\citeauthoryear{{Toomre} \& {Toomre}}{{Toomre} \& {Toomre}}{1972}]{Toomre_and_Toomre_1972}
{Toomre} A.,  {Toomre} J.,  1972, \mn@doi [\apj] {10.1086/151823}, \href {https://ui.adsabs.harvard.edu/abs/1972ApJ...178..623T} {178, 623}

\bibitem[\protect\citeauthoryear{{Tully} \& {Fisher}}{{Tully} \& {Fisher}}{1977}]{Tully_and_Fisher_1977}
{Tully} R.~B.,  {Fisher} J.~R.,  1977, \aap, \href {https://ui.adsabs.harvard.edu/abs/1977A&A....54..661T} {54, 661}

\bibitem[\protect\citeauthoryear{{Valenzuela} \& {Remus}}{{Valenzuela} \& {Remus}}{2024}]{Valenzuela_and_Remus_2024}
{Valenzuela} L.~M.,  {Remus} R.-S.,  2024, \mn@doi [\aap] {10.1051/0004-6361/202244758}, \href {https://ui.adsabs.harvard.edu/abs/2024A&A...686A.182V} {686, A182}

\bibitem[\protect\citeauthoryear{{Varghese}, {Ibata}  \& {Lewis}}{{Varghese} et~al.}{2011}]{Varghese_2011}
{Varghese} A.,  {Ibata} R.,   {Lewis} G.~F.,  2011, \mn@doi [\mnras] {10.1111/j.1365-2966.2011.19097.x}, \href {https://ui.adsabs.harvard.edu/abs/2011MNRAS.417..198V} {417, 198}

\bibitem[\protect\citeauthoryear{{V{\'a}zquez-Mata} et~al.,}{{V{\'a}zquez-Mata} et~al.}{2022}]{Vazquez-Mata_et_al_2022}
{V{\'a}zquez-Mata} J.~A.,  et~al., 2022, \mn@doi [\mnras] {10.1093/mnras/stac635}, \href {https://ui.adsabs.harvard.edu/abs/2022MNRAS.512.2222V} {512, 2222}

\bibitem[\protect\citeauthoryear{{Vera-Casanova} et~al.,}{{Vera-Casanova} et~al.}{2022}]{Vera-Casanova_et_al_2022}
{Vera-Casanova} A.,  et~al., 2022, \mn@doi [\mnras] {10.1093/mnras/stac1636}, \href {https://ui.adsabs.harvard.edu/abs/2022MNRAS.514.4898V} {514, 4898}

\bibitem[\protect\citeauthoryear{{Walder}, {Erkal}, {Collins}  \& {Martinez-Delgado}}{{Walder} et~al.}{2024}]{Walder_2024}
{Walder} M.,  {Erkal} D.,  {Collins} M.,   {Martinez-Delgado} D.,  2024, \mn@doi [arXiv e-prints] {10.48550/arXiv.2402.13314}, \href {https://ui.adsabs.harvard.edu/abs/2024arXiv240213314W} {p. arXiv:2402.13314}

\bibitem[\protect\citeauthoryear{{Walmsley}, {Ferguson}, {Mann}  \& {Lintott}}{{Walmsley} et~al.}{2019}]{Walmsley_et_al_2019}
{Walmsley} M.,  {Ferguson} A. M.~N.,  {Mann} R.~G.,   {Lintott} C.~J.,  2019, \mn@doi [\mnras] {10.1093/mnras/sty3232}, \href {https://ui.adsabs.harvard.edu/abs/2019MNRAS.483.2968W} {483, 2968}

\bibitem[\protect\citeauthoryear{{Walmsley} et~al.,}{{Walmsley} et~al.}{2022a}]{Walmsley_et_al_2022}
{Walmsley} M.,  et~al., 2022a, \mn@doi [\mnras] {10.1093/mnras/stab2093}, \href {https://ui.adsabs.harvard.edu/abs/2022MNRAS.509.3966W} {509, 3966}

\bibitem[\protect\citeauthoryear{{Walmsley} et~al.,}{{Walmsley} et~al.}{2022b}]{Walmsley_et_al_2022b}
{Walmsley} M.,  et~al., 2022b, \mn@doi [\mnras] {10.1093/mnras/stac525}, \href {https://ui.adsabs.harvard.edu/abs/2022MNRAS.513.1581W} {513, 1581}

\bibitem[\protect\citeauthoryear{{Wechsler} \& {Tinker}}{{Wechsler} \& {Tinker}}{2018}]{Wechsler_and_Tinker_2018}
{Wechsler} R.~H.,  {Tinker} J.~L.,  2018, \mn@doi [\araa] {10.1146/annurev-astro-081817-051756}, \href {https://ui.adsabs.harvard.edu/abs/2018ARA&A..56..435W} {56, 435}

\bibitem[\protect\citeauthoryear{{White} \& {Frenk}}{{White} \& {Frenk}}{1991}]{White_and_Frenk_1991}
{White} S. D.~M.,  {Frenk} C.~S.,  1991, \mn@doi [\apj] {10.1086/170483}, \href {https://ui.adsabs.harvard.edu/abs/1991ApJ...379...52W} {379, 52}

\bibitem[\protect\citeauthoryear{{White} \& {Rees}}{{White} \& {Rees}}{1978}]{White_and_Rees_1978}
{White} S.~D.~M.,  {Rees} M.~J.,  1978, \mn@doi [\mnras] {10.1093/mnras/183.3.341}, \href {https://ui.adsabs.harvard.edu/abs/1978MNRAS.183..341W} {183, 341}

\bibitem[\protect\citeauthoryear{{Wright}}{{Wright}}{2006}]{Wright_2006}
{Wright} E.~L.,  2006, \mn@doi [\pasp] {10.1086/510102}, \href {https://ui.adsabs.harvard.edu/abs/2006PASP..118.1711W} {118, 1711}

\bibitem[\protect\citeauthoryear{{Yoon} \& {Lim}}{{Yoon} \& {Lim}}{2020}]{Yoon_and_Lim_2020}
{Yoon} Y.,  {Lim} G.,  2020, \mn@doi [\apj] {10.3847/1538-4357/abc621}, \href {https://ui.adsabs.harvard.edu/abs/2020ApJ...905..154Y} {905, 154}

\bibitem[\protect\citeauthoryear{{Yoon}, {Park}, {Chung}  \& {Lane}}{{Yoon} et~al.}{2022}]{Yoon_2022}
{Yoon} Y.,  {Park} C.,  {Chung} H.,   {Lane} R.~R.,  2022, \mn@doi [\apj] {10.3847/1538-4357/ac415d}, \href {https://ui.adsabs.harvard.edu/abs/2022ApJ...925..168Y} {925, 168}

\bibitem[\protect\citeauthoryear{{Yoon}, {Ko}  \& {Kim}}{{Yoon} et~al.}{2023}]{Yoon_et_al_2023}
{Yoon} Y.,  {Ko} J.,   {Kim} J.-W.,  2023, \mn@doi [\apj] {10.3847/1538-4357/acbcc5}, \href {https://ui.adsabs.harvard.edu/abs/2023ApJ...946...41Y} {946, 41}

\bibitem[\protect\citeauthoryear{{Yoon}, {Kim}  \& {Ko}}{{Yoon} et~al.}{2024}]{Yoon_et_al_2024}
{Yoon} Y.,  {Kim} J.-W.,   {Ko} J.,  2024, \mn@doi [\apj] {10.3847/1538-4357/ad7816}, \href {https://ui.adsabs.harvard.edu/abs/2024ApJ...974..299Y} {974, 299}

\bibitem[\protect\citeauthoryear{{York} et~al.,}{{York} et~al.}{2000}]{York_2000}
{York} D.~G.,  et~al., 2000, \mn@doi [\aj] {10.1086/301513}, \href {https://ui.adsabs.harvard.edu/abs/2000AJ....120.1579Y} {120, 1579}

\bibitem[\protect\citeauthoryear{{Zhu}, {Marinacci}, {Maji}, {Li}, {Springel}  \& {Hernquist}}{{Zhu} et~al.}{2016}]{Zhu+2016}
{Zhu} Q.,  {Marinacci} F.,  {Maji} M.,  {Li} Y.,  {Springel} V.,   {Hernquist} L.,  2016, \mn@doi [\mnras] {10.1093/mnras/stw374}, \href {https://ui.adsabs.harvard.edu/abs/2016MNRAS.458.1559Z} {458, 1559}

\bibitem[\protect\citeauthoryear{{van Dokkum}}{{van Dokkum}}{2005}]{van_Dokkum_2005}
{van Dokkum} P.~G.,  2005, \mn@doi [\aj] {10.1086/497593}, \href {https://ui.adsabs.harvard.edu/abs/2005AJ....130.2647V} {130, 2647}

\makeatother
\end{thebibliography}


\newpage
\appendix

\section{Visual classification} \label{section:appendix_classification}

We detail in this section the mutually exclusive categories that were used to classify the galaxies between the different types of features. They are defined as follows and are illustrated in Figure \ref{figure:appendix-classes}. Figure \ref{fig:appendix-montage_png} illustrates the type of images used for the visual classification.

We classify each galaxy based on the presence and morphology of LSB features. If no tidal disturbance or perturbed stellar halo is observed, the galaxy has \textit{no feature}. \textit{Streams} are formed in minor mergers \citep[e.g.,][]{Bullock_and_Johnston_2005,Belokurov_2006,Martinez-Delgado_2010} and are stellar material tidally stripped from a satellite galaxy, sometimes already fully disrupted. They generally have narrow, faint, long, often curved morphologies and can be classified as orphan streams when no visible progenitor remains. \textit{Tidal tails/arms} originate from major mergers \citep[e.g.,][]{Arp_1966,Toomre_and_Toomre_1972} but can also be triggered by non-merging flybys \citep[e.g.,][]{Duc_and_Renaud_2013}. The stellar material comes from the target galaxy, and generally has a wide, elongated, antennae-like shape. It also includes cases where the distinction between tidal tails and spiral arms is ambiguous. \textit{Plumes} are tidal tails with rounder and broader morphologies \citep[e.g.,][]{Bilek_et_al_2020}. \textit{Shells}, which are arc-shaped features with sharp outer edges, often forming concentric patterns or umbrella-like structures (e.g., \citealt{Prieur_1990, Ebrova_2013, Pop_2018}). In addition, we include \textit{disturbed or asymmetric stellar haloes}, which are diffuse extended envelopes exhibiting strong disturbances, multiple tidal features, or asymmetry. Two optional categories are also used: \textit{grand-design spirals}, where prominent spiral arms are present (not to be confused with tidal tails), and \textit{curiosities}, which refer to galaxies with rare or particularly peculiar morphologies. We sometimes included faint dwarf galaxies in this category. 

\begin{figure}
    \centering
    \includegraphics[width=\linewidth]{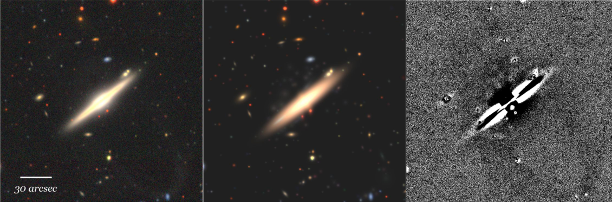}
    \caption{Example of a montage PNG image used during the classification process, directly available at \url{https://portal.nersc.gov/project/cosmo/data/sga/2020/html/}. \textit{Left}: original $grz$ data image. \textit{Middle}: $grz$ \texttt{Tractor} model image of all astronomical sources. \textit{Right}: residual image (median of the $g,r,z$ residual images), stretched between 35-90\% percentiles of the pixel distribution for a better visualisation of the faint features. White regions indicate areas where the model overestimates the light, while black regions show an excess of light in the data. Streams appear as black regions in the residuals. }
    \label{fig:appendix-montage_png}
\end{figure}

\begin{figure}
	\includegraphics[width=\linewidth]{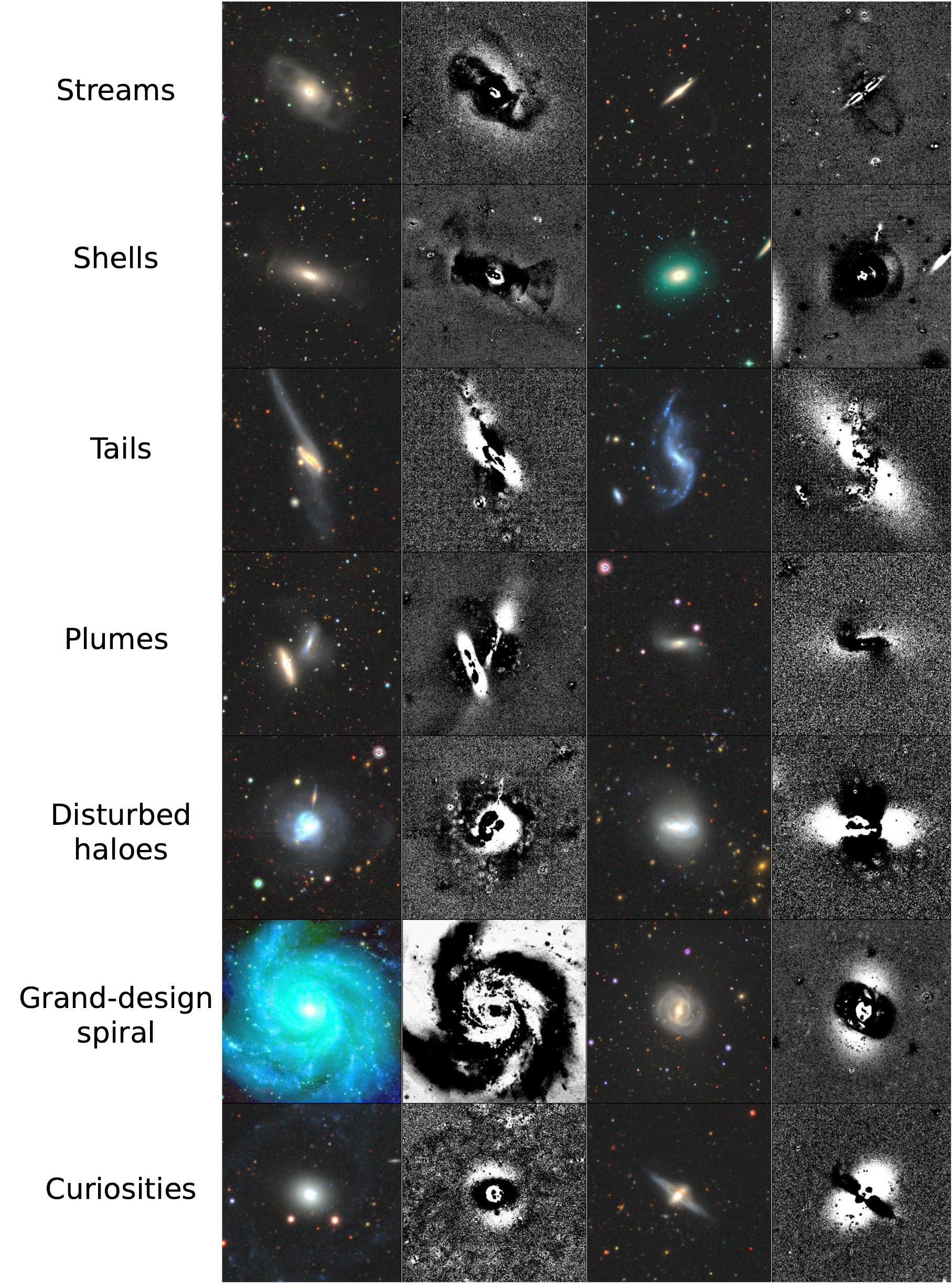}
    \caption{Illustration of the mutually exclusive categories used for visual inspection. From top to bottom: streams, shells, tails/arms, plumes, disturbed/asymmetric haloes, grand-design spirals, curiosities. In each row, each object is displayed once in $grz$ original data image and once in stretched residual image. North is up, East is left.}
    \label{figure:appendix-classes}
\end{figure}

After completing the classification, we compiled the individual votes according to the voting procedure outlined in Section \ref{sec:method_voting}. We analyse the variability of the confidence score, which reflects the degree of agreement among classifiers. Figure \ref{fig:appendix-results_pie_chart_scores} shows the distribution of the scores as a function of the feature type. Among the streams, 266 (30.9\% of all streams) have a score higher than 0.5, and only 59 (6.9\%) have a score higher than 0.75. Tails with scores higher than 0.5 represent 39.6\% of tails, and respectively 39.3\%, 21.9\% and 18.5\% for shells, plumes and disturbed haloes for the same threshold. The disagreement among classifiers can be significant, highlighting the challenge of distinguishing between tidal feature categories, even for expert users, due to sometimes subtle and ambiguous boundaries. 
 
\begin{figure*}
	\includegraphics[width=0.9\textwidth]{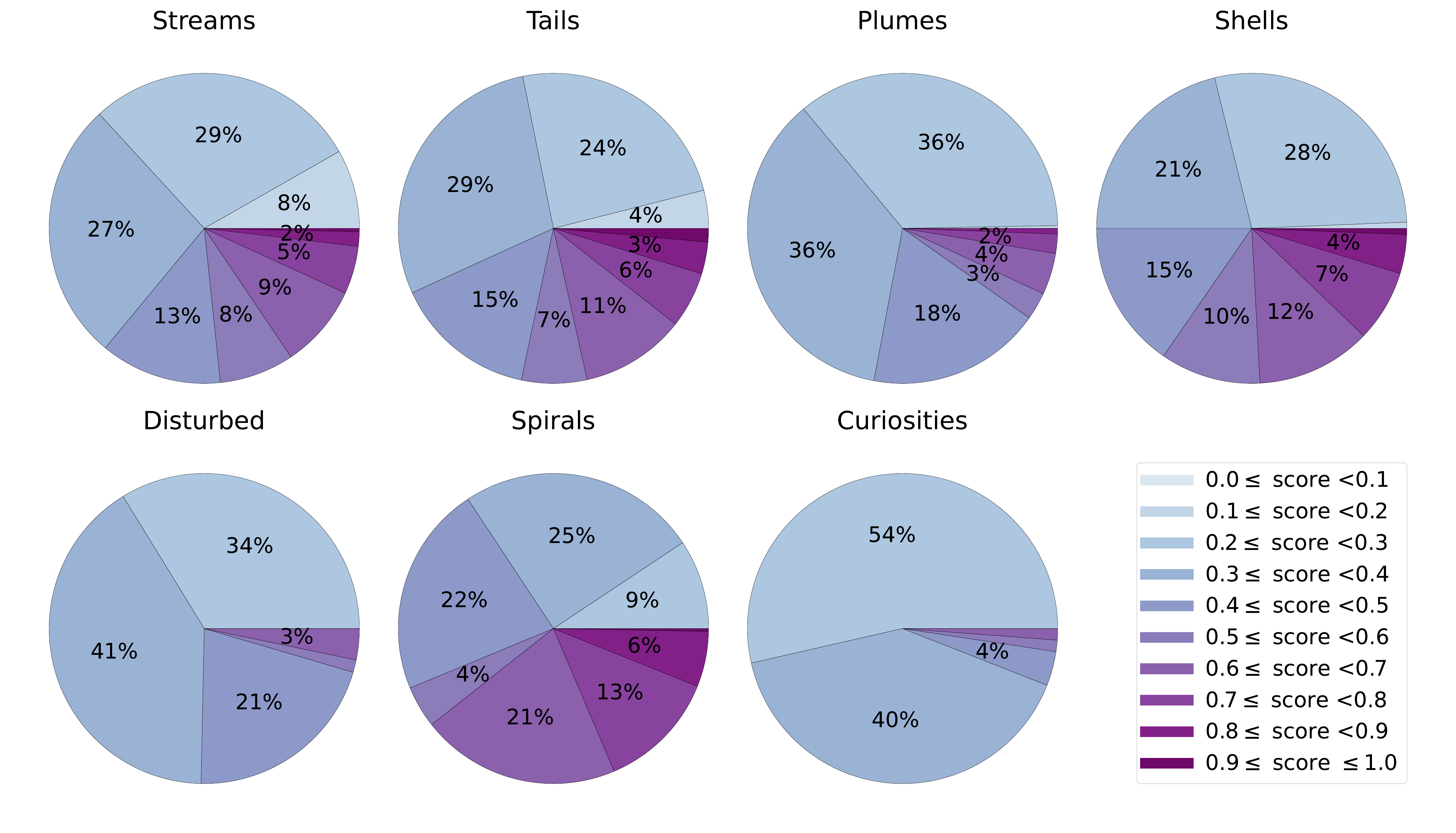}
    \caption{ Pie charts of the confidence score for each feature (stream, tails, plumes, shells, disturbed halo, grand-design spirals and curiosities). Darker shades indicate higher confidence scores. The fractions of features of a given type in each bin of score divided by the total number of features of that type are overlaid on the plot (for better visibility, only the fractions higher than 2\% are overlaid). There is a high variability in the confidence scores, with only a small fraction of the votes with a high score. This reflects both the expertise of the classifiers and the ambiguity between some feature categories.}
    \label{fig:appendix-results_pie_chart_scores}
\end{figure*}

As a complement to Section \ref{section:results}, we investigate the evolution of the fraction of galaxies with tidal features as a function of stellar mass and of the morphological type in Figure \ref{fig:appendix_fraction_features_vs_mass_morphtype}. The fraction of galaxies with features increases with galaxy mass for ETGs and LTGs, with a sharper increase after a mass of about $3\times 10^{10} M_\odot$ for LTGs and around  $10^{10} M_\odot$ for ETGs. The fraction of galaxies with any tidal features ranges from 2\% to 48.5\% for ETGs and from 2.5\% to 26.7\% for LTGs. For ETGs, the increase is driven by streams and shells, that reaches 20\% in the higher mass bin, while the fraction of tails and plumes remains below 4\%. For LTGs, the trend is dominated by streams and tails (reaching 14\% and 9\%, respectively, in the highest mass bin), while shells and plumes remain below 3\%. 

\begin{figure*}
    \centering
    \includegraphics[width=0.8\linewidth]{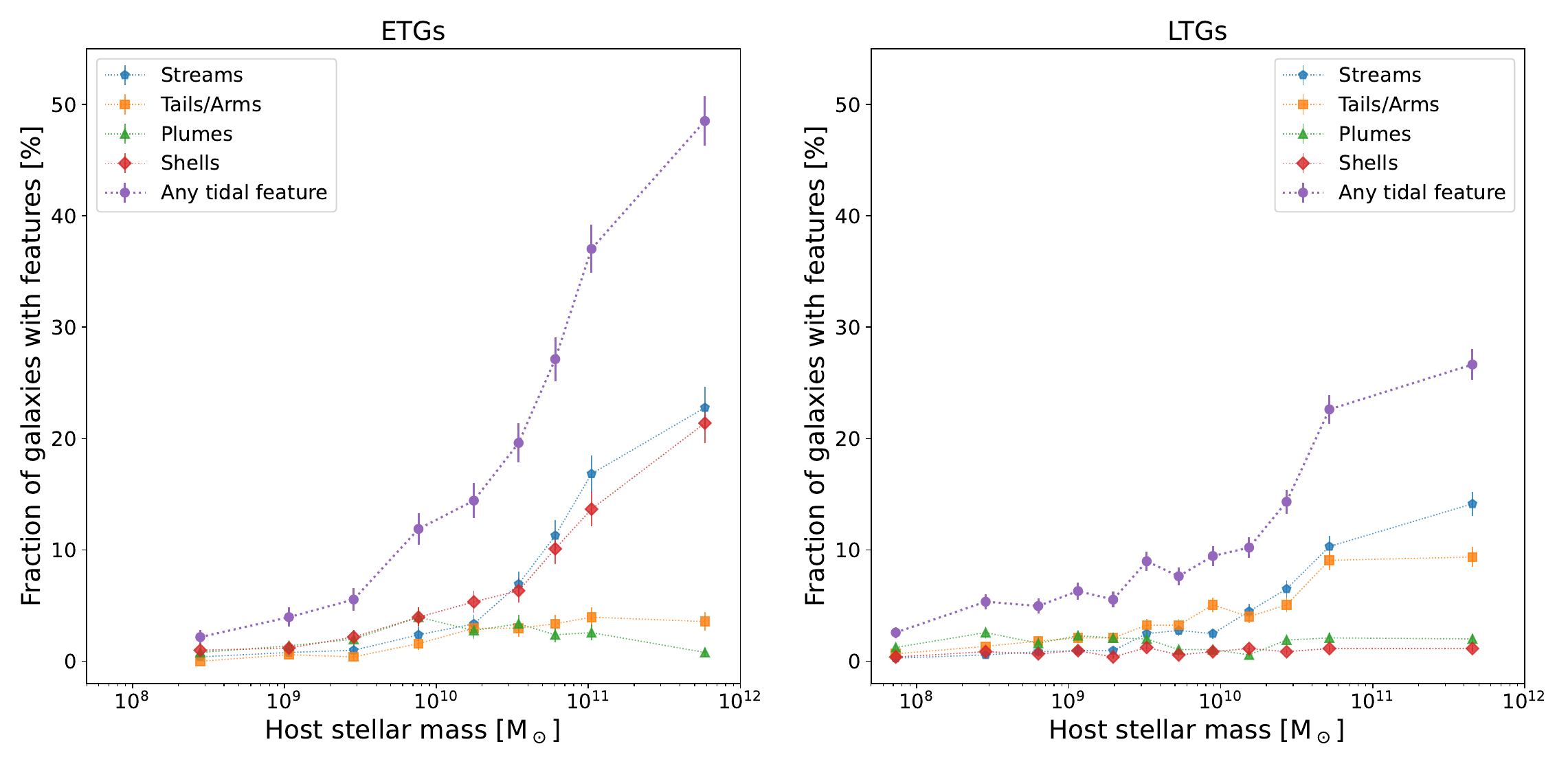}
    \caption{Tidal feature fractions as a function of the host galaxy's stellar mass (in $M_\odot$) for ETGs \textit{(left)} and LTGs \textit{(right)}. The mass bins contains approximately the same number of galaxies (about 505 for ETGs and 1,048 for LTGs). The errorbars represent the $1\sigma$ standard error on proportions in each bin. Streams are plotted as blue pentagons, tails as orange squares, plumes as green triangles, shells as red rhombuses and any tidal feature as purple circles. The increase of the tidal feature fraction with mass is visible for both morphological types. ETGs host more tidal features than LTGs, with a sharper increase in the trend with mass for stellar masses above $\sim 5\times 10^{10} M_\odot$.}
    \label{fig:appendix_fraction_features_vs_mass_morphtype}
\end{figure*}


\section{STRRINGS   sample}\label{section:appendix_streams_segmentation}
\subsection{STRRINGS properties}\label{sec:appendix-strrings_properties}

In this section, we detail our stream track fitting method and we provide the STRRINGS streams' properties. We start by explaining our stream track fitting.

We begin by segmenting the stream on the residual image using \texttt{Jafar}. For the purpose of our method, when the stream wraps around or crosses itself, we split the segmentation into multiple masks, each assigned to a distinct angular region. The order in which these non-overlapping masks are defined is preserved, which later enables unambiguous reconstruction of the stream path.
Once the segmentation masks are defined, we proceed to extract the stream track by fitting Gaussians in radial profiles taken from angular bins. For each angle bin, we measure the residual light profile as a function of radius. The segmentation masks act as priors: they constrain the position of the Gaussian peaks within the expected region of the stream.
If two masks fall within the same angular bin, two Gaussians are fitted accordingly. Figure \ref{fig:illustration_track} illustrates these two cases, one with an angular overlap of the stream on itself (in dotted cyan) and a bin without overlap (in dashed purple). The ordering of the masks ensures that we can correctly associate each Gaussian component with its corresponding segments. This processed is repeated across all angular bins to reconstruct the full, unwrapped stream track.

\begin{figure*}
    \centering
    \includegraphics[width=0.8\linewidth]{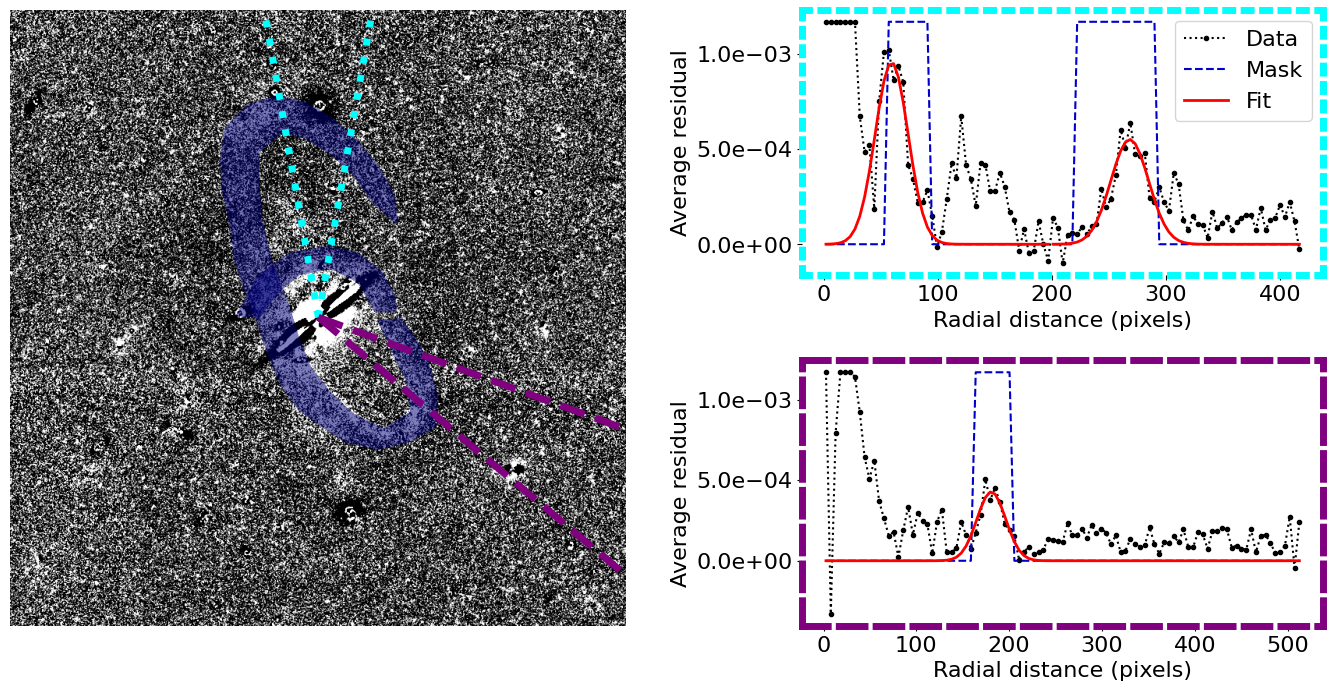}
    \caption{Illustration of the track fitting process for one stream. \textit{Left}: residual image with the stream segmentation overlaid in shaded blue regions. Two angle bins are represented by the dotted cyan and dashed purple lines. \textit{Top, right}: for the first angle bin (dotted cyan), average flux in the residual image (in nanomaggies) as a function of the galactocentric radial distance (in pixels). Data is plotted as black dots, the delineation of the mask is in dashed blue while the Gaussian fit is in solid red line. \textit{Bottom, right}: for the second angle bin (dashed purple), average flux in the residual image (in nanomaggies) as a function of the galactocentric radial distance (in pixels). Data is plotted as black dots, the delineation of the mask is in dashed blue while the Gaussian fit is in solid red line.  }
    \label{fig:illustration_track}
\end{figure*}

Then, we provide the geometry, photometry, colour and stellar mass estimates of our segmented STRRINGS streams in Table \ref{tab:best_streams}. The original, model and residual images of the streams, along with their track and contours, are displayed in Figure \ref{fig:appendix-best_stream_segmentation_all-part1}. 
Note that in some cases, the fitted track appears to deviate from the visually identified track. This discrepancy is often caused by very bright pixels (potentially resulting from local overdensities or imperfect \texttt{Tractor} modelling) which can bias the Gaussian fit and pull parts of the track away from the actual stream. Furthermore, in regions where the stream overlaps itself, the algorithm may mistakenly fit a single component instead of two, which can also lead to misalignment.

The coordinates of the contours of the segmented streams are provided in a format compatible with SAOImageDS9 and Aladin, as well as in a csv file in the online version of the paper. A truncated example is shown below, while the full file will be available in the online version.
\begin{table*}
\centering
\scriptsize
\caption{Geometric and photometric properties of the STRRINGS sample.  (1): galaxy name, (2): galaxy distance  (Mpc, computed from \textit{Z\_LEDA}), (3): Area (kpc$^2$), (4): length of the stream track (kpc), (5): width obtained from the stream track (kpc), (6): angular extent of the stream (deg), (7): median surface brightness in $g$-band (mag$\,$arcsec$^{-2}$), (8): median surface brightness in $r$-band (mag$\,$arcsec$^{-2}$), (9): median surface brightness in $z$-band (mag$\,$arcsec$^{-2}$), (10): median $g-r$ colour (mag), (11): stream stellar mass (M$_\odot)$, (12): stellar mass ratio between the stream and the host galaxy.}
  \label{tab:best_streams} 
\begin{tabular}{cccccccccccc}
\hline 
Name & D  & Area &Length  & Width & Angular extent & SB$_{g}$ & SB$_{r}$ & SB$_{z}$& $g-r$ & M$_\star,\texttt{stream}$ &  $\frac{\texttt{M}_{\star,\texttt{stream}}}{\texttt{M}_{\star,\texttt{host}}}$\\
    & [Mpc]  & [kpc$^{2}$] & [kpc]&[kpc] & [deg] & [mag$\,$arcsec$^{-2}$] & [mag$\,$arcsec$^{-2}$]  & [mag$\,$arcsec$^{-2}$] & [mag]  & [M$_\odot$] &\\
(1) & (2)  & (3) & (4) &(5)  & (6) & (7)  & (8) & (9)  &(10)  & (11) & (12)\\

\hline
ESO079-003\_GROUP & 38.7 & 676.9 & 142.4 & 6.1 & 110.2 & 26.77 $\pm$ 0.06 & 26.46 $\pm$ 0.06 & 25.36 $\pm$ 0.05 & 0.31 $\pm$ 0.09 & 2.8E+08 & 0.003 \\
ESO298-016 & 76.6 & 1538.6 & 106.2 & 10.2 & 110.4 & 26.94 $\pm$ 0.08 & 26.51 $\pm$ 0.08 & 25.3 $\pm$ 0.06 & 0.43 $\pm$ 0.11 & 9.3E+08 & 0.004 \\
ESO356-012 & 75.2 & 324.8 & 55.4 & 4.3 & 147.1 & 26.8 $\pm$ 0.17 & 26.36 $\pm$ 0.16 & 25.31 $\pm$ 0.13 & 0.47 $\pm$ 0.23 & 2.1E+08 & 0.022 \\
IC0120 & 70.9 & 1102.8 & 165.3 & 5.9 & 352.7 & 26.86 $\pm$ 0.1 & 26.58 $\pm$ 0.09 & 25.33 $\pm$ 0.07 & 0.28 $\pm$ 0.13 & 3.8E+08 & 0.009 \\
IC1124 & 77.3 & 270.2 & 45.7 & 5.0 & 81.2 & 26.52 $\pm$ 0.17 & 25.75 $\pm$ 0.14 & 24.77 $\pm$ 0.11 & 0.77 $\pm$ 0.22 & 4.8E+08 & 0.007 \\
IC1657\_GROUP & 52.6 & 1040.9 & 210.3 & 4.5 & 274.2 & 26.87 $\pm$ 0.07 & 26.66 $\pm$ 0.07 & 25.17 $\pm$ 0.05 & 0.21 $\pm$ 0.1 & 3.1E+08 & 0.004 \\
NGC0804 & 77.5 & 2540.1 & 468.3 & 6.9 & 396.6 & 26.5 $\pm$ 0.06 & 25.72 $\pm$ 0.05 & 24.75 $\pm$ 0.04 & 0.78 $\pm$ 0.07 & 5.0E+09 & 0.050 \\
NGC1084\_GROUP & 21.0 & 723.1 & 293.9 & 2.4 & 477.7 & 27.32 $\pm$ 0.04 & 26.73 $\pm$ 0.04 & 25.33 $\pm$ 0.03 & 0.58 $\pm$ 0.06 & 3.5E+08 & 0.008 \\
NGC1121 & 38.2 & 2146.5 & 1199.3 & 5.4 & 713.6 & 27.01 $\pm$ 0.04 & 26.58 $\pm$ 0.03 & 25.32 $\pm$ 0.02 & 0.43 $\pm$ 0.05 & 1.3E+09 & 0.040 \\
NGC1162 & 58.0 & 2466.6 & 248.7 & 7.9 & 352.7 & 26.95 $\pm$ 0.05 & 26.53 $\pm$ 0.05 & 25.26 $\pm$ 0.03 & 0.41 $\pm$ 0.07 & 1.2E+09 & 0.006 \\
NGC2769 & 70.9 & 1624.4 & 125.8 & 10.2 & 102.8 & 27.03 $\pm$ 0.07 & 26.44 $\pm$ 0.05 & 24.97 $\pm$ 0.04 & 0.6 $\pm$ 0.08 & 1.1E+09 & 0.006 \\
NGC3041 & 20.7 & 90.7 & 20.0 & 3.4 & 51.2 & 26.59 $\pm$ 0.08 & 25.94 $\pm$ 0.07 & 24.97 $\pm$ 0.06 & 0.64 $\pm$ 0.11 & 1.1E+08 & 0.004 \\
NGC4256\_GROUP & 37.3 & 2449.0 & 241.0 & 8.8 & 299.9 & 26.94 $\pm$ 0.03 & 26.25 $\pm$ 0.02 & 24.86 $\pm$ 0.02 & 0.68 $\pm$ 0.04 & 2.2E+09 & 0.017 \\
NGC4388\_GROUP & 37.1 & 505.1 & 66.1 & 7.2 & 109.6 & 26.88 $\pm$ 0.07 & 26.32 $\pm$ 0.07 & 25.13 $\pm$ 0.07 & 0.55 $\pm$ 0.1 & 4.6E+08 & 0.003 \\
NGC4632 & 25.4 & 129.0 & 33.8 & 2.8 & 117.2 & 26.53 $\pm$ 0.08 & 27.4 $\pm$ 0.09 & 25.12 $\pm$ 0.06 & -0.87 $\pm$ 0.12 & 3.5E+06 & 0.000 \\
NGC4686 & 73.9 & 817.0 & 134.7 & 8.7 & 95.6 & 26.78 $\pm$ 0.09 & 26.0 $\pm$ 0.07 & 25.06 $\pm$ 0.07 & 0.79 $\pm$ 0.12 & 1.1E+09 & 0.004 \\
NGC5055\_GROUP & 7.5 & 187.2 & 57.8 & 2.6 & 127.6 & 26.86 $\pm$ 0.03 & 25.96 $\pm$ 0.03 & 25.06 $\pm$ 0.03 & 0.89 $\pm$ 0.04 & 3.2E+08 & 0.007 \\
NGC5263 & 71.1 & 897.0 & 214.5 & 5.4 & 286.3 & 26.31 $\pm$ 0.08 & 25.67 $\pm$ 0.07 & 24.93 $\pm$ 0.06 & 0.62 $\pm$ 0.11 & 1.2E+09 & 0.020 \\
NGC5387 & 76.8 & 92.8 & 24.5 & 2.1 & 93.0 & 26.29 $\pm$ 0.26 & 26.15 $\pm$ 0.27 & 25.35 $\pm$ 0.23 & 0.13 $\pm$ 0.38 & 3.8E+07 & 0.001 \\
NGC5513\_GROUP & 73.8 & 890.1 & 106.4 & 5.9 & 152.2 & 26.11 $\pm$ 0.08 & 25.55 $\pm$ 0.07 & 24.6 $\pm$ 0.05 & 0.59 $\pm$ 0.1 & 1.4E+09 & 0.005 \\
NGC5907 & 9.9 & 308.6 & 71.7 & 4.6 & 212.9 & 26.43 $\pm$ 0.02 & 26.13 $\pm$ 0.02 & 24.87 $\pm$ 0.02 & 0.28 $\pm$ 0.03 & 1.5E+08 & 0.004 \\
PGC000902 & 79.1 & 2915.5 & 1151.5 & 7.6 & 567.5 & 26.85 $\pm$ 0.06 & 26.59 $\pm$ 0.06 & 25.37 $\pm$ 0.05 & 0.26 $\pm$ 0.09 & 9.8E+08 & 0.027 \\
PGC006791 & 79.2 & 188.4 & 41.6 & 4.5 & 58.6 & 26.7 $\pm$ 0.22 & 26.25 $\pm$ 0.21 & 25.13 $\pm$ 0.15 & 0.5 $\pm$ 0.3 & 1.6E+08 & 0.002 \\
PGC021008 & 83.8 & 248.3 & 65.2 & 2.3 & 331.7 & 26.57 $\pm$ 0.17 & 25.76 $\pm$ 0.14 & 24.92 $\pm$ 0.13 & 0.8 $\pm$ 0.22 & 6.2E+08 & 0.007 \\
PGC039258 & 85.1 & 470.8 & 82.7 & 3.4 & 238.6 & 27.45 $\pm$ 0.22 & 27.19 $\pm$ 0.21 & 24.84 $\pm$ 0.1 & 0.25 $\pm$ 0.3 & 9.9E+07 & 0.001 \\
PGC1001085 & 80.0 & 1490.3 & 407.4 & 5.4 & 426.5 & 26.35 $\pm$ 0.07 & 25.58 $\pm$ 0.06 & 24.88 $\pm$ 0.05 & 0.76 $\pm$ 0.09 & 3.0E+09 & 0.058 \\
PGC1092512 & 76.3 & 908.0 & 128.8 & 5.4 & 227.7 & 26.7 $\pm$ 0.1 & 26.11 $\pm$ 0.09 & 25.05 $\pm$ 0.07 & 0.59 $\pm$ 0.14 & 9.0E+08 & 0.015 \\
PGC3092153 & 87.8 & 116.1 & 44.2 & 2.2 & 141.9 & 26.45 $\pm$ 0.28 & 26.14 $\pm$ 0.27 & 24.85 $\pm$ 0.2 & 0.33 $\pm$ 0.39 & 7.6E+07 & 0.009 \\
PGC430221 & 78.3 & 233.5 & 45.9 & 4.0 & 162.2 & 27.2 $\pm$ 0.28 & 26.73 $\pm$ 0.26 & 25.19 $\pm$ 0.15 & 0.49 $\pm$ 0.38 & 9.5E+07 & 0.009 \\
PGC938075 & 1.8 & 0.6 & 4.3 & 0.1 & 404.0 & 27.02 $\pm$ 0.1 & 26.57 $\pm$ 0.1 & 25.28 $\pm$ 0.07 & 0.43 $\pm$ 0.14 & 3.2E+05 & 0.124 \\
UGC01245 & 74.8 & 1475.8 & 86.0 & 11.1 & 161.1 & 26.58 $\pm$ 0.07 & 25.83 $\pm$ 0.06 & 24.9 $\pm$ 0.05 & 0.75 $\pm$ 0.1 & 2.6E+09 & 0.020 \\
UGC01424 & 75.8 & 304.2 & 98.0 & 2.4 & 205.7 & 26.87 $\pm$ 0.18 & 26.41 $\pm$ 0.17 & 25.28 $\pm$ 0.13 & 0.47 $\pm$ 0.24 & 2.2E+08 & 0.005 \\
UGC01970 & 28.2 & 774.5 & 138.3 & 5.8 & 227.8 & 26.74 $\pm$ 0.04 & 25.94 $\pm$ 0.04 & 25.12 $\pm$ 0.03 & 0.81 $\pm$ 0.05 & 1.5E+09 & 0.217 \\
UGC08717 & 68.6 & 1643.5 & 333.7 & 5.1 & 496.8 & 26.6 $\pm$ 0.06 & 25.88 $\pm$ 0.06 & 25.03 $\pm$ 0.04 & 0.71 $\pm$ 0.09 & 2.3E+09 & 0.028 \\
UGC09239 & 81.9 & 2060.9 & 234.4 & 8.7 & 227.9 & 26.39 $\pm$ 0.06 & 25.85 $\pm$ 0.06 & 24.96 $\pm$ 0.04 & 0.53 $\pm$ 0.08 & 2.1E+09 & 0.035 \\
\hline
\end{tabular}
 
\end{table*}

\begin{verbatim}
# Region file format: DS9 version 4.1
global color=yellow dashlist=8 3 width=1 
font="helvetica 10 normal roman" select=1 highlite=1 
dash=0 fixed=0 edit=1 move=1 delete=1 include=1 
source=1 icrs
# Stream ESO079-003_GROUP
polygon(7.93,-64.25,...,7.93,-64.25) # color=red
# Stream IC1124
polygon(232.51,23.63,...,232.51,23.63) # color=red
...
\end{verbatim}

\begin{figure*}
    \centering
    \includegraphics[width=\linewidth]{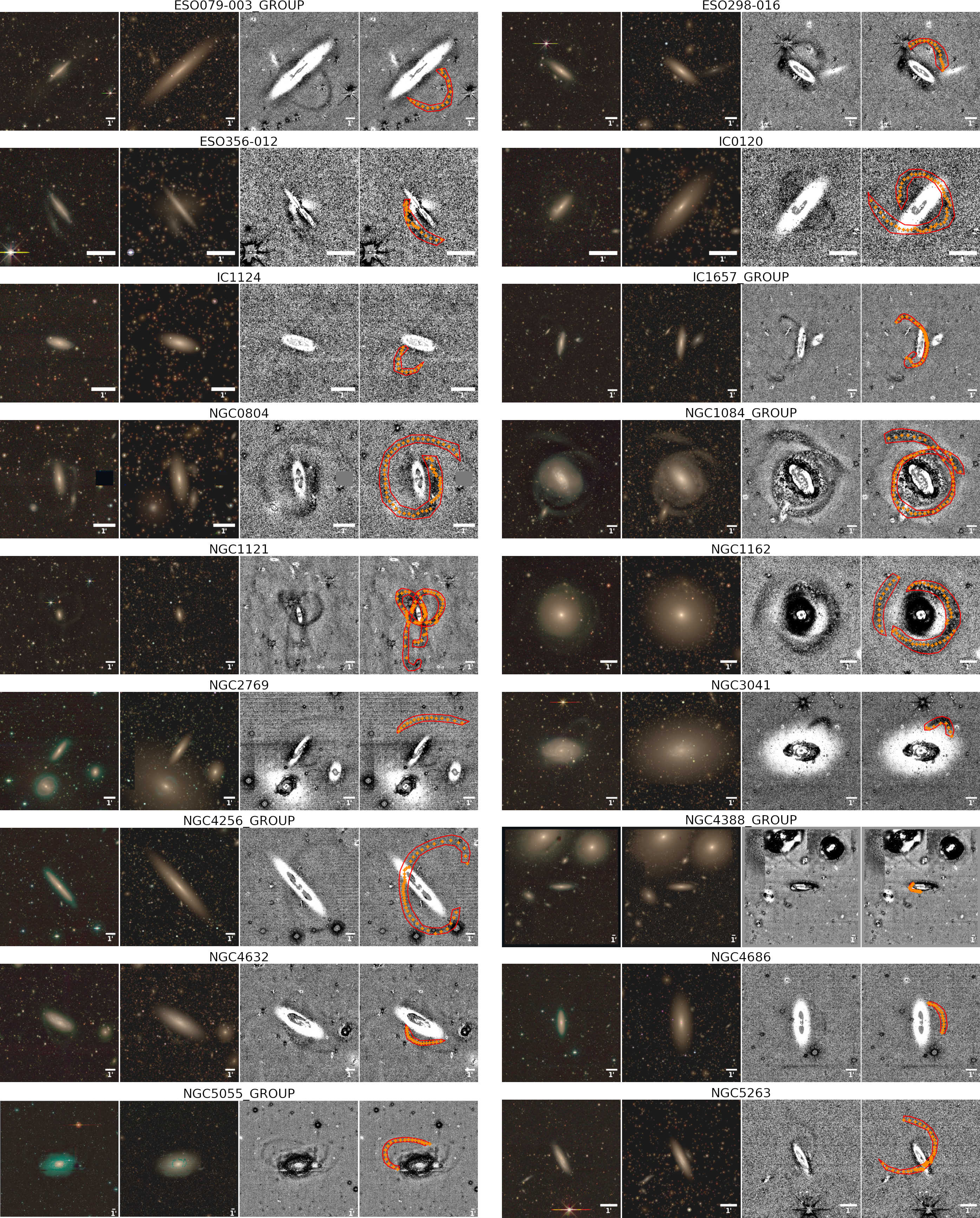}
 \caption{Illustration of our 35 STRRINGS streams. A scalebar of $1'$ is indicated at the bottom of each plot, while the galaxy name is indicated on top. North is up, East is left. Four images per galaxy are displayed. \textit{Left:} original $grz$ data image. \textit{Middle, left}: \texttt{Tractor} $grz$ model of all astronomical sources. \textit{Middle, right}: residual image (median of $g$,$r$,$z$-band residual images, smoothed and stretched for better visualisation of faint features). White regions indicate model excess, black regions are excess of light in the data. Streams appear as black regions. \textit{Right}: residual image with stream segmentation (red contours) from \texttt{Jafar} and stream track (orange crosses), as described in Section \ref{section:method-segmentation}. }
    \label{fig:appendix-best_stream_segmentation_all-part1}
\end{figure*}
\begin{figure*}\ContinuedFloat
    \centering
    \includegraphics[width=\linewidth]{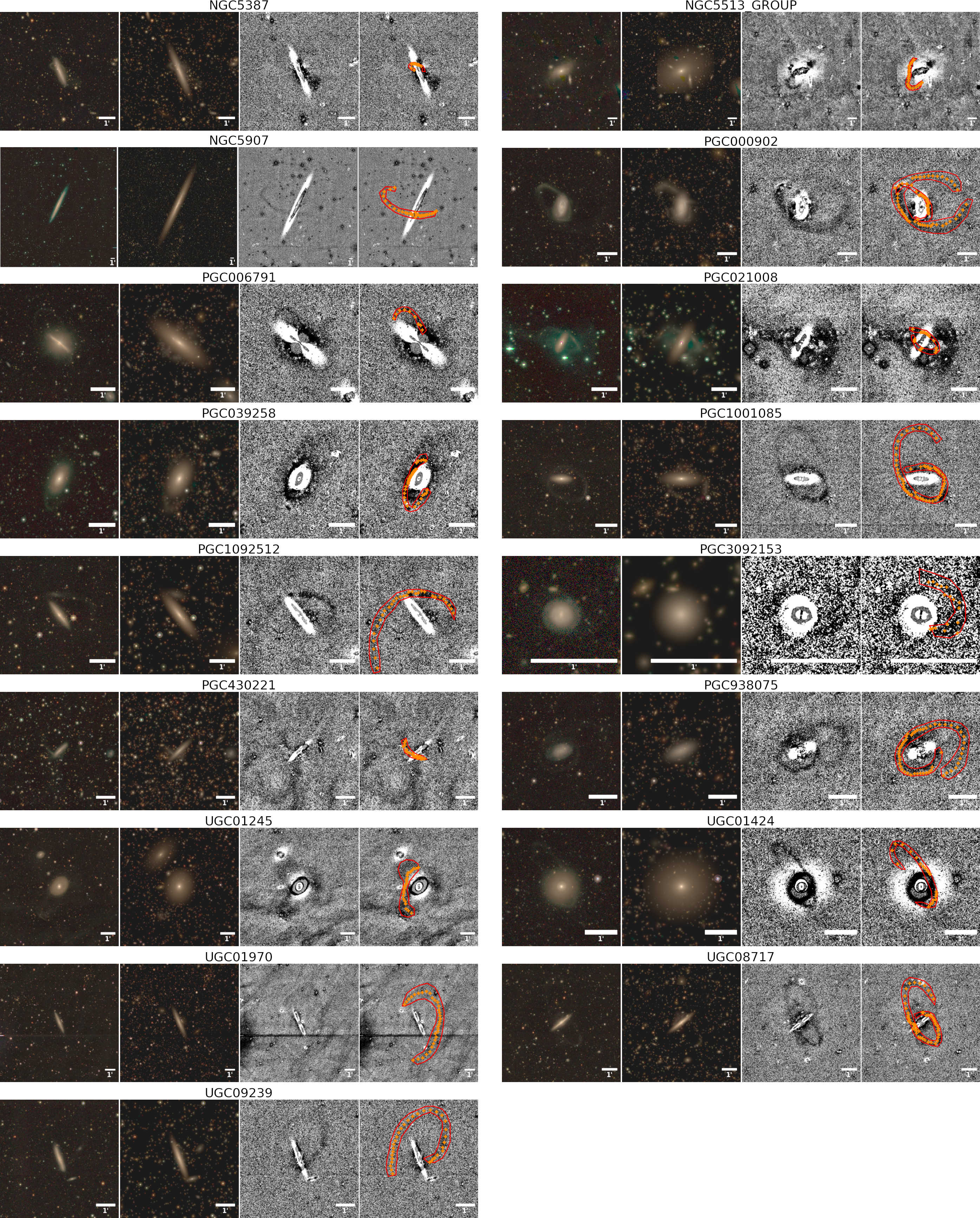}
    \caption{Continued}
   
\end{figure*}

The errors on SB and colour measurements must be taken with caution. The error of SB $\Delta\mu$ is computed as:
\begin{equation*}
    \Delta\mu =  \frac{2.5}{\ln(10)} \times \frac{1}{SNR}
 ;\qquad  SNR =  \frac{\sqrt{N} (F - S_a)}{\sqrt{F + \sigma_s^{2}} } 
\end{equation*}
where SNR is the signal-to-noise ratio from  \cite{Akhlaghi_and_Ichikawa_2015}, $F$ is the average flux value of the feature of interest, $N$ is the number of pixel inside the feature, $\sigma_s$ is the variation of the background and $S_a$ is the average of the background. We compute the average and variation of the background from the residual image where we mask all sources and the galaxy, clip the values between 5\%-95\% of the pixel distribution then fit these clipped background values by a Gaussian whose mean is taken as $S_a$ and standard deviation as $\sigma_s$. 
The error on colours $\Delta (g-r)$ is taken as
\begin{equation*}
 \Delta (g-r) = \sqrt{(\Delta\mu_{g})^{2} + (\Delta\mu_{r})^{2}}
\end{equation*}

As discussed in Section \ref{sec:sb_colors_streams}, there are large variations of the SB distribution along the streams due to the fact that we are working with residual images. A bad fit from the \texttt{Tractor} model can result in small residual values that do not have a physical meaning. Likewise, negative residuals from model oversubtraction are not physical. Therefore, we `truncate' the distribution of the flux by only keeping positive values.

Finally, we investigate the evolution of the SB and width variations along the tracks of the streams, as illustrated in Figure \ref{fig:sb_variations_track}. We could not identify obvious trends for each stream, and neither could we see increased SB values where the candidate dwarf galaxy progenitors are located. This is because we are retrieving the SB from the residual images in which the dwarf galaxies have been partly modelled and removed. In addition, some low SB values actually coincide with the location of the host galaxy (when the stream is looping around or behind), for which the residuals can be negative. 
\begin{figure*}
    \centering
    \includegraphics[width=\linewidth]{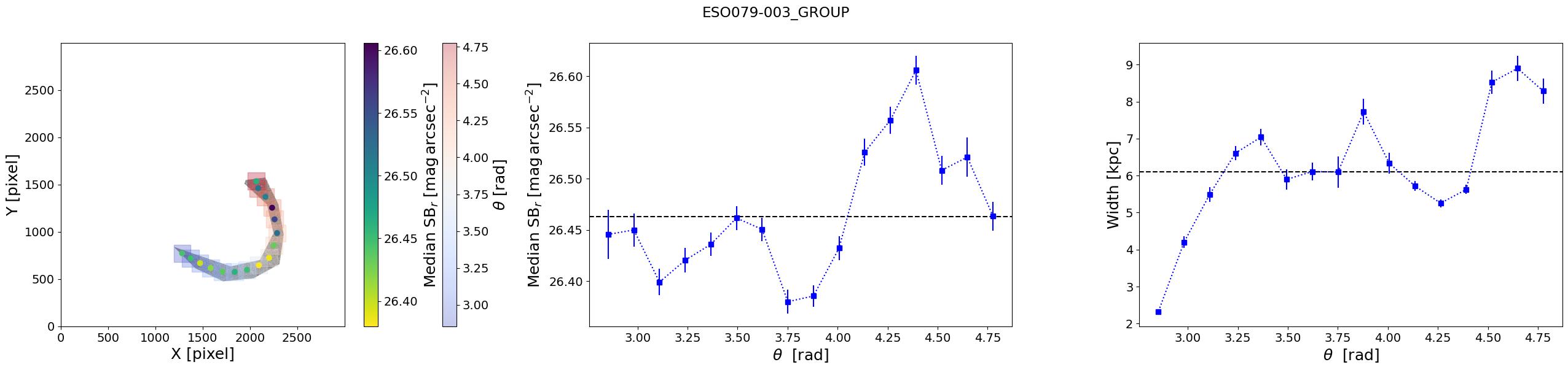}
    \caption{Illustration of the variations of surface brightness in the $r-$band SB $_r$ and width along the track of one stream (around the host galaxy ESO079-003\_GROUP). The angular distance $\theta$ (in rad) is used to parametrise the stream track. \textit{Left}: In each bin of angular distance  $\theta$ (in rad) (coloured boxes) along the stream track, the median SB$_r$ (in mag$\,$arcsec$^{-2}$) value is retrieved and plotted as a coloured dot on top of the image of the stream. \textit{Middle}: Median SB$_r$ (in mag$\,$arcsec$^{-2}$) as a function of $\theta$ (in rad). The errorbars correspond to the standard error on the mean in each $\theta$ bin. The black dashed line corresponds to the median SB value of the whole stream. \textit{Right}: Width (in kpc) of the stream as a function of $\theta$ (in rad). The errorbars are the width uncertainties from the Gaussian fit of the track. The black dashed line corresponds to the median width of the whole stream. Variations of SB and width are visible along the track, while the medians values are the most representative overall.}
    \label{fig:sb_variations_track}
\end{figure*}

\subsection{Comparison to literature streams catalogues}\label{sec:comparison_streams_catalogues}
In Section \ref{sec:discussion-streams_literature}, we compare the classification of our galaxy sample to catalogues of streams from the literature based on DESI-LS images, namely \cite{Martinez-Delgado_et_al_2023,Miro-Carretero_et_al_2023,Miro-Carretero_et_al_2024} (hereafter \citetalias{Martinez-Delgado_et_al_2023}, \citetalias{Miro-Carretero_et_al_2023}, and \citetalias{Miro-Carretero_et_al_2024}, respectively). We provide in Table \ref{tab:appendix-compare_streams} the galaxies identified as having streams by \citetalias{Martinez-Delgado_et_al_2023}, \citetalias{Miro-Carretero_et_al_2023}, and \citetalias{Miro-Carretero_et_al_2024} and we compare it to our classification, including the individual votes, the majority feature and the score. We illustrate in Figure \ref{fig:comparison_literature_streams} some galaxies for which our tidal feature classification differs from these literature streams catalogues. 
\begin{table*}
\tiny
\centering
\caption{Comparison between our tidal feature catalogue and the stream catalogues of \protect\citetalias{Martinez-Delgado_et_al_2023, Miro-Carretero_et_al_2023, Miro-Carretero_et_al_2024} using DESI-LS images for the 75 galaxies we have in common. (1): galaxy name, (2): galaxy's right ascension (in deg), (3): galaxy's declination (in deg), (4): individual votes from our tidal feature catalogue, (5): majority feature from our voting system, (6): confidence score on the majority feature from our voting system, (7): Reference of the stream identified in the literature.   }
\label{tab:appendix-compare_streams}
\begin{tabular}{ccccccc}
\hline
Name & RA & Dec & Votes & Majority feature & Score & Stream reference \\
(1) & (2) & (3) & (4) &(5) & (6) &(7) \\
\hline
ESO485-011 & 71.637 & -26.411 & {Curiosities: 2, No: 4, Streams: 1, Plumes: 1} & Curiosities & 0.25 & MC24 \\
NGC0171 & 9.340 & -19.934 & {No: 3, Spiral: 4, Shells: 1} & Spiral & 0.50 & MD23 \\
NGC1136\_GROUP & 42.724 & -54.976 & {Plumes: 1, Streams: 1, No: 2, Tails: 1, Shells: 1, Spiral: 2} & Spiral & 0.25 & MC24 \\
ESO287-004 & 319.526 & -46.301 & {Curiosities: 1, No: 4, Spiral: 2, Streams: 1} & Spiral & 0.25 & MC24 \\
NGC4378 & 186.325 & 4.925 & {No: 3, Spiral: 3, Tails: 2} & Spiral & 0.37 & MC23 \\
NGC3614\_GROUP & 169.589 & 45.748 & {Streams: 2, No: 2, Spiral: 4} & Spiral & 0.50 & MC23 \\
NGC5604 & 216.178 & -3.212 & {No: 4, Tails: 1, Spiral: 2, Streams: 1} & Spiral & 0.25 & MC23 \\
NGC2543 & 123.241 & 36.255 & {Spiral: 3, Streams: 2, Tails: 2, No: 1} & Spiral & 0.38 & MC23 \\
NGC2460 & 119.218 & 60.349 & {Spiral: 4, Tails: 2, No: 1, Streams: 1} & Spiral & 0.50 & MC23 \\
NGC4793\_GROUP & 193.669 & 28.939 & {Spiral: 5, No: 2, Streams: 1} & Spiral & 0.62 & MC23 \\
PGC127531 & 48.731 & -62.987 & {No: 2} & No & 1.00 & MC24 \\
NGC4799 & 193.815 & 2.897 & {No: 2} & No & 1.00 & MC23 \\
NGC2701 & 134.774 & 53.772 & {No: 2} & No & 1.00 & MC23 \\
IC1904 & 48.753 & -30.708 & {No: 2} & No & 1.00 & MC24 \\
ESO243-006 & 12.988 & -43.476 & {No: 2} & No & 1.00 & MC24 \\
ESO081-008 & 35.403 & -64.610 & {Curiosities: 1, No: 5, Disturbed: 1, Spiral: 1} & No & 0.62 & MC24 \\
ESO242-007 & 6.093 & -45.509 & {No: 2} & No & 1.00 & MC24 \\
PGC009063 & 35.765 & -1.749 & {No: 2} & No & 1.00 & MC24 \\
PGC128506 & 19.274 & -55.959 & {No: 2} & No & 1.00 & MC24 \\
PGC007743\_GROUP & 30.555 & -6.079 & {No: 2} & No & 1.00 & MC24 \\
NGC7400 & 343.587 & -45.347 & {No: 2} & No & 1.00 & MC24 \\
PGC452979 & 60.478 & -51.753 & {No: 2} & No & 1.00 & MC24 \\
PGC134111 & 62.889 & -27.275 & {Shells: 1, No: 5, Streams: 1, Disturbed: 1} & No & 0.62 & MC24 \\
ESO483-012 & 62.594 & -23.617 & {Disturbed: 1, No: 4, Plumes: 2, Spiral: 1} & Plumes & 0.25 & MC24 \\
PGC199568 & 44.641 & -33.836 & {Plumes: 3, No: 3, Streams: 2} & Plumes & 0.37 & MC24 \\
NGC1209 & 46.513 & -15.611 & {Plumes: 2, Streams: 1, No: 2, Disturbed: 2, Shells: 1} & Plumes & 0.25 & MC23 \\
ESO287-051 & 325.494 & -44.091 & {Plumes: 3, Streams: 1, No: 2, Shells: 1, Disturbed: 1} & Plumes & 0.38 & MC24 \\
NGC0681\_GROUP & 27.295 & -10.427 & {Shells: 8, No: 1} & Shells & 0.89 & MD23 \\
NGC5971 & 233.904 & 56.462 & {Shells: 3, Streams: 2, No: 2, Spiral: 1} & Shells & 0.38 & MD23 \\
PGC597851 & 85.889 & -39.471 & {Shells: 6, No: 2} & Shells & 0.75 & MC24 \\
NGC4390 & 186.461 & 10.459 & {Shells: 6, No: 2, Spiral: 1} & Shells & 0.67 & MD23 \\
IC1833 & 40.411 & -28.171 & {Shells: 4, Streams: 3, No: 1} & Shells & 0.50 & MC24 \\
NGC5631\_GROUP & 216.639 & 56.583 & {Shells: 5, Streams: 2, No: 2} & Shells & 0.56 & MC23 \\
NGC5493 & 212.872 & -5.044 & {Shells: 4, Streams: 1, No: 3} & Shells & 0.50 & MC23 \\
ESO476-004 & 20.281 & -26.726 & {Shells: 3, No: 4, Streams: 2} & Shells & 0.34 & MC24 \\
PGC069613 & 341.042 & -57.939 & {Shells: 8, Streams: 1} & Shells & 0.89 & MC24 \\
NGC0922 & 36.269 & -24.788 & {Shells: 3, Streams: 2, No: 2, Spiral: 1} & Shells & 0.38 & MC24 \\
PGC131085 & 32.639 & -37.344 & {Shells: 7, Disturbed: 1, Streams: 1} & Shells & 0.78 & MC24 \\
NGC0788 & 30.277 & -6.816 & {Shells: 5, Streams: 2, No: 1} & Shells & 0.63 & MD23 \\
NGC0636 & 24.777 & -7.513 & {Shells: 6, Streams: 2, No: 1} & Shells & 0.67 & MC23 \\
NGC1076 & 40.872 & -14.754 & {Shells: 4, Streams: 1, No: 1, Plumes: 1, Disturbed: 1, Tails: 1} & Shells & 0.45 & MD23 \\
NGC0823 & 31.834 & -25.442 & {Shells: 4, Streams: 4, No: 1} & Streams & 0.45 & MC24 \\
NGC1121 & 42.663 & -1.734 & {Streams: 6, No: 2, Tails: 1} & Streams & 0.67 & MC24 \\
ESO340-043 & 308.770 & -41.684 & {Streams: 5, No: 2, Shells: 2} & Streams & 0.56 & MC24 \\
ESO114-001 & 24.419 & -60.865 & {Tails: 2, Streams: 2, No: 1, Plumes: 1, Spiral: 2} & Streams & 0.26 & MC24 \\
ESO147-006 & 340.828 & -59.414 & {Tails: 1, No: 4, Spiral: 1, Streams: 2} & Streams & 0.25 & MC24 \\
ESO160-002\_GROUP & 87.812 & -53.575 & {Streams: 5, No: 1, Plumes: 1, Tails: 1} & Streams & 0.63 & MC24 \\
ESO476-008\_GROUP & 21.641 & -23.226 & {Tails: 1, Streams: 3, Spiral: 3, No: 1} & Streams & 0.38 & MC24 \\
NGC7506 & 347.921 & -2.160 & {Shells: 3, Streams: 5, Disturbed: 1} & Streams & 0.56 & MC24 \\
ESO287-046 & 325.002 & -44.095 & {Streams: 4, Spiral: 3, No: 1} & Streams & 0.50 & MC24 \\
IC1816 & 37.963 & -36.672 & {Streams: 4, No: 1, Disturbed: 1, Spiral: 1, Shells: 1} & Streams & 0.50 & MC24 \\
IC1657\_GROUP & 18.529 & -32.651 & {Streams: 7, No: 1, Tails: 1} & Streams & 0.78 & MC24 \\
NGC1309 & 50.527 & -15.400 & {Shells: 2, Streams: 3, No: 2, Tails: 1} & Streams & 0.38 & MC23 \\
NGC7721 & 354.703 & -6.518 & {Streams: 4, No: 3, Spiral: 1} & Streams & 0.50 & MC23 \\
IC0174 & 29.067 & 3.762 & {Tails: 1, Streams: 4, Disturbed: 2, Shells: 2} & Streams & 0.44 & MD23 \\
UGC06023 & 163.587 & 27.240 & {Shells: 1, Streams: 4, No: 2, Plumes: 1, Tails: 1} & Streams & 0.45 & MD23 \\
UGC08717 & 206.842 & 30.334 & {Streams: 6, No: 2, Tails: 1} & Streams & 0.67 & MD23 \\
NGC1309 & 50.527 & -15.400 & {Shells: 2, Streams: 3, No: 2, Tails: 1} & Streams & 0.38 & MD23 \\
NGC3131\_GROUP & 152.152 & 18.231 & {Streams: 4, No: 1, Tails: 2, Plumes: 1} & Streams & 0.50 & MD23 \\
NGC3689 & 172.046 & 25.661 & {Streams: 6, No: 2} & Streams & 0.75 & MD23 \\
PGC1001085 & 187.189 & -8.642 & {Streams: 7, No: 1, Tails: 1} & Streams & 0.78 & MD23 \\
NGC5750 & 221.546 & -0.223 & {Shells: 1, Streams: 3, No: 2, Disturbed: 1, Spiral: 1} & Streams & 0.38 & MC23 \\
NGC5297\_GROUP & 206.599 & 43.872 & {Spiral: 1, Tails: 1, Streams: 4, No: 1, Disturbed: 1} & Streams & 0.50 & MC23 \\
NGC3689 & 172.046 & 25.661 & {Streams: 6, No: 2} & Streams & 0.75 & MC23 \\
NGC5812\_GROUP & 225.232 & -7.457 & {Streams: 6, No: 1, Tails: 2} & Streams & 0.67 & MC23 \\
NGC2782 & 138.521 & 40.114 & {Shells: 4, Streams: 4, Tails: 1} & Streams & 0.44 & MC23 \\
NGC2648\_GROUP & 130.666 & 14.286 & {Plumes: 2, No: 3, Tails: 1, Streams: 2} & Streams & 0.25 & MC23 \\
IC0160 & 26.623 & -13.248 & {Tails: 2, Streams: 3, No: 1, Plumes: 1, Disturbed: 1} & Streams & 0.38 & MD23 \\
NGC1079 & 40.935 & -29.003 & {Curiosities: 2, Streams: 2, No: 1, Tails: 2, Spiral: 1} & Streams & 0.26 & MC23 \\
NGC0259 & 12.014 & -2.775 & {Streams: 3, No: 3, Disturbed: 1, Tails: 1} & Streams & 0.37 & MD23 \\
NGC0577 & 22.670 & -1.994 & {Streams: 7, Shells: 1} & Streams & 0.88 & MD23 \\
NGC4385 & 186.428 & 0.573 & {Tails: 5, Plumes: 1, Streams: 2} & Tails & 0.62 & MD23 \\
NGC0095 & 5.556 & 10.492 & {Tails: 2, No: 3, Streams: 1, Spiral: 2} & Tails & 0.25 & MD23 \\
PGC007995 & 31.453 & -5.294 & {Plumes: 2, Streams: 2, Tails: 3, Disturbed: 1} & Tails & 0.38 & MD23 \\
UGC04132 & 119.804 & 32.915 & {Tails: 4, No: 1, Streams: 2, Plumes: 1} & Tails & 0.50 & MD23 \\
PGC007995 & 31.453 & -5.294 & {Plumes: 2, Streams: 2, Tails: 3, Disturbed: 1} & Tails & 0.38 & MC24 \\
NGC4750 & 192.530 & 72.874 & {No: 4, Streams: 1, Disturbed: 1, Tails: 2} & Tails & 0.25 & MC23 \\
ESO549-023 & 57.241 & -22.131 & {Spiral: 3, No: 2, Tails: 3} & Tails & 0.38 & MC24 \\
\hline
\end{tabular}
\end{table*}

\begin{figure*}
    \centering
    \includegraphics[width=\linewidth]{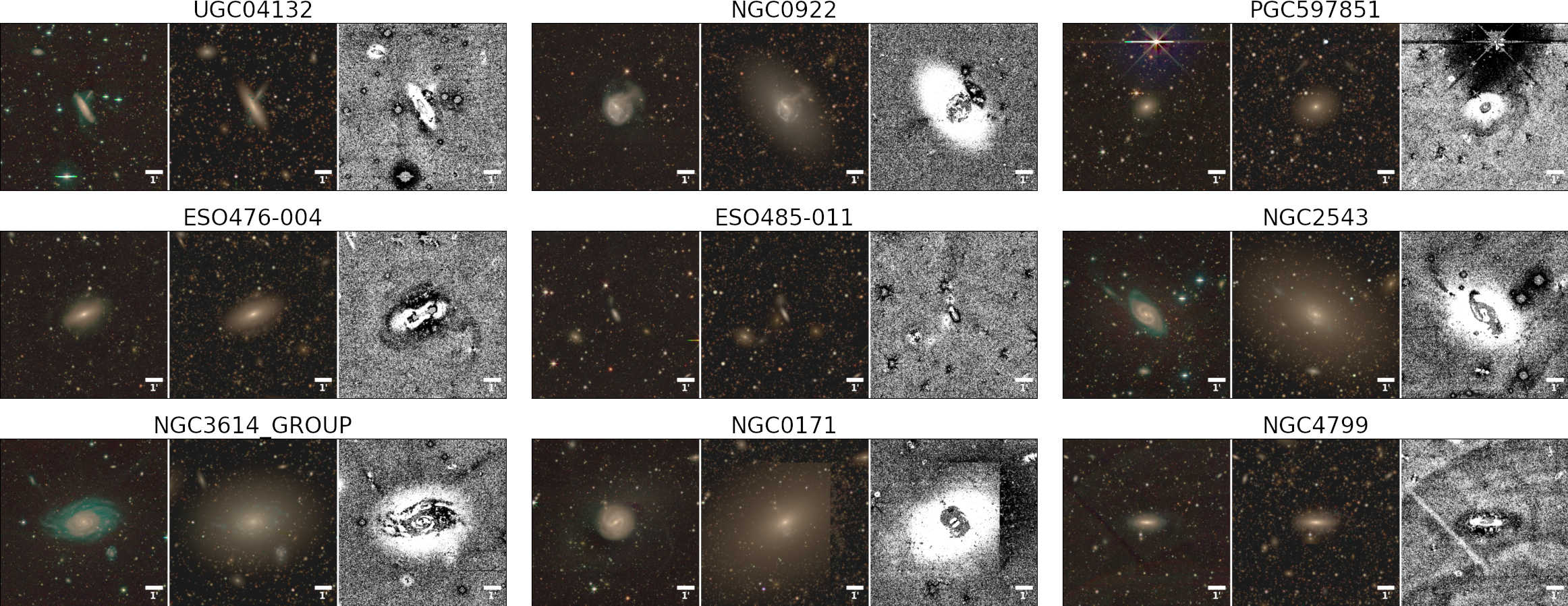}
    \caption{Comparison of a few galaxies identified as having streams in the literature \citetalias{Martinez-Delgado_et_al_2023,Miro-Carretero_et_al_2023, Miro-Carretero_et_al_2024} to the classifications in our tidal feature catalogue. Three images are displayed for each galaxy, from left to right: original $grz$ data image, \texttt{Tractor} $grz$ model image, and residual image (median of $g,r,z$-band residuals images, smoothed and stretched for better visualisation of faint features). A scalebar of $1'$ is indicated at the bottom of each panel. North is up, East is left.
    We classified UGC04132 as having tails, NGC0922 as an umbrella, PGC597851 as having inner shells (although a stream is also visible). For ESO0476-004 we have three votes for shells and two for streams, hence it was classified as shells. ESO485-011 has a dwarf galaxy we classified as curiosity. NGC2543 and NGC3614 do show streams, but they were classified as grand-design spirals due to having a majority of spiral versus streams votes. NGC0171 and NGC4799 do have a long stream but the presence of large artefacts polluting the residual images and our initial smaller field-of-view prevented us from detecting them.    }
    \label{fig:comparison_literature_streams}
\end{figure*}

\section{Photometry from original and residual images}\label{section:appendix-photometry_coadd_vs_residual}
We investigate the differences between the photometry of the stream estimated from the original images using a standard method versus from the residual images. For the original images, we download the DECaLS images in $g,r,z$ bands at a pixel size of 0.262$\arcsec$ from the DESI Legacy survey files\footnote{\href{https://portal.nersc.gov/cfs/cosmo/data/legacysurvey/dr10/}{DESI Legacy survey files}}. For each of our 35 galaxies, we retrieve all the images in a field-of-view large enough around the galaxy, combine them using SWarp and crop the result to be centred on the object. We then perform source extraction and masking using SEP \citep{Barbary_et_al_2016}, the Python implementation of SExtractor \citep{Bertin_and_Arnouts_1996}, with a detection threshold of 2.5 and a background mesh size of 128$\times$128 pixels. We then apply the stream segmentation mask from \texttt{Jafar}. We retrieve the values of the flux, surface brightness and colours inside the mask. We compare these values to the ones obtained directly from the residual images (as described in Section \ref{section:method-segmentation}). 

Figure \ref{fig:appendix-photometry_coadd_vs_residual} shows the comparison between the median surface brightness of the  streams in $g,r,z$-band computed from the original coadded images SB$_{\textrm{coadd}}$ and from the residual images SB$_{\textrm{resid}}$. We can see that the order of magnitudes of the SB values are similar. The residual photometry is slightly deeper than the one from the original coadded images. The median (mean) difference between SB$_{\textrm{coadd}}$ and SB$_{\textrm{resid}}$ is 0.07 mag$\,$arcsec$^{-2}$ (0.14) in the $g$-band, 0.09 mag$\,$arcsec$^{-2}$ (0.19) in the $r$-band and 0.09 mag$\,$arcsec$^{-2}$ (0.15) in the $z$-band. The slightly deeper values of the residual images were expected as they are supposed to be clean from any source of light contamination. This test supports our approach to use the residuals for the photometry of the streams.
\begin{figure*}
    \centering
    \includegraphics[width=0.7\linewidth]{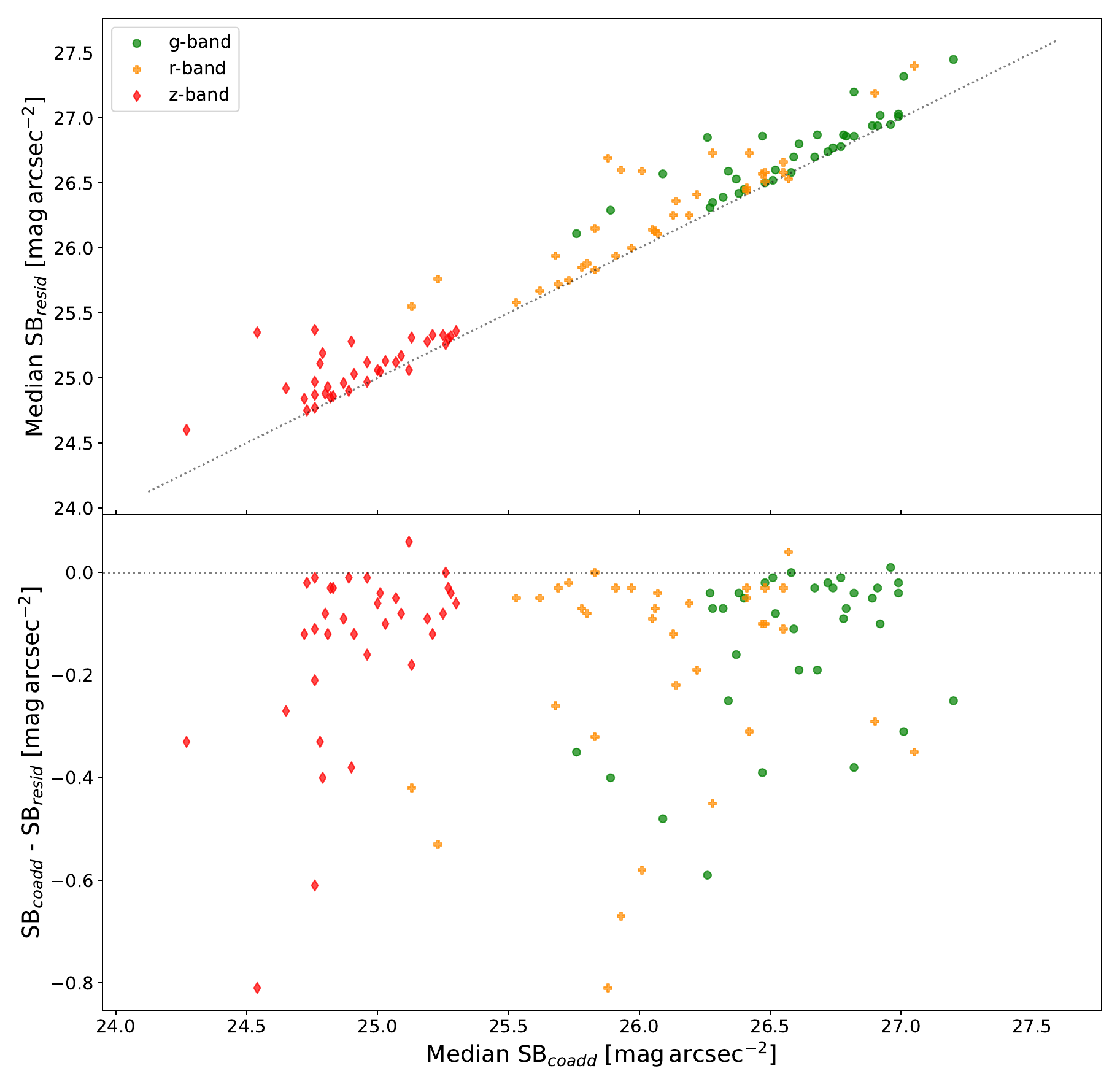}
    \caption{Comparison of the median surface brightness (in mag$\,$arcsec$^{-2}$) of streams in $g$- (green circles), $r$- (yellow crosses) and $z$-band (red diamonds) computed from the original coadded images SB$_{\textrm{coadd}}$ and from the residual images SB$_{\textrm{resid}}$. \textit{Top}: median SB$_{\textrm{resid}}$ as a function of the median SB$_{\textrm{coadd}}$. The black dotted line is the 1:1 line. \textit{Bottom}: difference between the median SB$_{\textrm{coadd}}$ and SB$_{\textrm{resid}}$ as a function of the median SB$_{\textrm{coadd}}$. The black dotted line represents a difference of zero. The residual photometry is slightly deeper than for the original coadded images. }
    \label{fig:appendix-photometry_coadd_vs_residual}
\end{figure*}

\bsp	
\label{lastpage}
\end{document}